\documentclass[prd,aps,superscriptaddress,nofootinbib,showpacs,twocolumn]{revtex4}

\usepackage[intlimits]{amsmath}
\usepackage{amsfonts}
\usepackage{amssymb,amscd}
\usepackage{graphicx}
\usepackage{placeins}      

\graphicspath{{./Figs/}}

\newcommand{\One}{1\kern-4.5pt1}

\newcommand{\om}{\omega}

\newcommand{\ncfg}{N_{\text{cfg}}}

\begin{document}

\title{Bayesian study of relativistic open and hidden charm in anisotropic lattice QCD}

\author{Aoife Kelly}
\affiliation{Department of Theoretical Physics, National University of
  Ireland Maynooth, Maynooth, Co Kildare, Ireland}
\author{Alexander~Rothkopf}
\affiliation{Institut f\"{u}r Theoretische Physik, Universit\"{a}t Heidelberg, Philosophenweg 12, 69120 Heidelberg, Germany}
\author{Jon-Ivar Skullerud}
\affiliation{Department of Theoretical Physics, National University of
  Ireland Maynooth, Maynooth, Co Kildare, Ireland}
\affiliation{School of Mathematics, Trinity College, Dublin 2, Ireland}

\date{\today}
\begin{abstract}
We present the first combined study of correlators and spectral
properties of charmonium and open-charm-mesons at finite temperature
using a fully relativistic lattice QCD approach. The QCD medium is
captured by second generation anisotropic $24^3\times N_\tau$ lattices
from the FASTSUM collaboration, including 2+1 flavors of clover
discretized quarks with $m_\pi\approx 380$MeV. Two Bayesian methods
are deployed to reconstruct the spectral functions, the recent BR
method as well as the Maximum Entropy Method with Fourier basis. We
take particular care to disentangle genuine in-medium effects from
method artifacts with the help of the reconstructed
correlator. Consistent with the direct inspection of correlators, we
observe no significant in-medium modification for $J/\Psi$ and
$\eta_c$ around the crossover, while the $\chi_c$ states on the other
hand show clear changes around the transition. At the highest
temperature, $T=352$MeV, $J/\Psi$ and $\eta_c$ also exhibit
discernible changes compared to the vacuum. For $D$ mesons around
$T_c$, no significant modifications are observed, but we find clear
indications that no bound state survives at the highest temperature of $T=352$MeV. Above $T_c$ we discover a significant difference between $D$ and $D^*$ mesons, the latter being much more strongly affected by the medium.
\end{abstract}

\maketitle

Heavy quarks and the mesons formed by binding these to pairs with either light (open heavy-flavor) or other heavy quarks 
(hidden heavy-flavor, heavy quarkonium) are an essential part of our toolset to probe the physics of the quark-gluon plasma in relativistic heavy-ion
collisions at RHIC and LHC. A close collaboration between experiment \cite{Andronic:2015wma} and 
theory \cite{Aarts:2016hap,Prino:2016cni} has emerged to shed light on the various physics aspects which need to be understood
in order to produce a comprehensive understanding of the heavy particle yields measured in experiment.

In the early stages of the collision, the production of heavy quark--antiquark pairs from partonic processes, such
as gluon splitting, needs to be described. Due to the large momentum transfers involved, QCD perturbation theory
provides important insights here (see \cite{Andronic:2015wma} for a more detailed overview). 

Subsequently the heavy quarks may propagate through 
the collision center, interacting strongly with the bulk matter and in turn lose a significant amount of their 
original kinetic energy via radiative or collisional processes. The physics of in-medium propagation and energy loss has
been investigated via a multitude of models often based on a Boltzmann equation \cite{Uphoff:2012gb} or, if further simplified, on a
Langevin or equivalently Fokker--Plank equation \cite{Ozvenchuk:2014rpa,Xu:2017obm}. The transport coefficients characterizing the evolution have been estimated either via
resummed perturbation theory \cite{CaronHuot:2007gq,Alberico:2013bza}, strong coupling holography \cite{Kovtun:2003wp,Horowitz:2012cf} or lattice QCD simulations \cite{Francis:2015daa}. Direct perturbative
analysis of energy loss in a thermal medium \cite{Djordjevic:2013xoa}, computations based on a
Dyson--Schwinger like T-matrix resummation \cite{Riek:2010py}, as well as modeling with microscopic kinetic equations \cite{Song:2015sfa} have provided further insightful results.

Just after the production of a quark--antiquark pair these two particles may actually form a quarkonium bound state.
Even in the absence of a QCD medium, this binding is a genuinely non-perturbative process,
and its theoretical description has benefitted greatly from the separation of scales between the heavy quark mass and other relevant
scales in the system, such as the (local) temperature or the intrinsic scale of quantum fluctuations in QCD, $\Lambda_{\rm QCD}$.
It allows the construction of non-relativistic field theories (EFT) \cite{Brambilla:2004jw}, which provide a simplified language of the 
physics relevant to heavy quark binding and quarkonium production. In turn the in-medium modification of heavy quarkonium properties can be related to
both screening effects in the thermal medium, captured in the Debye mass of the
system, as well as scattering effects, such as e.g. Landau damping \cite{Laine:2006ns} or color singlet to
octet transitions \cite{Brambilla:2008cx}.

While for beauty quarks the scale separation is indeed pronounced and EFT approaches are well established, the situation for 
charm quarks is less straight forward. Estimates of initial temperatures in the collision center from relativistic hydrodynamics
give values of $T_{\rm init}^{\rm LHC}\approx 600$~MeV, the charm quark mass being barely double this. Thus for charmonium
and its open heavy-flavor cousins $D$ and $D^*$ a genuinely relativistic description is usually preferred. Standard lattice QCD
simulations, which routinely provide vital insight into the properties of light hadrons at zero and finite temperature, can be directly applied
to the charm sector too, as the required lattice spacings, while small, remain computationally manageable.

We may ask what insight can be gained from performing a thermal lattice simulation of charmonium and open-charm mesons?
As the simulated system is in genuine equilibrium, we may learn about the probability of heavy-quark binding
being able to sustain bound states, either open or hidden, in the late stages of the collision. And since in most models 
the production of open heavy-flavor occurs via a hadronization prescription at freezeout, from the theory point of view
this is just the relevant regime to look at. The success of the statistical model of hadronization to predict charmonium
yields from RHIC to LHC energies \cite{Andronic:2008gm} further supports a focus on the late stages of the collision 
to understand not only open, but also hidden charm production.

Considering the binding properties of both quarkonium and open-charm
together is valuable, since the latter system provides the natural baseline for the measurements of suppression in the former.
In model computations in a hot pion medium e.g., D meson modifications were found to strongly influence $J/\psi$ production during
the expansion of the fireball \cite{Fuchs:2004fh}.

Recent measurements of a finite D meson flow \cite{Abelev:2013lca} and
even $J/\Psi$ \cite{ALICE:2013xna,Acharya:2017tgv} flow have unambiguously
shown that the charm quarks at the LHC are in at least partial kinetic
equilibration with the surrounding medium. In turn the charm quarks
may have already shed most of their knowledge about the earlier stages
of the collision and follow the bulk in its expansion. Hence 
a thermal description seems to be a good starting point.

Thermal lattice simulations obviously cannot tell us about possible cold nuclear matter effects that may affect the
production of heavy mesons. Recent measurements by the ALICE Collaboration of the nuclear modification 
factor $R_{pPb}$ for D mesons in proton--lead collisions at the LHC have however shown that the strong 
suppression of heavy quarks they observed at transverse momenta above $p_T>2$GeV is not due to cold nuclear matter effects, but to the strong
coupling of the charm quarks to the QGP medium \cite{Abelev:2014hha,De:2016wmf}.

What we can measure on the lattice are meson correlation functions, which are distinct
from the actual particle correlation functions determined in experiment. They provide us with
 insight on global in-medium modification if one considers e.g. the appropriate ratios of vacuum
 correlators with those at $T>0$. In order to extract information about the in-medium modification
 of individual meson states, e.g. the $J/\Psi$ or $D^*$, the spectral function encoded in the
 correlator needs to be computed via an unfolding procedure, which, as will be discussed, presents
 a formidable numerical challenge. Once we have access to the spectra, we can learn about
 medium induced mass shifts and thermal broadening, which also for heavy flavors 
 can be translated to changes in the production yields and thus to measurable observables (see e.g. 
 \cite{Burnier:2015tda,Burnier:2016kqm}).

The study of quarkonium spectral functions in a thermal medium is an active field of research with several different
strategies present on the market. On the one hand there are efforts that have successfully used 
non-relativistic effective field theories to extract bottomonium spectral information in-medium 
\cite{Aarts:2010ek,Aarts:2011sm,Aarts:2012ka,Aarts:2013kaa,Aarts:2014cda,Kim:2014iga}. 
Recently also charmonium at finite temperature has been considered in this fashion \cite{Kim:2015rdi,Kim:2017aio}. 
As a plus, the non-relativistic setting allows one to access more relevant datapoints on the
lattice, since the corresponding correlators are not symmetric in Euclidean time. At the same time the effective field 
theory approach introduces additional uncertainties, related e.g. to the (non-perturbative) matching to QCD
and the absence of a naive continuum limit. Non-relativistic quarkonium spectra have also recently been
computed based on new results on the complex lattice QCD heavy-quark potential \cite{Burnier:2014ssa,Burnier:2015tda,Burnier:2016kqm}.

On the other hand there are several groups that aim at computing quarkonium in a fully relativistic framework.
Those who focus on beauty require extremely finely spaced lattices, rendering the computation feasible
only in the quenched approximation \cite{Ding:2017rty,Shu:2015tva}, which has also been used recently to study 
charmonium in \cite{Ikeda:2016czj}. The charm mass however is already light enough that it is possible to consider 
charmonium also in a full QCD medium, as has been done previously in Refs.~\cite{Aarts:2007pk,Borsanyi:2014vka}.

In contrast to quarkonium, so far there have been only very few lattice studies carried out on in-medium open charm mesons 
\cite{Bazavov:2014yba,Bazavov:2014cta,Mukherjee:2015mxc}, as they require a relativistic formulation
at least for the light quark partner. Previous works used spatial correlators and cumulants 
to gain insight into the survival and melting patterns of $D$ mesons around the transition temperature $T_c$.
In contrast to spatial correlation functions, which are related to the spectral function of the system via a double
integral transform, the temporal correlators connect via a single integral transform. Thus spectral information manifests
itself more directly on the latter than the former. In the present
study we will, for the first time, present results for
the temporal correlators and spectral functions of open charm mesons. Preliminary results have been presented
in \cite{Kelly:2016apt,Kelly:2017opi}. Let us also note an alternative QCD based
approach to thermal in-medium D-meson properties, the sum-rule approach \cite{Hilger:2008jg, Wang:2015uya,Suzuki:2015est}.

In the following section we briefly discuss the lattice QCD setup (Sec.~\ref{lattice}) and introduce
the Bayesian approach to spectral function reconstruction in (Sec.~\ref{Bayes}). Our presentation
of numerical results start at $T=0$ in Sec.~\ref{zeroT}. The in-medium correlator and
spectral analysis for charmonium is contained in Sec.~\ref{QuarkoniumFiniteT} and for open charm in Sec.~\ref{OpenCharmFiniteT}.
We close with a summary and conclusion in Sec.~\ref{summary}.

\section{Methods}
\label{methods}
\subsection{Lattice Setup}
\label{lattice}

In this study we deploy lattices generated by the FASTSUM collaboration to describe
the QCD medium in which the open and hidden heavy flavor mesons are immersed.
These second generation ensembles ~\cite{Aarts:2014cda,Aarts:2014nba} feature 
2+1 flavors of anisotropic clover fermions and a mean-field improved anisotropic
Szymanzik gauge action. With an anisotropy parameter of $\xi=a_s/a_\tau=3.5$
the spatial lattice spacing is $a_s=0.123$~fm. While the strange quark mass
is tuned to its physical value, the pion mass takes on a value
of $m_\pi\approx380$~MeV. The temperature is changed in the fixed scale manner, by
varying the number of points in the Euclidean time direction. The available temperatures
and the corresponding number of configurations are listed in
table~\ref{tab:lattices}.  The configurations were sampled every 10
HMC trajectories.  The integrated Polyakov loop autocorrelation time
for the ensembles above $T_c$ was about 20 trajectories, while the
plaquette autocorrelation time below $T_c$ was found to be about 30
trajectories.  The pseudocritical temperature was determined from the
inflection point of the Polyakov loop to be $T_c=185(4)$~MeV, for more
details see Ref.~\cite{Aarts:2014nba}.

Our finite temperature simulations utilize the same bare action parameters as those 
by the Hadron Spectrum Collaboration~\cite{Edwards:2008ja}, who also kindly provided
the zero temperature ensemble used for calibration at $N_\tau=128$. The relativistic charm quark 
action and its parameters are taken over from the vacuum study~\cite{Liu:2012ze, Moir:2013ub} and
both the configurations and the correlators were generated using the
Chroma software package~\cite{Edwards:2004sx}.
\begin{table}
\centering
\begin{tabular}{|ccccr|}\hline
$N_s$&   $N_\tau$ & $T$ (MeV) & $T/T_c$ & $\ncfg$ \\\hline
16 & 128 & \,44 & 0.24 &  500 \\
24 &  40 & 141  & 0.76 &  500 \\
24 &  36 & 156  & 0.84 &  500\\
24 &  32 & 176  & 0.95 & 1000 \\
24 &  28 & 201  & 1.09 & 1000 \\
24 &  24 & 235  & 1.27 & 1000 \\
24 &  20 & 281  & 1.52 & 1000 \\
24 &  16 & 352  & 1.90 & 1000 \\ \hline
\end{tabular}
\caption{Lattice volumes $N_s^3\times N_\tau$, temperatures $T$ and
  number of configurations $\ncfg$ used in this work.  The
  configurations were separated by 10 units of molecular dynamics
  time.  The pseudocritical temperature $T_c$ was determined from the
  inflection point of the Polyakov loop~\cite{Aarts:2014nba}.}
\label{tab:lattices}
\end{table}

The Euclidean correlators $G(\tau;T)$ we compute in the lattice simulation are
related to the spectral function $\rho(\om;T)$ via the integral relation
\begin{align}
\label{eq:spectral}&G(\tau;T) =
\int\rho(\omega;T)K(\tau,\omega;T)d\omega\,,\\
\intertext{where}
&K(\tau,\omega;T) = \frac{\cosh[\omega(\tau-1/2T)]}{\sinh(\omega/2T)}\,.
\end{align}
Information about the hadronic states, including their energies, widths and
thresholds in the medium, is encoded in the peak and continuum structure
of $\rho$. Our strategy to extract spectral functions from Euclidean time correlators
via Bayesian inference is detailed in the next subsection.  We note
that the correlators have not been renormalised, and we can therefore
only determine the shape of the spectral function, and not the
absolute value, which would be required to determine for example
transport coefficients.

We may already learn about the overall in-medium modification by considering
the changes induced by thermal effects in the correlators themselves. Comparing
correlators at different temperatures in the relativistic theory requires us to disentangle
two effects. From Eq.~\eqref{eq:spectral} we see that besides the spectral function, also the kernel
carries a temperature dependence. Hence we cannot simply truncate the low temperature
correlator and divide it with the one at higher temperature, as is done in non-relativistic 
settings, where the kernel does not contain a temperature dependence.

Instead we should consider the so-called reconstructed correlator
\begin{equation}
 G_r(\tau;T,T_r) = \int_0^\infty\rho(\om;T_r)K(\tau,\om,T)d\om\,,
\end{equation}
where $T_r$ denotes a reference temperature at which the spectral function can
be reliably constructed and which is usually chosen to be the lowest available
temperature. Then we can obtain the Euclidean correlator that ensues if the
low temperature spectral function were present at the higher temperature. I.e.
in the absence of in-medium modification we have
$\rho(\om;T)=\rho(\om;T_r)$, and hence
$G(\tau;T)=G_r(\tau;T,T_r)$. The converse is not necessarily true, since
the loss of information occurring in Eq.~\eqref{eq:spectral} may lead to a situation where
multiple changes in the underlying spectral function compensate each other 
in the Euclidean correlator. 

We will use the reconstructed correlator in the following to form direct ratios 
with the actual finite temperature correlators and as a benchmark for the systematic uncertainties
of the spectral reconstruction algorithm. It is important in this regard to note that the
reconstructed correlator can actually be computed directly from the
underlying correlator $G(\tau;T_r)$ without the need for a spectral reconstruction~\cite{Ding:2012sp}.  
Based on the following identity for hyperbolic functions
\begin{equation}
\frac{\cosh\big[\om(\tau-N/2)\big]}{\sinh(\om N/2)}
 = \sum_{n=0}^{m-1}\frac{\cosh\big[\om(\tau+nN+mN/2)\big]}{\sinh(\om mN/2)}\,,
\end{equation}
where
\begin{equation}
 T=\frac{1}{a_\tau N},\,\, T_r = \frac{1}{a_\tau N_r},\quad 
\frac{N_r}{N} = m \in \mathbb{N}\,,
\end{equation}
one finds
\begin{equation}
G_r(\tau;T,T_r) = \sum_{n=0}^{m-1}G(\tau+nN,T_r)\,.
\label{eq:Grec-direct}
\end{equation}

\subsection{Bayesian Spectral reconstruction}
\label{Bayes}

The extraction of spectral functions from \eqref{eq:spectral}
represents an ill-posed inverse problem. The data obtained from
lattice simulations are stochastic estimates of the correlator
$G(\tau)$ with finite precision $\Delta G/G$, evaluated at $N_\tau$
discrete points $G(\tau_j=a_t j)\equiv G_j$. For a numerical application we also need to discretize the spectrum using $N_\omega$ points,
\begin{align}
G_i =\sum_{l=1}^{N_\omega} \rho_l K_{il} \Delta \omega_l\,,\label{eq:numconv}
\end{align}
in anticipation of the many different structure we may find: narrow bound state peaks, wide resonances, continuous structures above the threshold and even a transport peak at low frequencies. This necessitates the use of $N_\omega \gg N_\tau$, so that attempting a naive $\chi^2$ fit of the parameters $\rho_l$ to $G_i$ leads to an infinite number of solutions reproducing the simulation data within their errors.

In this study we deploy Bayesian inference \cite{Jarrell:1996,Bishop:2007} to give meaning to the ill-posed inversion task. This well established branch of mathematics allows us to systematically incorporate additional, so called prior information about the spectrum into the reconstruction task. This in turn regularizes the naive $\chi^2$ fit and leads to a unique Bayesian answer. The starting point is Bayes' theorem,
\begin{align}
 P[\rho|G,I]= \frac{P[G|\rho,I]P[\rho|I]}{P[G|I]}\,,
\end{align}
which states that the probability for a test function $\rho$ to be the correct spectrum given simulation data and prior information is proportional to two $\rho$-dependent terms. The first, $P[G|\rho,I]={\rm exp}[-L]$, is called the likelihood and encodes how the data are generated. In the case of stochastically sampled data, such as in lattice QCD, the likelihood may be expressed in terms of the quadratic distance between the simulation data $G_i$ and the Euclidean correlator $G^\rho_i$ arising from the current test function $\rho$ inserted in \eqref{eq:numconv}
\begin{align}
 L[\rho]=\frac{1}{2}\sum_{ij}^{\ncfg}(G_i-G^\rho_i)C_{ij}(G_j-G^\rho_j)\,.
\end{align}
Here $C_{ij}$ denotes the usual covariance matrix. Note that in 
Monte Carlo simulations on the lattice individual correlators may not
be perfectly uncorrelated. This leads to residual autocorrelations, which reduces the effective number of available configurations $\ncfg$. A $\chi^2$ fit wishes to determine the maximum of the likelihood function, which however contains many degenerate extrema.

The second term in the numerator, the prior probability $P[\rho|I(m)]=\exp[\alpha S[\rho,m]]$, provides the necessary regularization of the likelihood in the Bayesian approach. It is conventionally formulated with a hyperparameter $\alpha$ which weights the influence of data and prior information and which has to be determined self-consistently. Prior information enters into the functional $S$ in two distinct ways. On the one hand the form of $S$ itself favors certain spectra. On the other hand $S$ depends on the so called default model $m$, which by definition is the correct spectrum in the absence of data. I.e., $m$ acts as the unique extremum of the regulator $S$.

In the end one carries out a numerical optimization on the posterior $P[\rho|G,I]$ to find the most probable spectrum given data and prior information
\begin{align}
 \left.\frac{\delta}{\delta \rho}P[\rho|G,I]\right|_{\rho=\rho^{\rm
     BR}}=0\,.
\label{BayesStationary}
\end{align}
Via the competition between $L$ and $S$ the reconstructed spectrum will be partially fixed by data and partially constrained by prior information.

In this study we will use two different implementations of the Bayesian strategy: the well known Maximum Entropy Method \cite{Asakawa:2000tr,Jakovac:2006sf,Nickel:2006mm,Rothkopf:2011}, as well as the more recent BR method \cite{Burnier:2013nla}. While both are Bayesian in nature they differ in the deployed prior functional, in the way how the hyperparameter $\alpha$ is handled and in the implementation of the optimization problem of Eq.~\eqref{BayesStationary}. Note that two Bayesian approaches in general will yield different outcomes as long $\Delta G/G$ and $N_t$ is finite. Only in the ``Bayesian continuum limit'' will both methods agree.

Based on arguments from two-dimensional image reconstruction the MEM proposes to use the Shannon--Jaynes entropy,
\begin{align}
 S_{\rm SJ}[\rho]=\sum_l\;\Big( \rho_l-m_l-\rho_l{\rm
   log}\Big[\frac{\rho_l}{m_l}\Big]\Big)\Delta \omega_l\,,
\end{align}
as regulator. Following the state of the art implementation by Bryan \cite{Bryan} we work with a restricted search space in order to reduce the computational burden of the optimization in Eq.~\eqref{BayesStationary}. While Bryan uses the $N_\tau$ dimensional SVD basis of the Kernel we here deploy a different set of basis functions, the so called Fourier basis, which has been shown to provide improved frequency independent resolution \cite{Rothkopf:2012vv}. The hyperparameter $\alpha$ is treated in the following fashion: One repeats the reconstruction based on $S_{\rm SJ}$ for many different values of $\alpha$ and then computes an approximation of the probability $P[\alpha|\rho,D,I]$, which is closely related to the evidence probability $P[D|I]$. The final spectrum is the average over all individual reconstructions weighted with $P[\alpha|\rho,D,I]$.

The second method we deploy is the BR method, which has been developed with one-dimensional reconstruction problems in mind. Combining besides positive definiteness a smoothness criterion and the requirement of independence of the units used, the BR method regulator reads
\begin{align}
 S_{\rm BR}[\rho]=\sum_l\;\Big( 1-\frac{\rho_l}{m_l}+{\rm
   log}\Big[\frac{\rho_l}{m_l}\Big]\Big)\Delta \omega_l\,.
\end{align}
The absence of the factor $\rho$ in front of the logarithm allows one
to analytically determine the $\alpha$-dependent normalization of the
corresponding prior probability $P[\rho|I]$. In turn it becomes
possible to marginalize the hyperparameter $\alpha$ \emph{a priori} by
assuming full ignorance of its values, i.e., $P[\alpha]=1$:
\begin{align}
 P[\rho|G,I(m)]=P[G|I]\int_0^\infty d\alpha P[\rho|m,\alpha]\,.
\end{align}
The optimization of Eq.~\eqref{BayesStationary} is then implemented with the integrated posterior $P[\rho|G,I(m)]$.

The numerical implementation for this study uses the MPFR library to
implement arbitrary precision datatypes, which in practice are set to
$512$ bits of precision. Due to the presence of the $\rho {\rm
  log}[\rho]$ term in $S_{\rm SJ}$ the MEM converges relatively slowly
compared to the BR method. Therefore in the latter, as is common
practice, we accept an extremum if Eq.~\eqref{BayesStationary} is satisfied to
an accuracy of $\Delta_{\rm MEM} =5\times10^{-8}$, while for the BR
method we use $\Delta_{\rm BR}=10^{-60}$.

The robustness of the outcome of a Bayesian reconstruction depends on both the precision of the input data $G_i$ and the reliability of the prior information $I$. To quantify the effect of the former we carry out a ten-bin Jackknife, where the reconstruction is repeated, each time with a different subset of simulation datasets removed. The dependence on the prior information can be assessed by repeating the reconstruction with different default models. As a proxy for both effects Bayesian methods provide an internal measure of robustness, which under certain assumptions is related to the curvature of the minimization functional $Q=\log[ P[\rho|G,I] ]$. In previous studies we have seen that this internal measure, possibly due to its reliance on assumptions, often underestimates the actual uncertainty of the reconstruction, as determined from the Jackknife and default model variation.

In the following we will use an equidistant frequency discretization
of $N_\omega^{\rm MEM}=1000$ and $N_\omega^{\rm BR}=3000$ for the
spectral reconstructions in the interval $\omega^{\rm num}\in[0,5]$. The reason why we can use a smaller number of points for the MEM is related to the fact that for a given number of datapoints the number of Fourier basis functions is fixed. As long as all features of the basis function are resolved, the result becomes virtually independent of the actual $N_\omega^{\rm MEM}$. The Euclidean times are taken to be $\tau=\{0,\dots,N_\tau\}$. Since the naive finite temperature kernel diverges at the origin we instead deploy the regularized variant $K(\omega,\tau)=\cosh[\omega(\tau-\beta/2)]/\cosh[\omega \beta/2]$, so that raw output of the reconstruction is $\rho/\operatorname{arctan}[\omega\beta/2]$, which we convert to $\rho/\omega$ afterward. This also means that the default model provided is the default model for $\rho/\operatorname{arctan}[\omega\beta/2]$. We will show reconstructions for the choice $m(\omega)=1$ but vary the default model both in amplitude $m_0=\{0.1,1\}$ and form $m(\omega)=m_0(1+\omega)^2$ to assess the residual influence of the default model.

\section{Numerical Results}

Before embarking on an investigation of the properties of open and hidden charm mesons at finite temperature, let us inspect first the situation at zero temperature, as obtained on the FASTSUM ensembles. Not only will the obtained vacuum results allow us to judge the accuracy of our computations but also provides us with the necessary baseline for interpreting the Bayesian spectral reconstructions at finite temperature. 

\subsection{Zero Temperature}
\label{zeroT}
\begin{figure}[th!]
\includegraphics[scale=0.4]{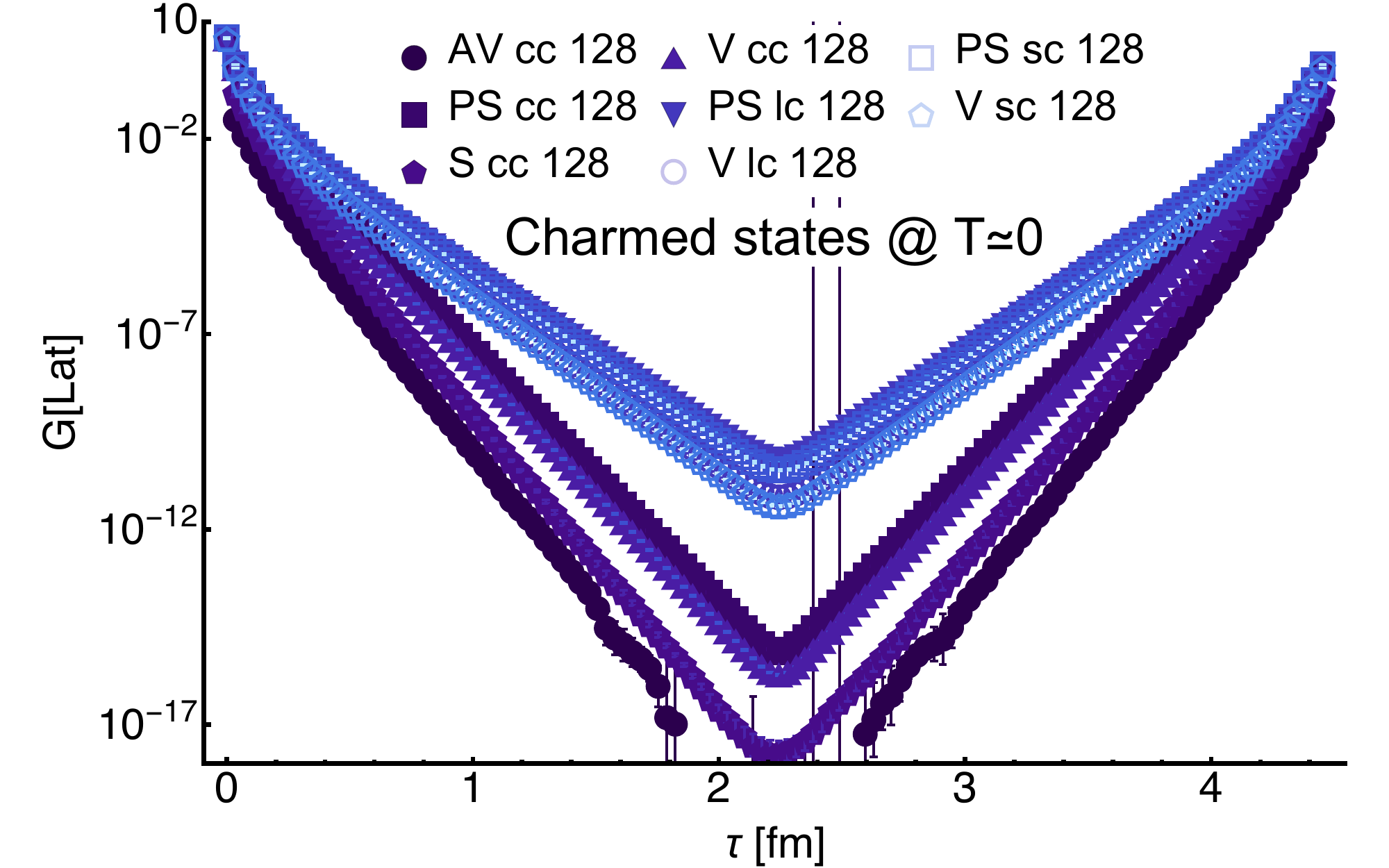}
\includegraphics[scale=0.4]{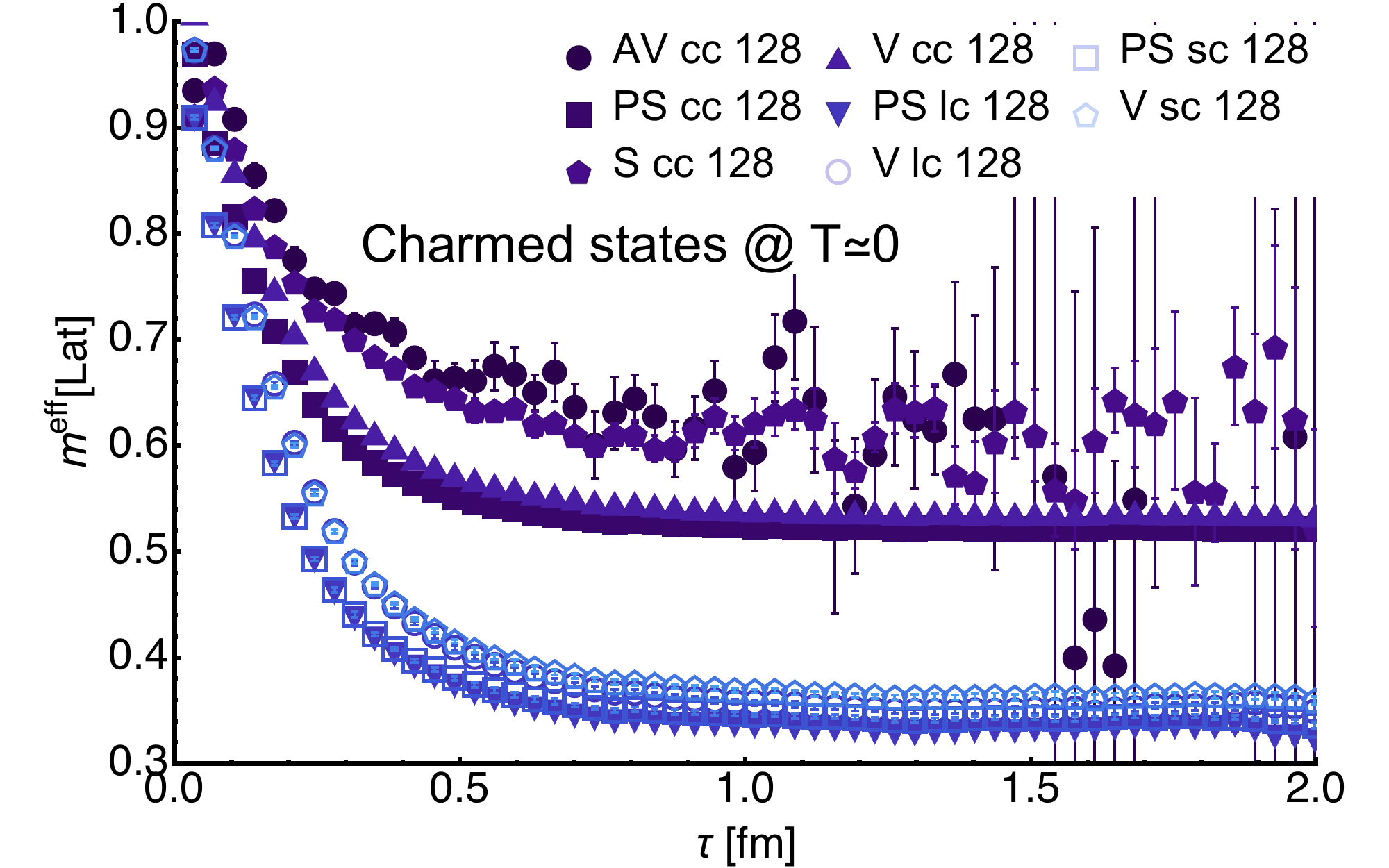}
\includegraphics[scale=0.4]{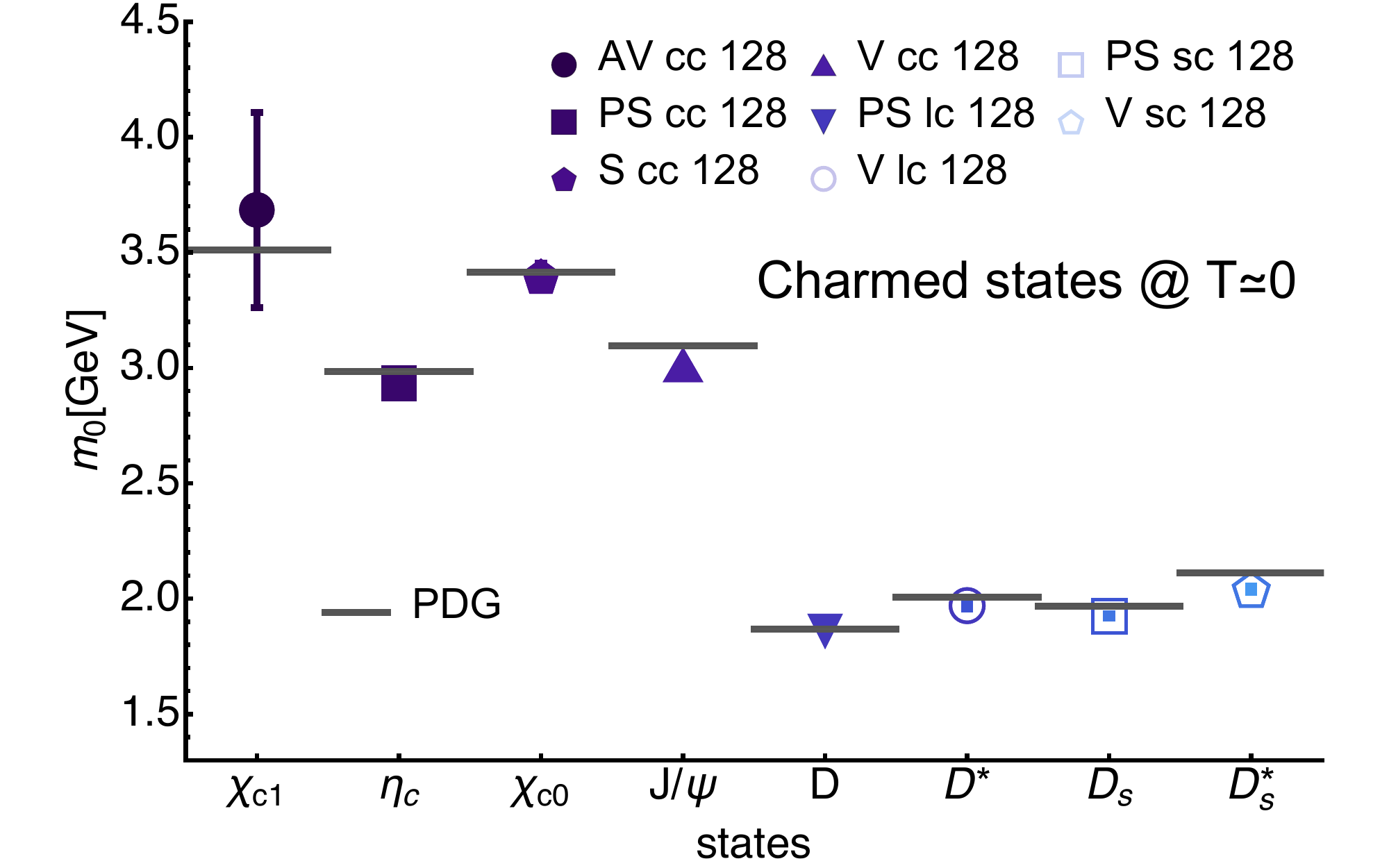}
\caption{(top) The raw Euclidean correlation functions for the states considered in this study. (middle) The corresponding time-dependent effective masses of the individual states. (bottom) Comparison of the asymptotic effective masses with the experimental values from the PDG (gray lines).}\label{Fig:ZeroTResults}
\end{figure}

We show the vacuum correlation functions at $N_\tau=128$ of all charmed states investigated in this study in the top panel of Fig.\ref{Fig:ZeroTResults}. The anticipated differences in the ground state masses manifest themselves visibly in the different slopes at intermediate Euclidean times. To access the value of the mass quantitatively we compute the so called effective mass 
\begin{align}
m^{\rm eff}(\tau)=\frac{1}{\tau}\log\Big[ G(\tau)/G(\tau+1) \Big],
\end{align}
using a ten-bin Jackknife resampling. From the middle panel of Fig.\ref{Fig:ZeroTResults} we see that it asymptotes to the ground state mass at late times. The higher the ground state mass, the lower the signal to noise ratio, i.e. in particular the scalar and axial-vector quarkonium channel still suffer from significant uncertainty. Fitting the late time behavior reveals the ground states masses listed in Tab.\ref{Tab:zeroTmass}.

\begin{table}[th!]
\centering
\begin{tabular}{|lllll|}\hline
Particle & $J^{PC}$  & $m$ (GeV) & HadSpec& PDG \\\hline
$\eta_c$ & $0^{-+}$ (PS) & $2.960(6)$ & 2.983* & 2.983 \\ 
$J/\Psi$ & $1^{--}$ (V) & $3.041(7)$ & 3.064 & 3.097 \\ 
$\chi_{c0}$ & $0^{++}$ (S) & $3.45(4)$ & 3.445 & 3.414 \\ 
$\chi_{c1}$ & $1^{++}$ (AV) & $3.9(6)$ & 3.505(1) & 3.511 \\ \hline
$D$ & $0^{-}$ (PS) & $1.88(1)$ & 1.895(1) & 1.868 \\ 
$D^*$ & $1^{-}$ (V) & $1.99(1)$ & 2.019(1) & 2.009 \\  \hline
$D_s$ & $0^{-}$ (PS) & $1.943(8)$ & 1.961(1) & 1.968 \\ 
$D^*_s$ & $1^{-}$ (V) & $2.06(1)$ & 2.081(5) & 2.112 \\ \hline
\end{tabular}
\caption{Vacuum ground state masses from the effective mass fit for
  all meson channels investigated in this study, together with the
  values determined by the HadSpec collaboration from the same
  ensemble \cite{Liu:2012ze,Moir:2013ub} and the experimental values from the
  Particle Data Group (PDG).  The $\eta_c$ mass was used by the
  HadSpec collaboration to set the charm quark mass.}
\label{Tab:zeroTmass}
\end{table}  

\begin{figure*}[th]
\centering
\includegraphics[scale=0.24]{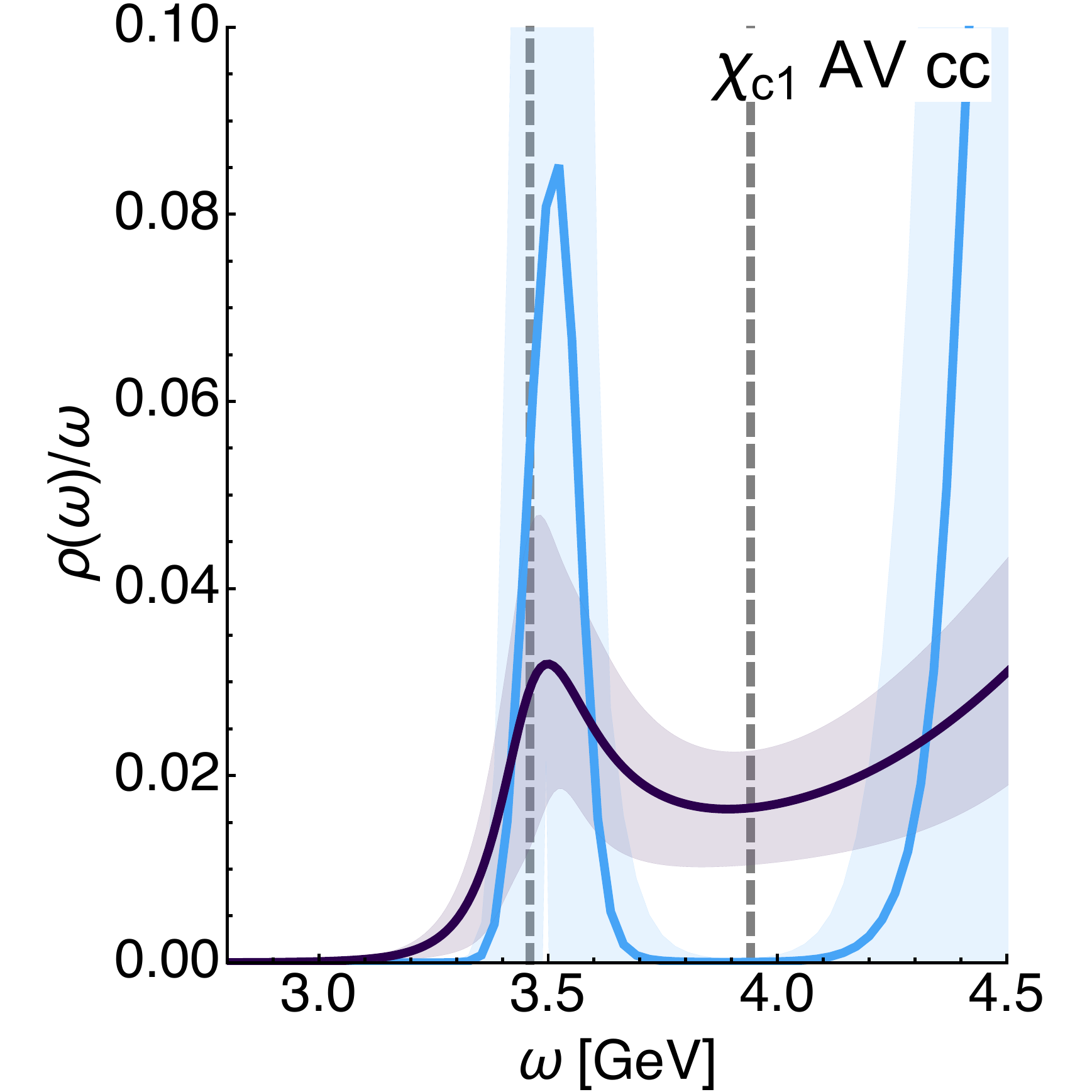}
\includegraphics[scale=0.24]{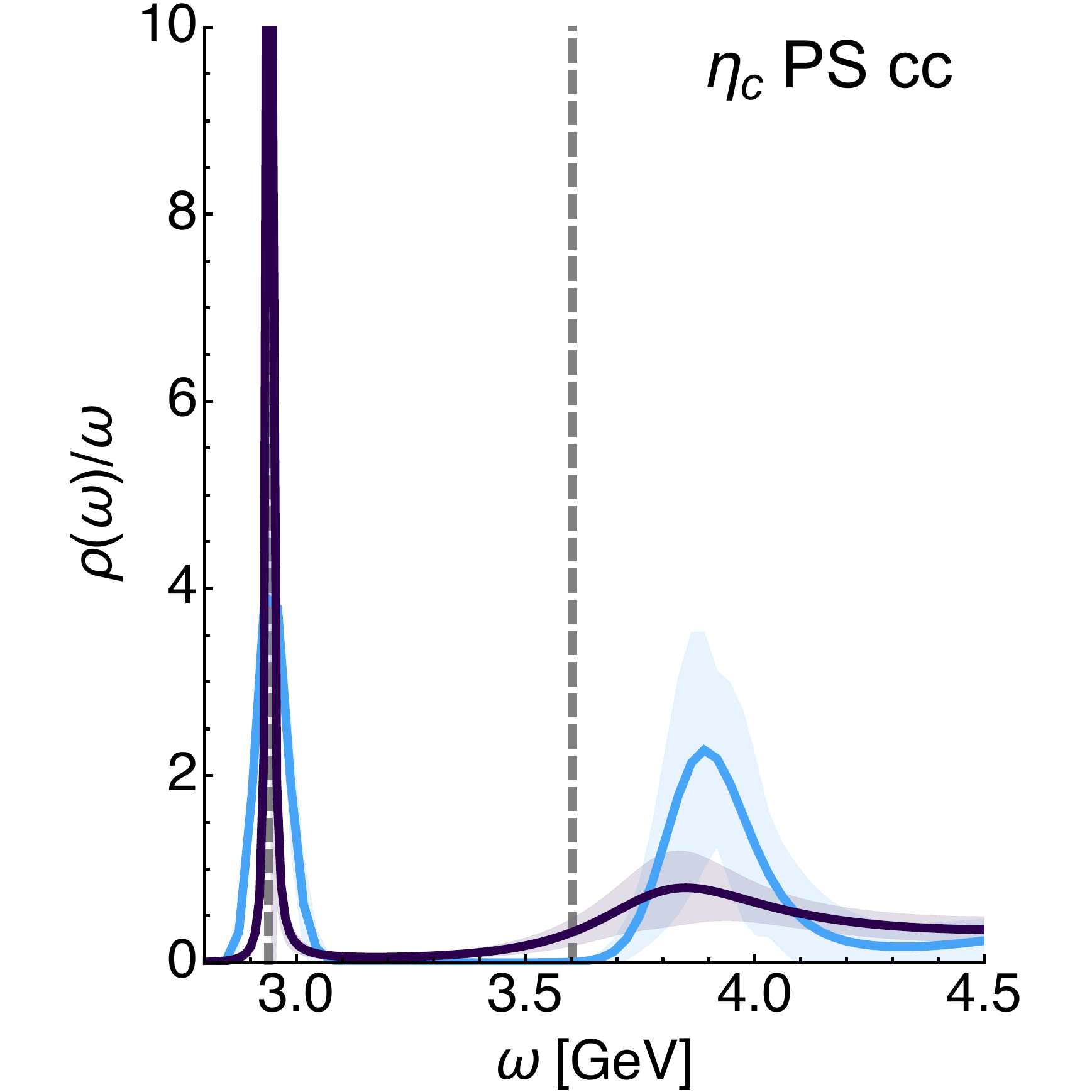}
\includegraphics[scale=0.24]{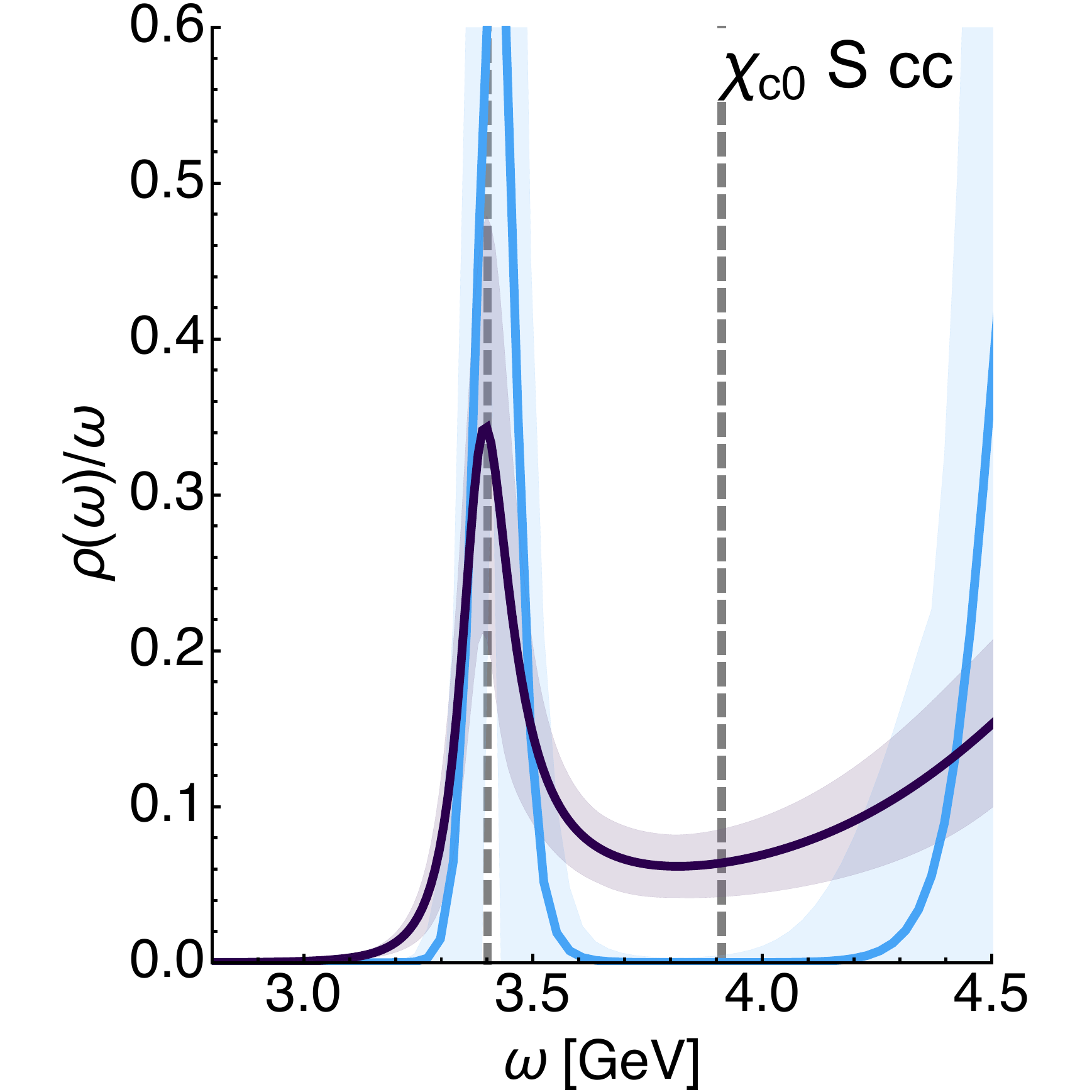}
\includegraphics[scale=0.24]{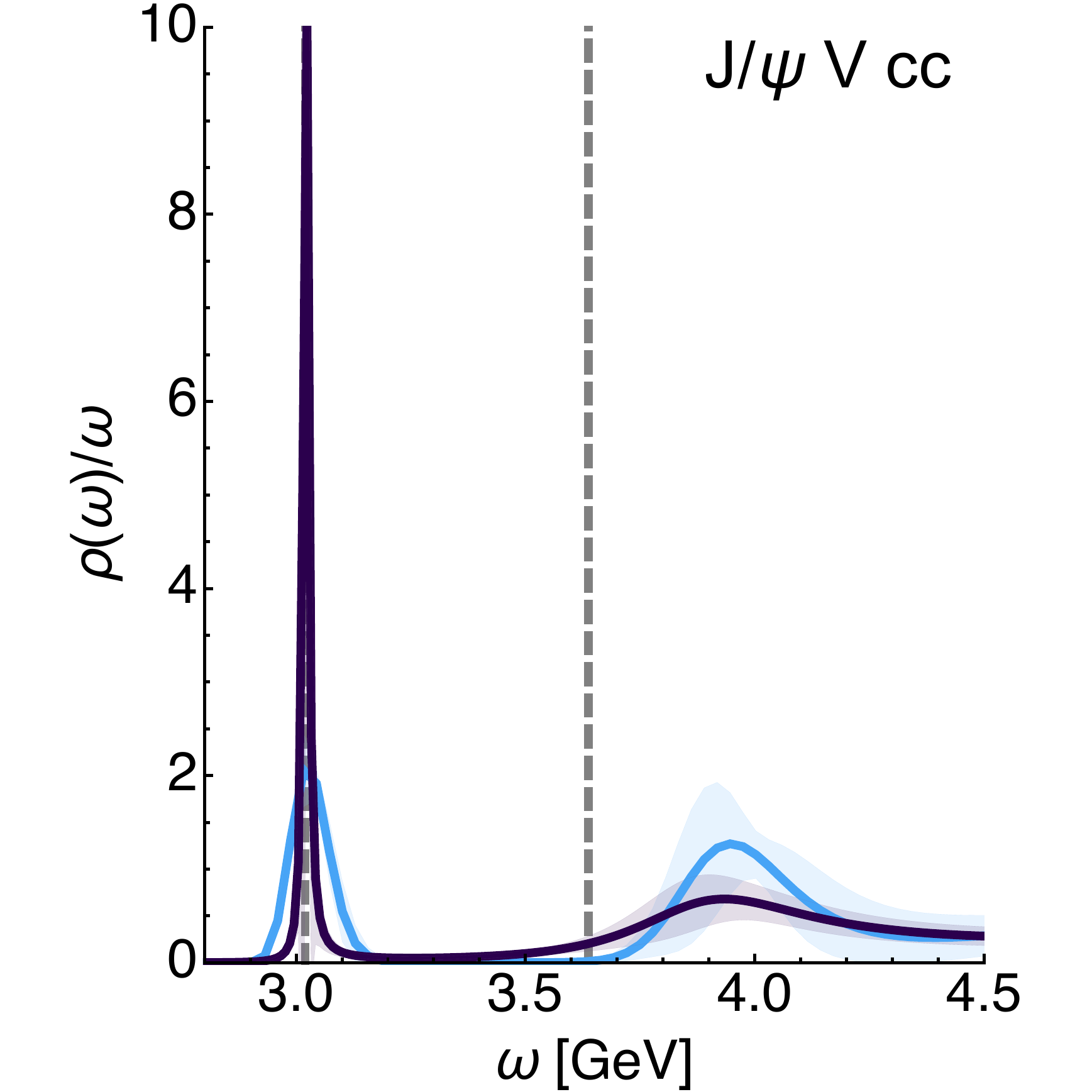}\\
\includegraphics[scale=0.24]{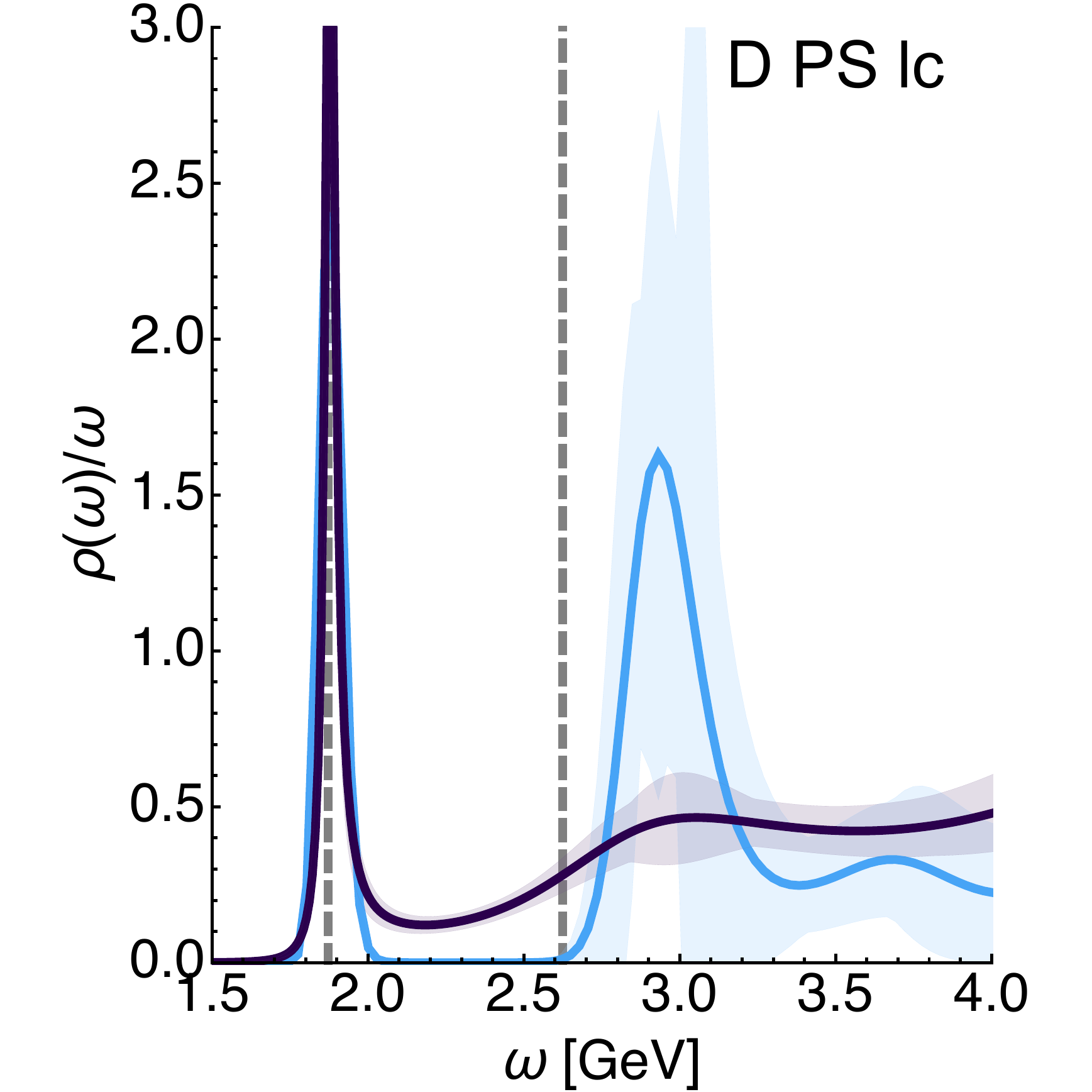}
\includegraphics[scale=0.24]{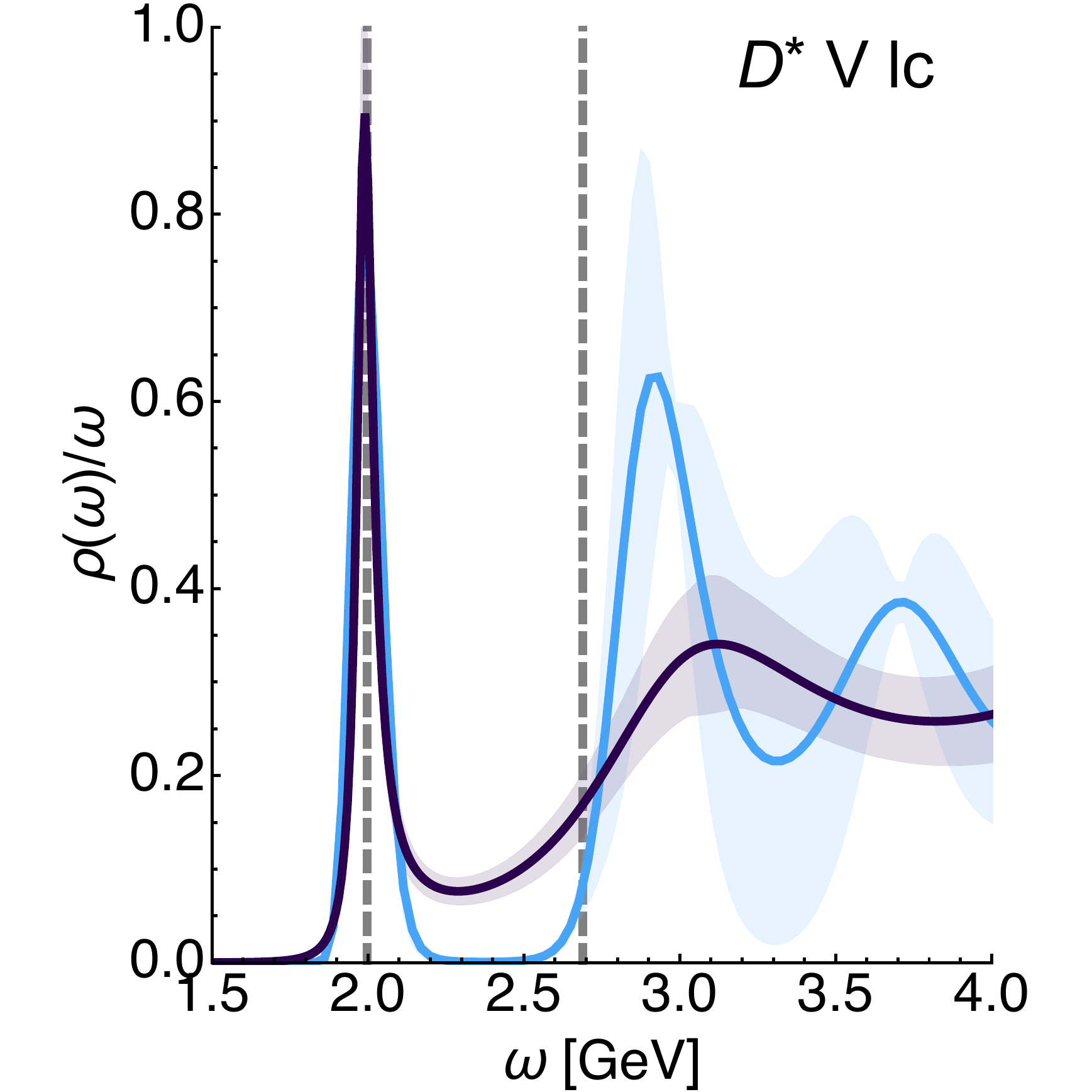}
\includegraphics[scale=0.24]{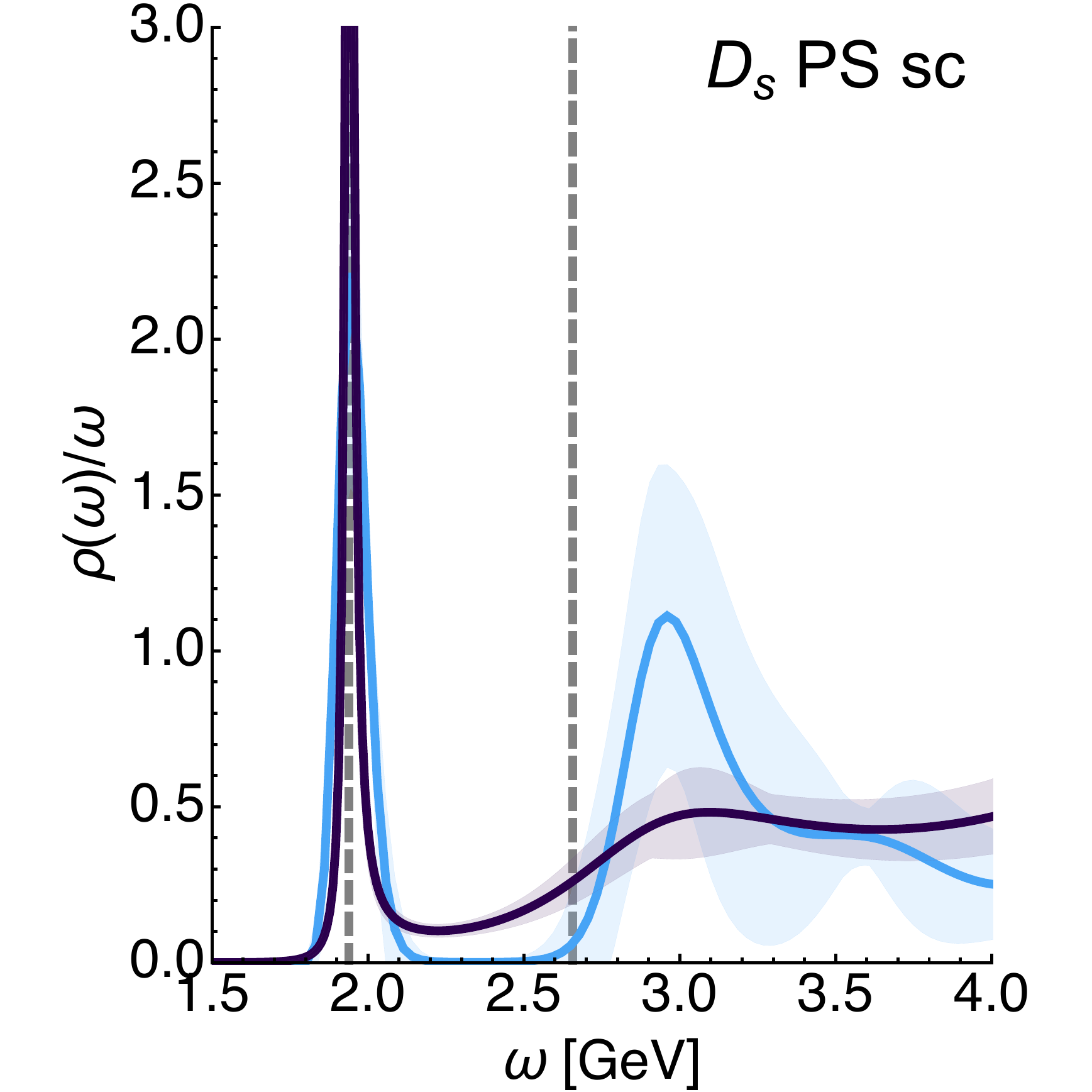}
\includegraphics[scale=0.24]{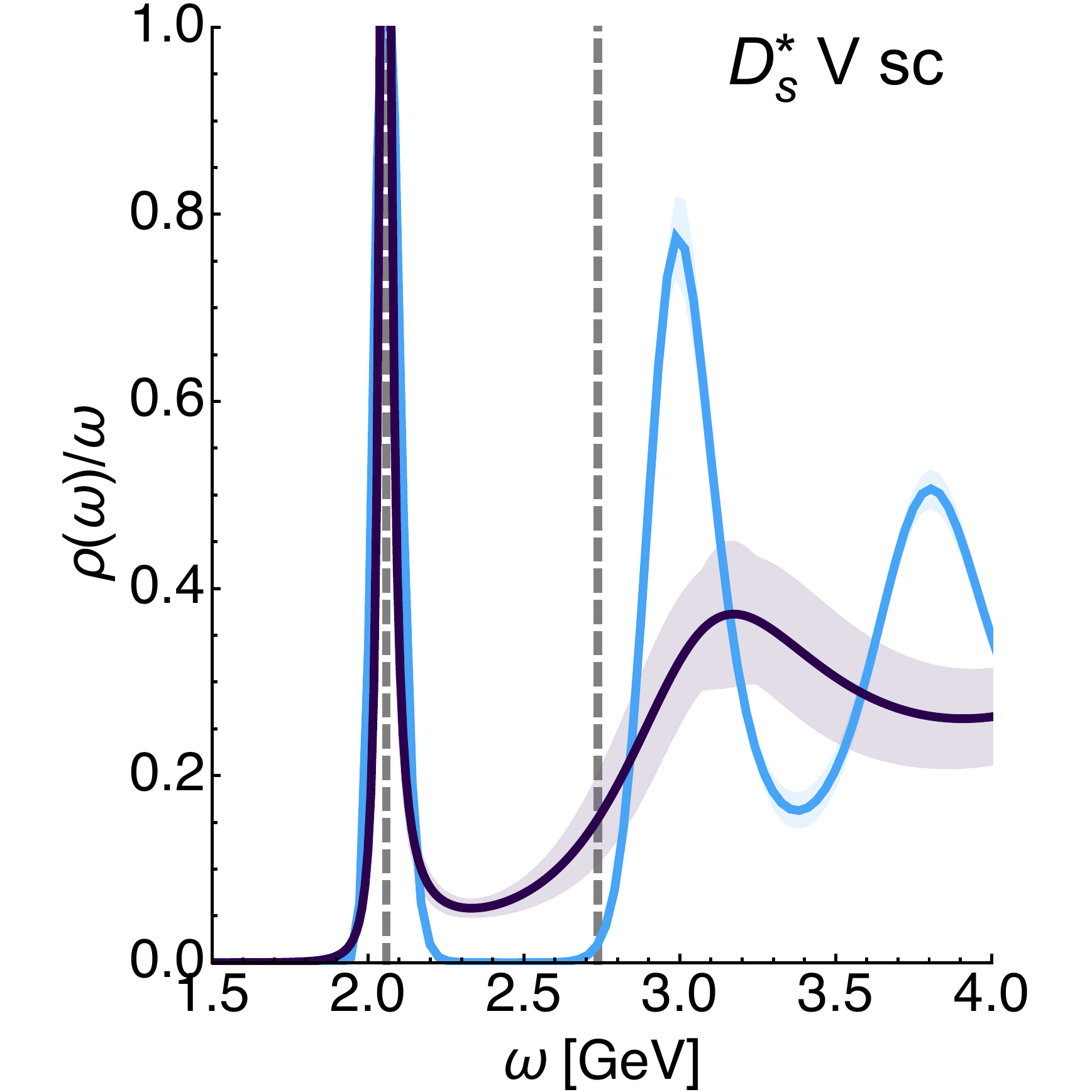}
\caption{Bayesian reconstruction of the frequency region around the
  ground state structure for all all meson channels investigated in
  this study.  Top row: charmonium; bottom row: open charm
  mesons. Dark solid lines denote the reconstruction result from the
  BR method, the lighter solid lines those from the Fourier basis
  MEM. The gray dashed lines indicate the position of the ground and
  first excited state determined from a variational computation \cite{Liu:2012ze,Moir:2013ub}.}\label{Fig:ZeroTSpectra}
\end{figure*}

As our study aims at investigating the in-medium spectral structure of charmed mesons, where information beyond masses, i.e. thermal widths, are of interest, we do not deploy distillation or smearing techniques for our correlators. Thus it is not surprising that a multi-exponential fit did not result in stable results for the masses of the first excited states, which usually require a more refined variational approach.

We continue by carrying out first Bayesian spectral reconstructions on
the $T=0$ correlators using both the BR method and the MEM. Results
are shown in Fig.~\ref{Fig:ZeroTSpectra}.  Errorbands in the plots arise from the combined variance obtained via a ten-bin Jackknife (statistical uncertainty), as well as from varying the underlying default model (systematic uncertainty).  At the lowest temperature a relatively large number of datapoints is available, i.e. here the statistical error dominates the uncertainty budget compared to the default model dependence. 

As shown in Fig.~\ref{Fig:ZeroTSpectra} both methods are able to locate the position of the lowest lying structure well from the currently available data. As expected from the use of naive interpolation operators, the overlap with the excited states is not strong enough for an accurate determination of the next higher lying structure. Even though the BR method in
general does better in reproducing isolated ground state peaks, with the current data quality neither method has a clear advantage in reproducing excited state
peaks. For all states except P-wave charmonium a second bump at higher frequencies is hinted at in the reconstructions, which however is not significant. 

Note that except for the small signal-to-noise case of the P-waves the
BR method produces a sharper ground state peak than the MEM. Furthermore the variance of the ground state structure is smaller in the BR method. Interestingly the MEM appears to show a more pronounced second structure than the BR method, which however also carries a larger uncertainty band. 

In preparation for the in-medium investigation we need to elucidate how increasing the temperature, i.e. the diminishing of available Euclidean datapoints, affects the spectral reconstruction. To this end we wish to take the lowest temperature result and encode it in a correlator which corresponds to a lattice at higher temperatures. In studies of non-relativistic mesons using the effective field theory NRQCD this is straightforwardly achieved by simply truncating the $T=0$ dataset to the number of points available at $T>0$. In a relativistic setting where the correlation function shows periodicity related to the inverse temperature a simple truncation is not appropriate and instead we have to turn to the reconstructed correlator, which takes into account the explicit temperature dependence of the integral kernel in Eq.~\eqref{eq:spectral}.

\begin{figure*}[th]
\centering
\includegraphics[scale=0.34]{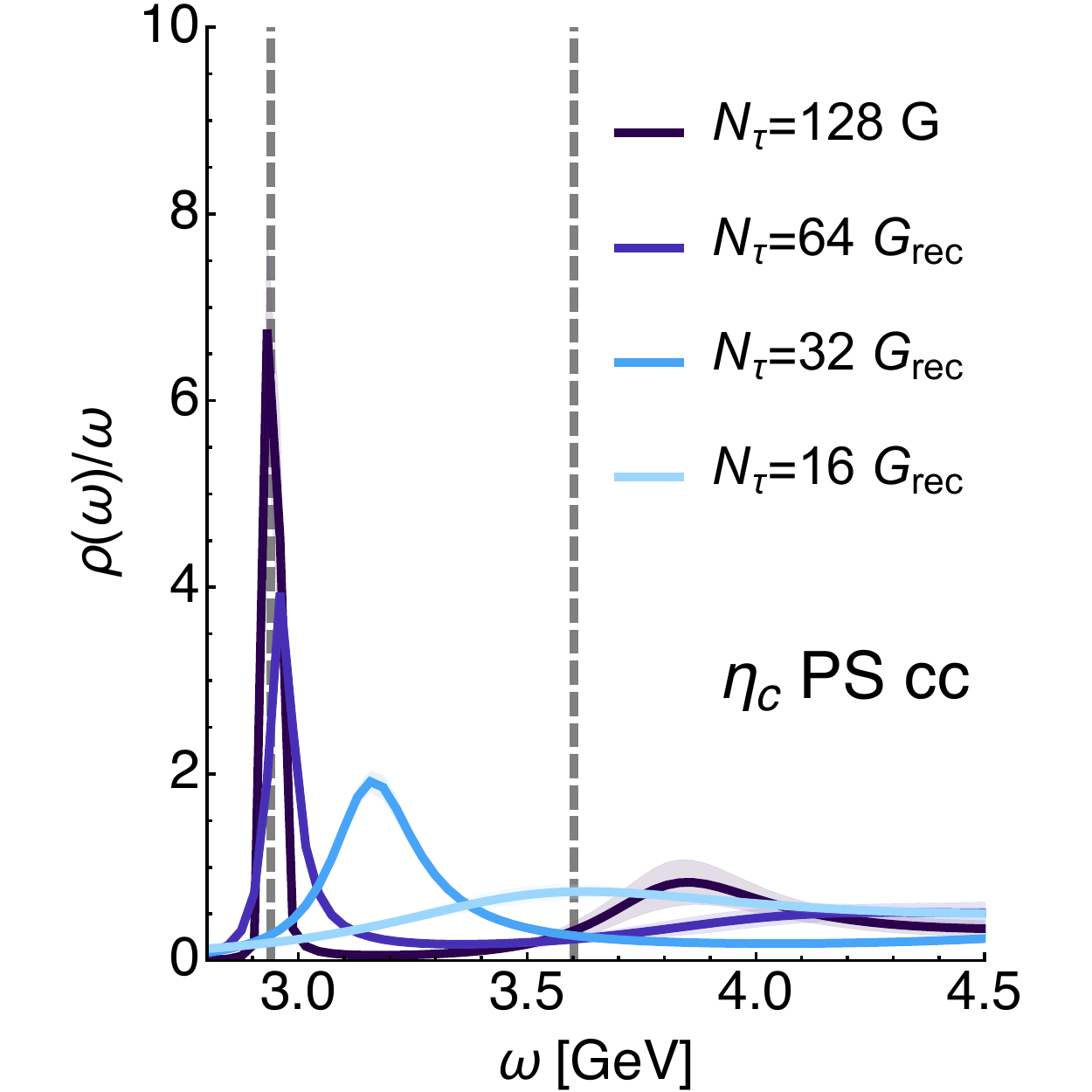}
\includegraphics[scale=0.24]{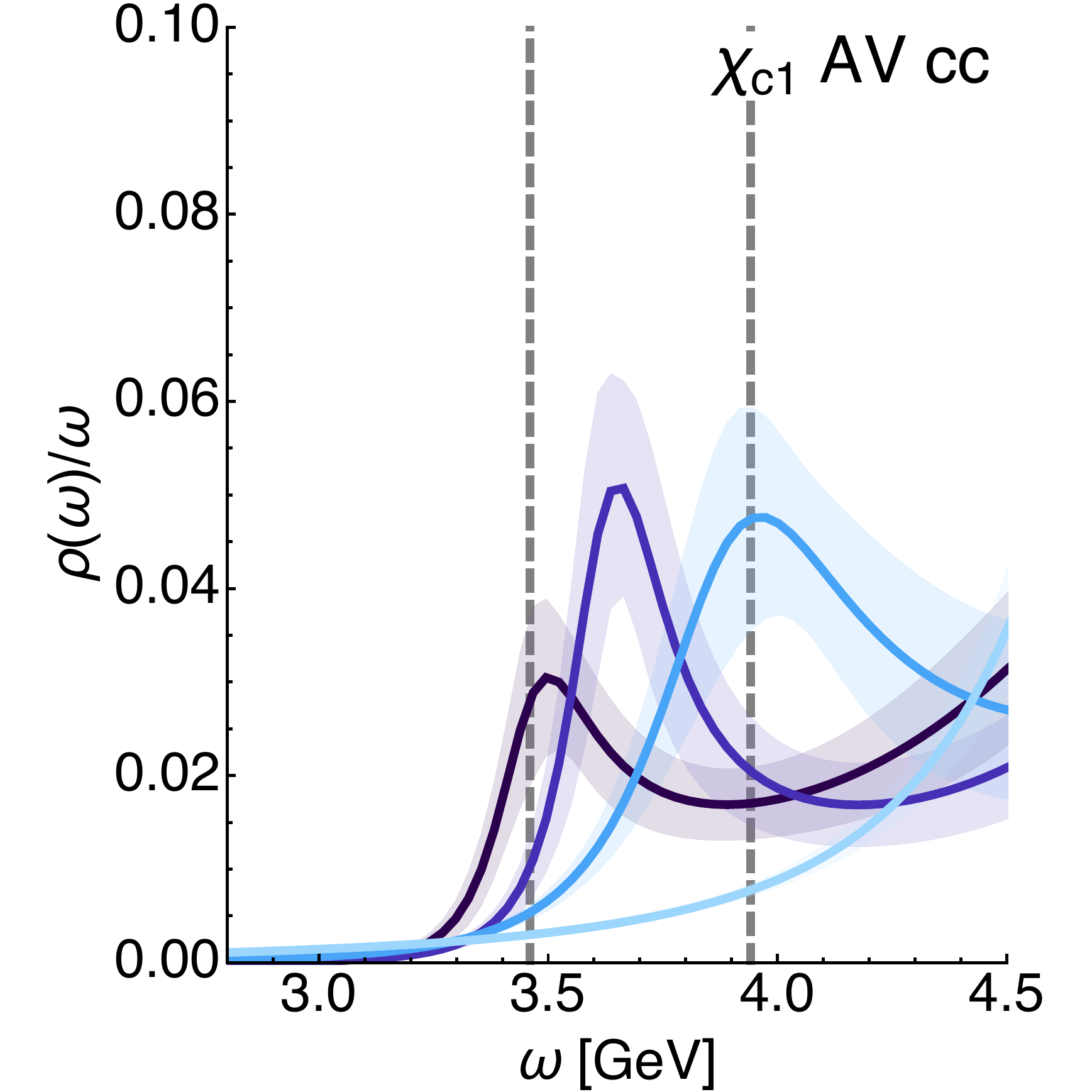}
\includegraphics[scale=0.24]{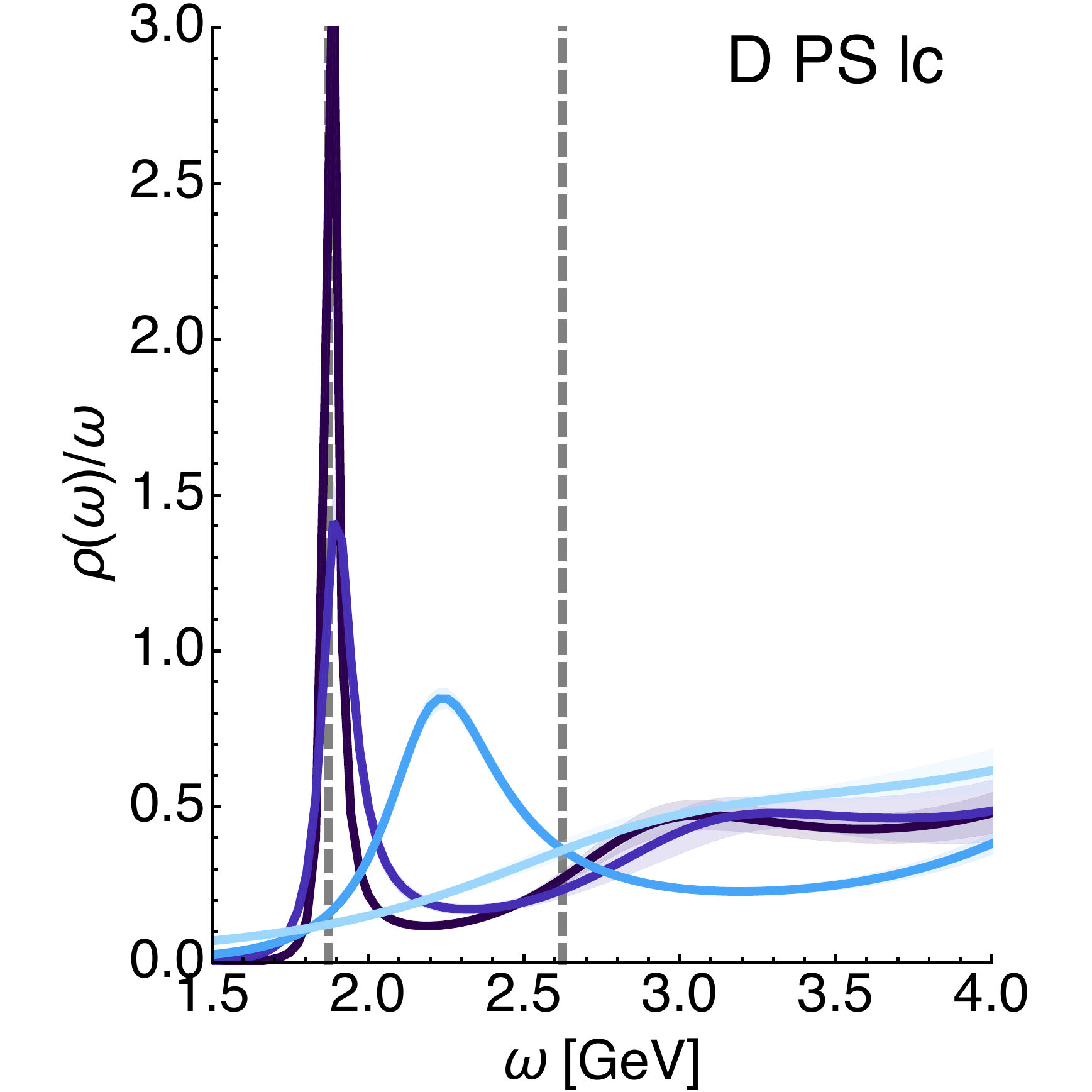}
\includegraphics[scale=0.24]{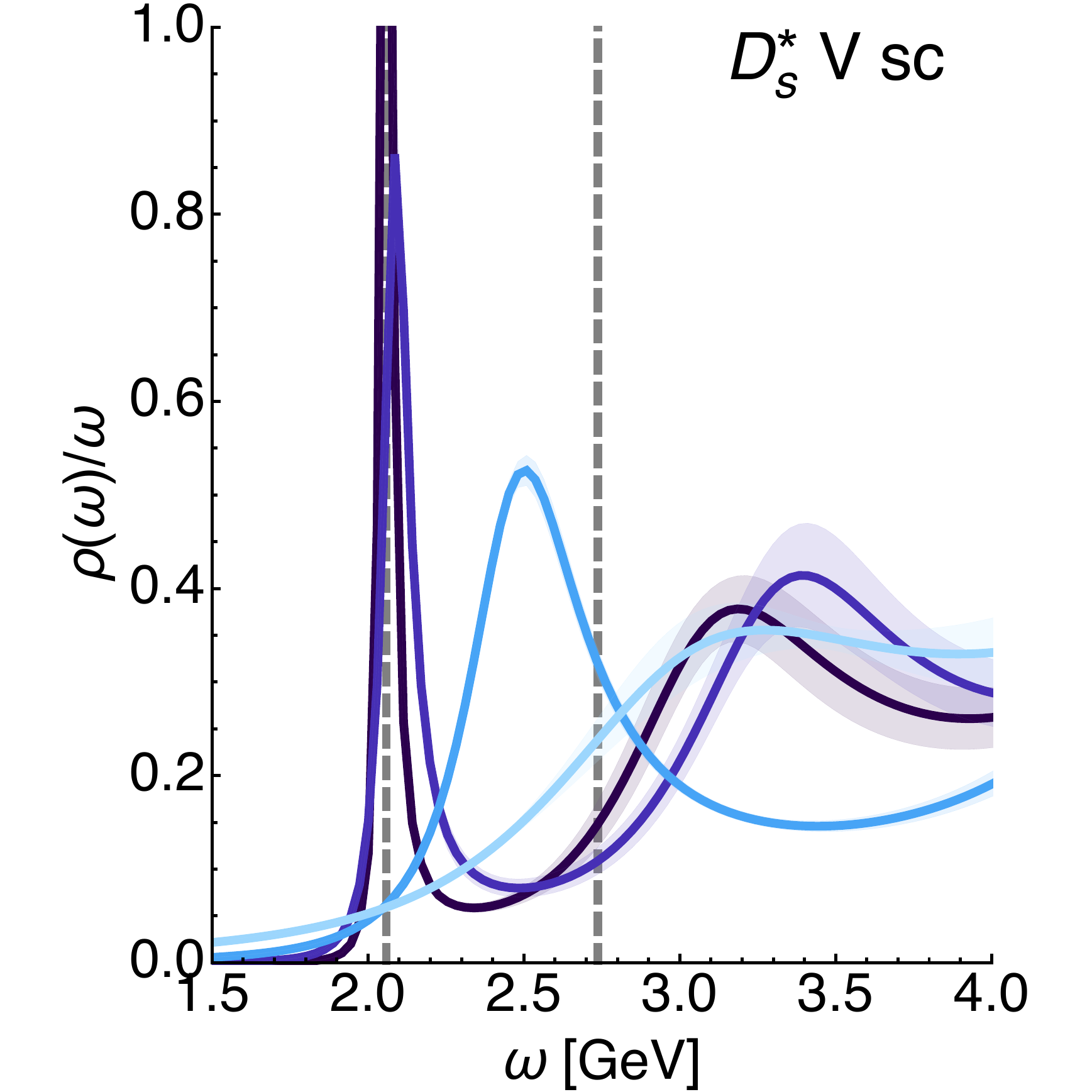}\\
\includegraphics[scale=0.34]{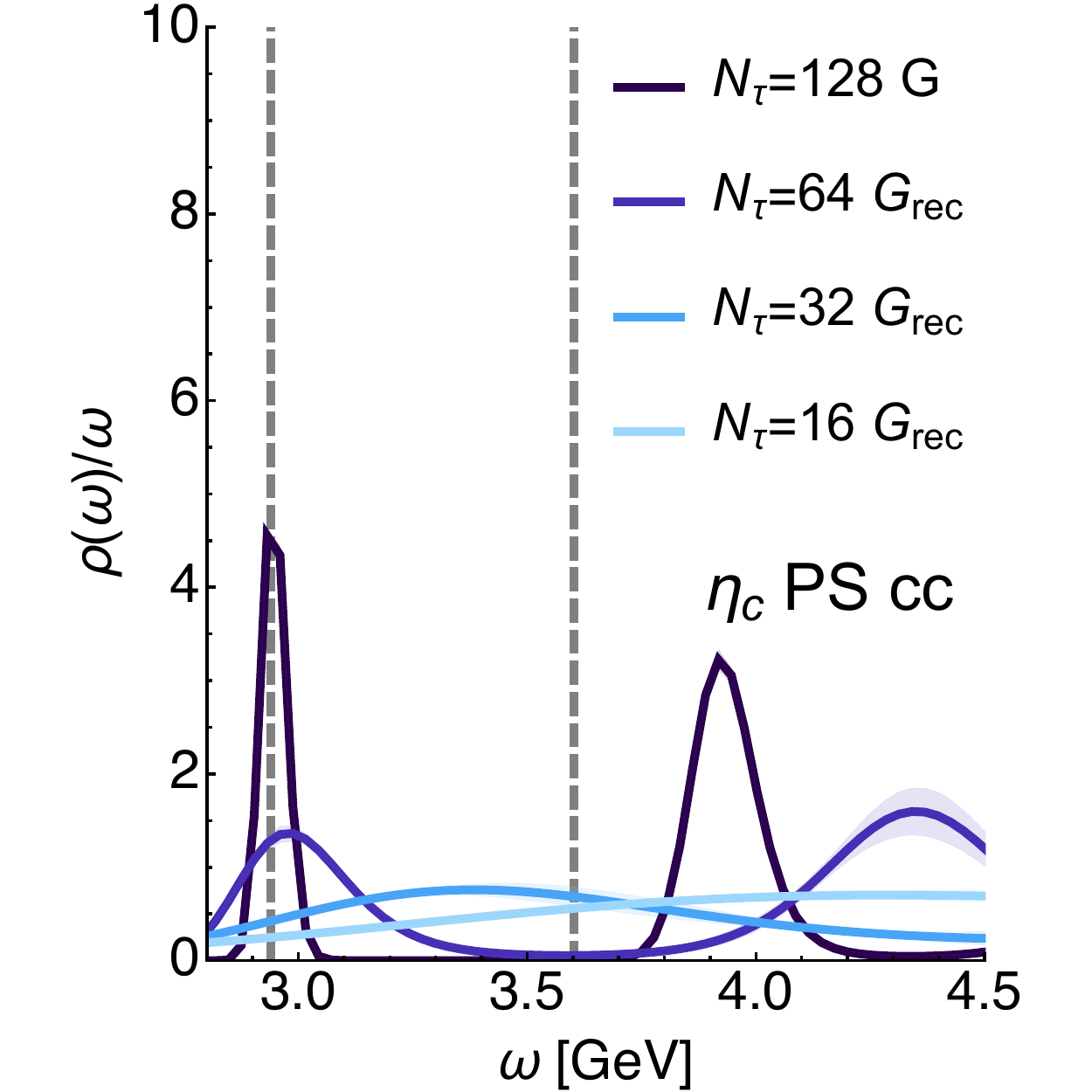}
\includegraphics[scale=0.24]{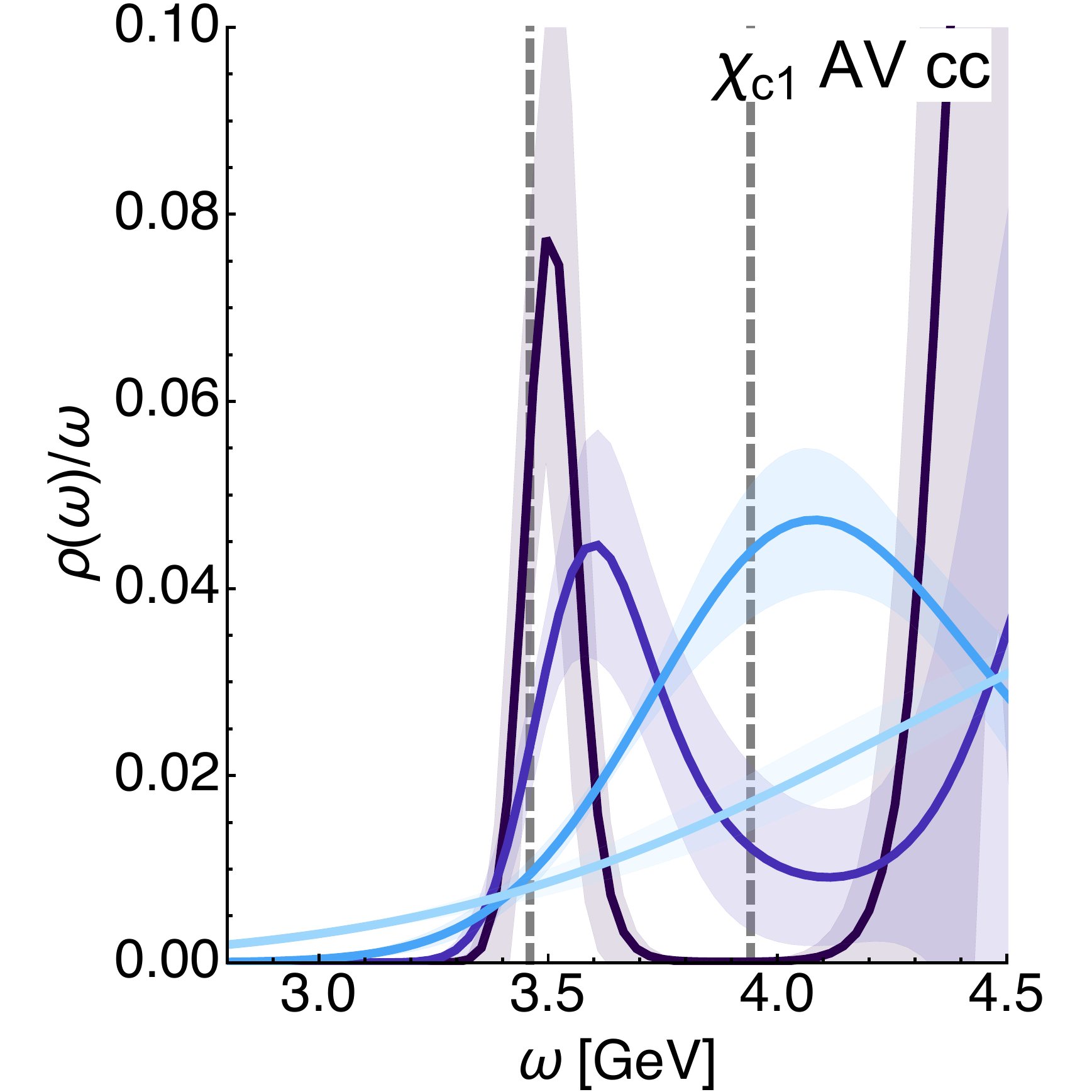}
\includegraphics[scale=0.24]{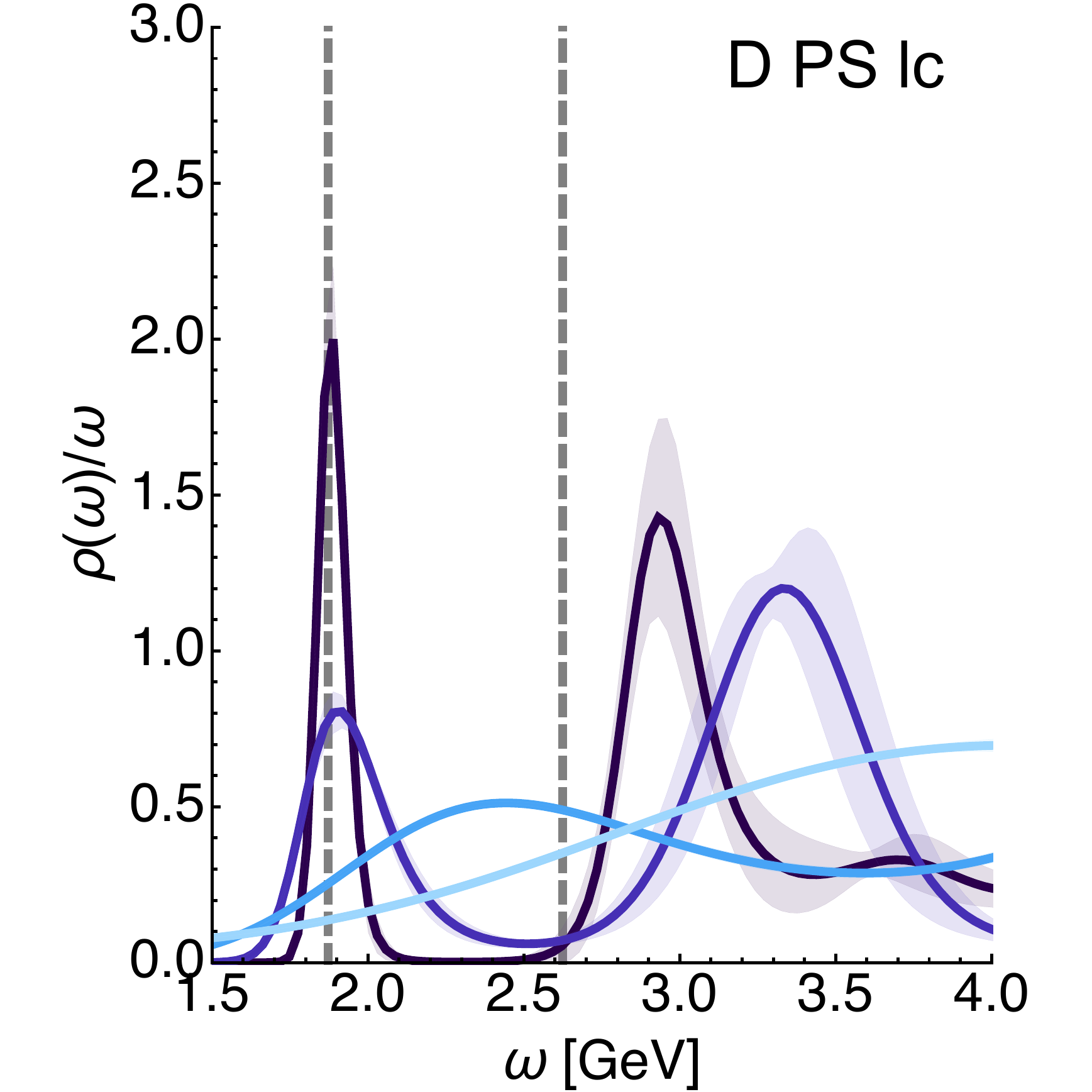}
\includegraphics[scale=0.24]{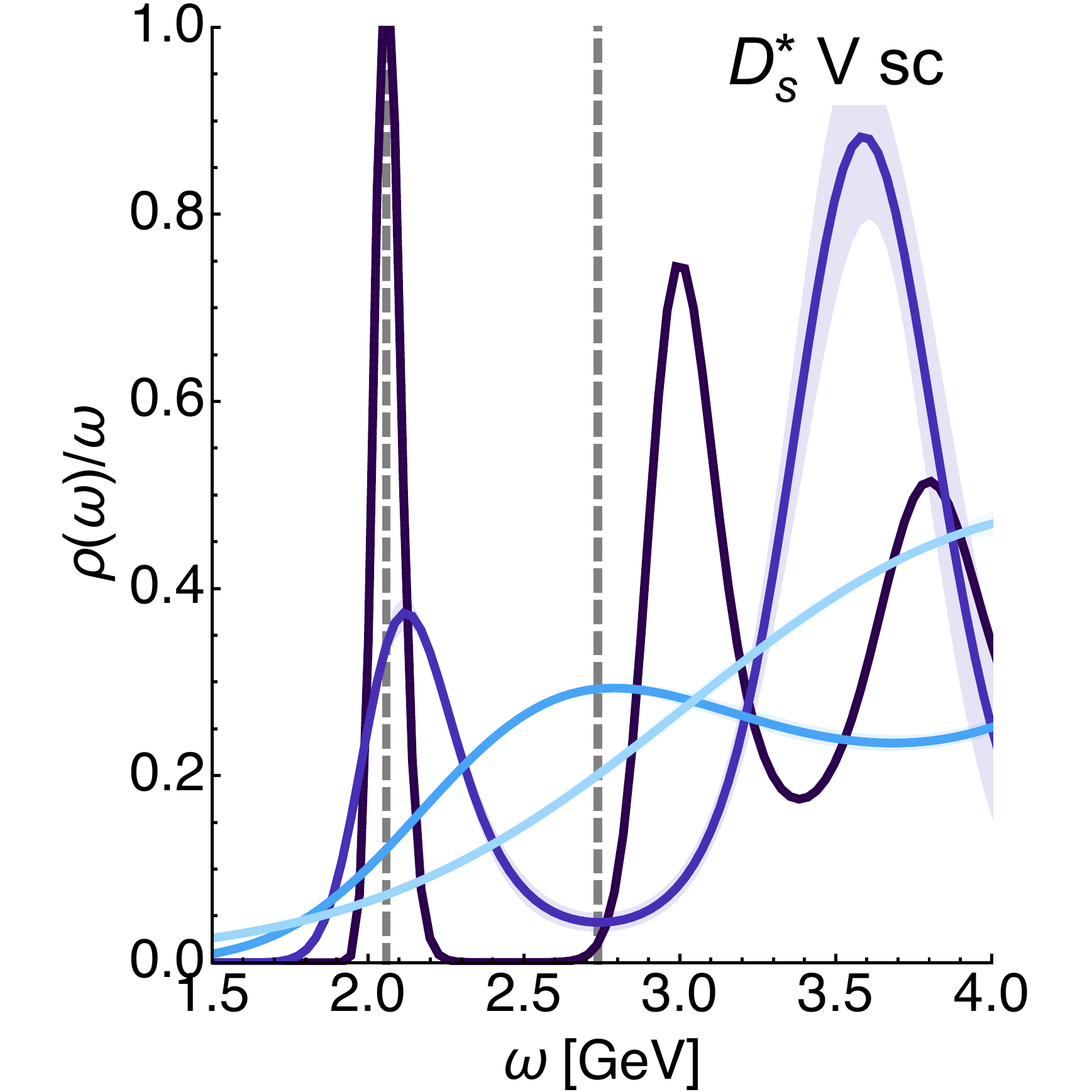}
\caption{Representative Bayesian spectra based on some of the reconstructed correlator datasets obtained from applying Eq.~\eqref{eq:Grec-direct} to low temperature data at $N_\tau=128$. The different colored curves correspond to different temporal extents of the reconstructed correlator $N_\tau=64,N_\tau=32, N_\tau=16$. The top row plots are obtained from the BR method; the bottom row plots from the MEM.}\label{Fig:ZeroTGRecSpec}
\end{figure*}

We first prepare reconstructed correlators of several different temporal extents by applying Eq.~\eqref{eq:Grec-direct} to the low temperature correlator at $N_\tau=128$. Since that equation is only well defined for an Euclidean extent which divides $N_\tau=128$ without remainder, $G_{\rm rec}$ is computed here for $N_\tau=64,N_\tau=32$ and $N_\tau=16$ to be used in the Bayesian reconstruction. For the subsequent comparison with finite temperature correlators, we have resorted to padding the $N_\tau=128$ correlators with zeros to compute also extents such as $N_\tau=40$.

In Fig.~\ref{Fig:ZeroTGRecSpec} several representative examples of the
outcome of the Bayesian analysis using the reconstructed correlator
are shown. In the top row the BR method has been deployed, while in
the bottom row the MEM was used. While quantitatively different, similar qualitative trends emerge. One finds that a change from $N_\tau=128$ to $N_\tau=64$ only induces a small broadening and a minor shift of the ground state structure to higher frequencies. Going to even smaller $N_\tau=32$ however already induces significant modifications to the low lying peak with $\Delta m=0.2-0.5$GeV. At $N_\tau=16$ the well defined ground state peak cannot be resolved anymore with the currently available statistics. In general the MEM produces more strongly washed out features than the BR method here. We have to keep in mind that the actual spectrum encoded in all the different reconstructed correlators is exactly the same and the differences in the outcome are simply a manifestation of the degradation of the resolution of the Bayesian method with limited available data.

Now that we understand the inherent method deficiency related to limiting the number of available datapoints, we can proceed to inspecting genuine finite temperature data.

\subsection{Quarkonium at finite temperature}
\label{QuarkoniumFiniteT}
\subsubsection{Correlation functions}

\begin{figure}[h]
\includegraphics[scale=0.35]{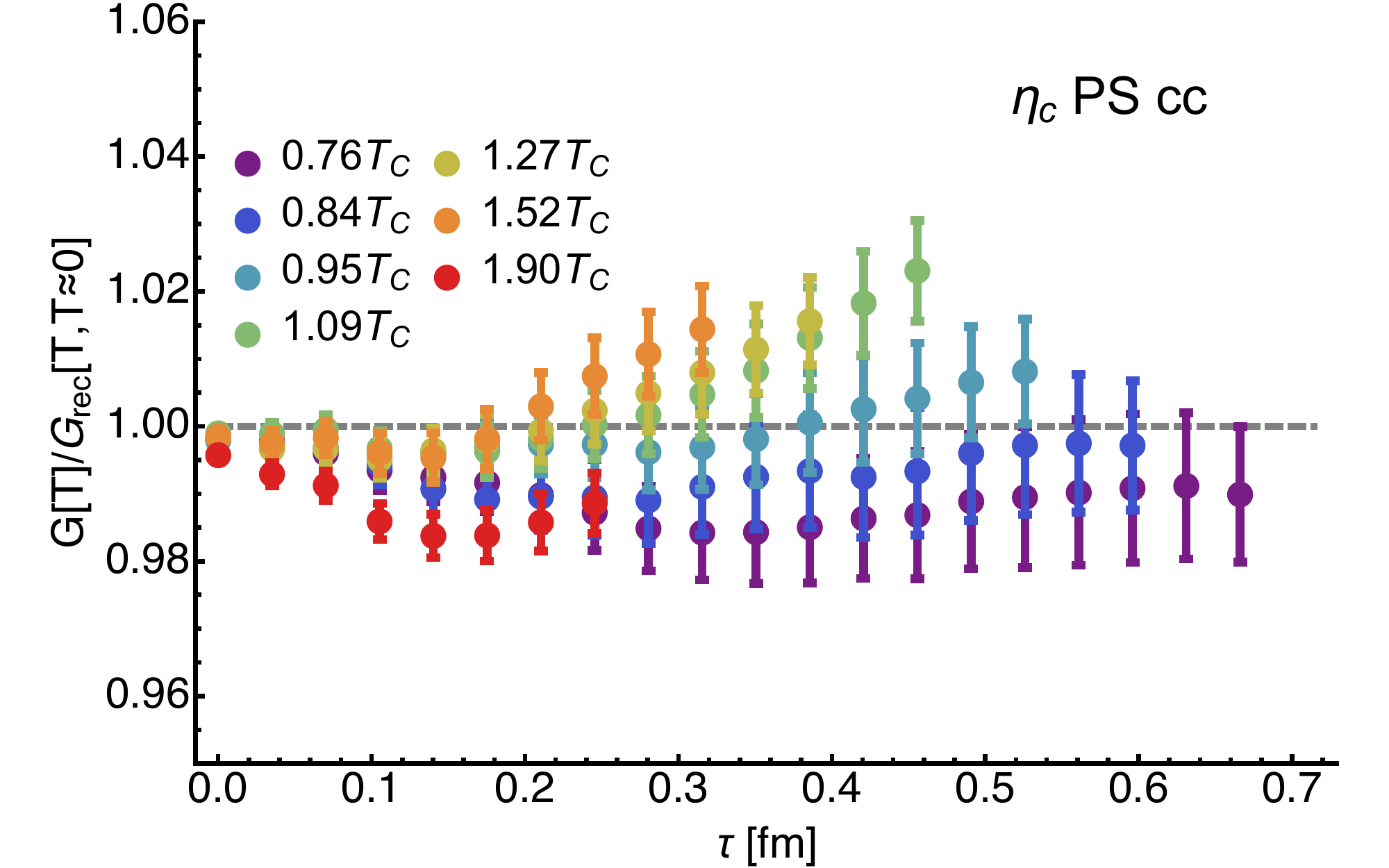}
\includegraphics[scale=0.35]{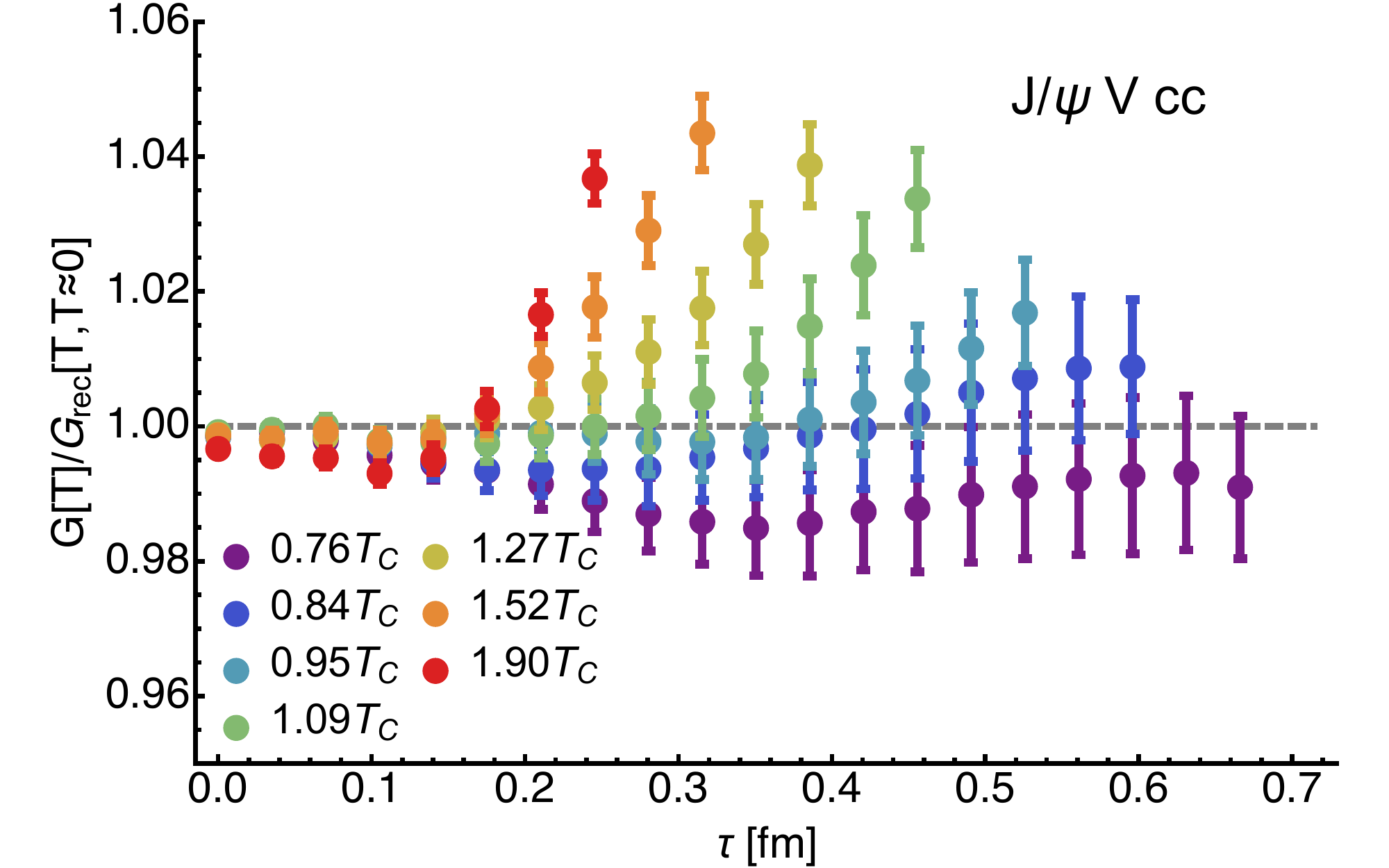}
\includegraphics[scale=0.35]{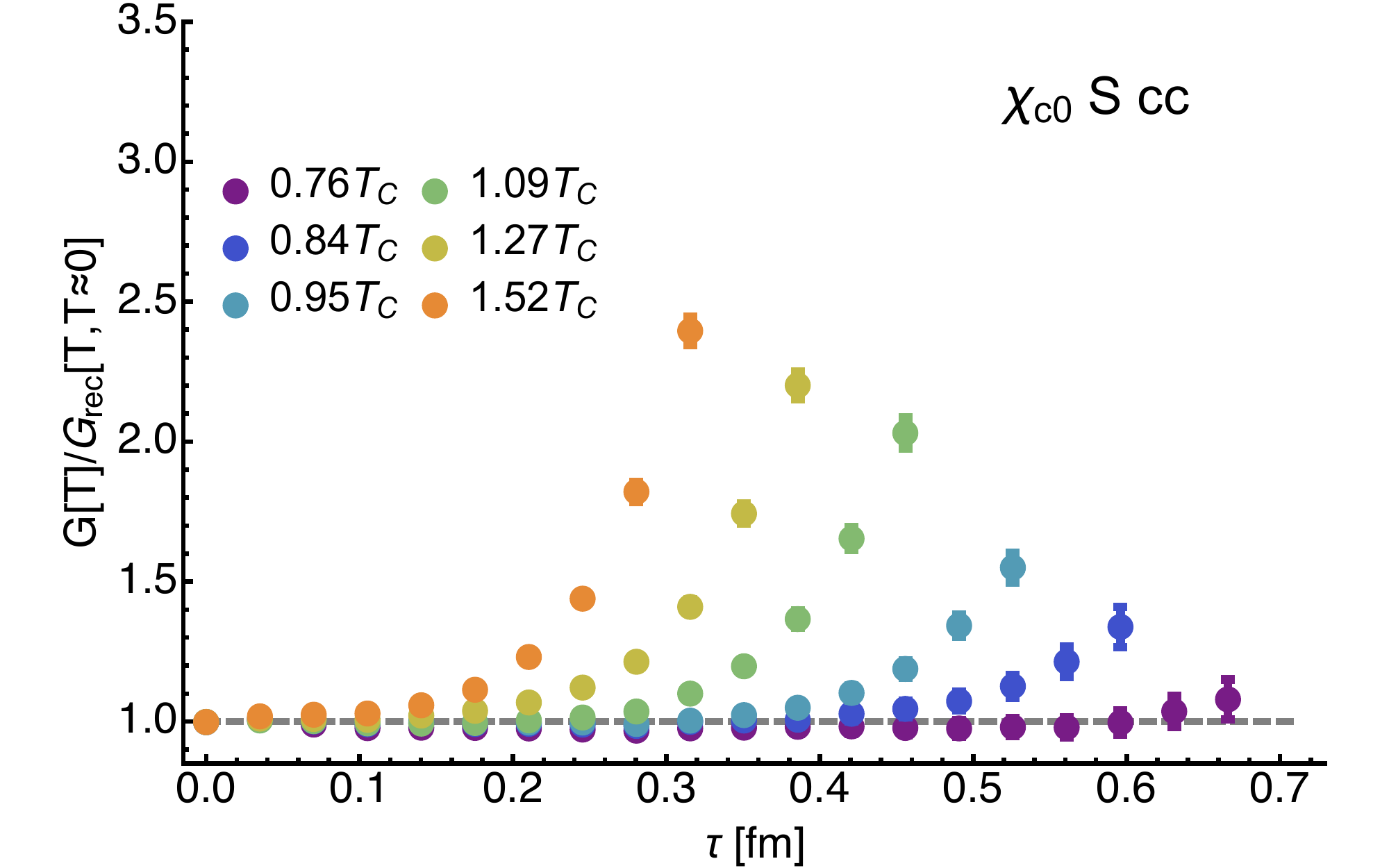}
\includegraphics[scale=0.35]{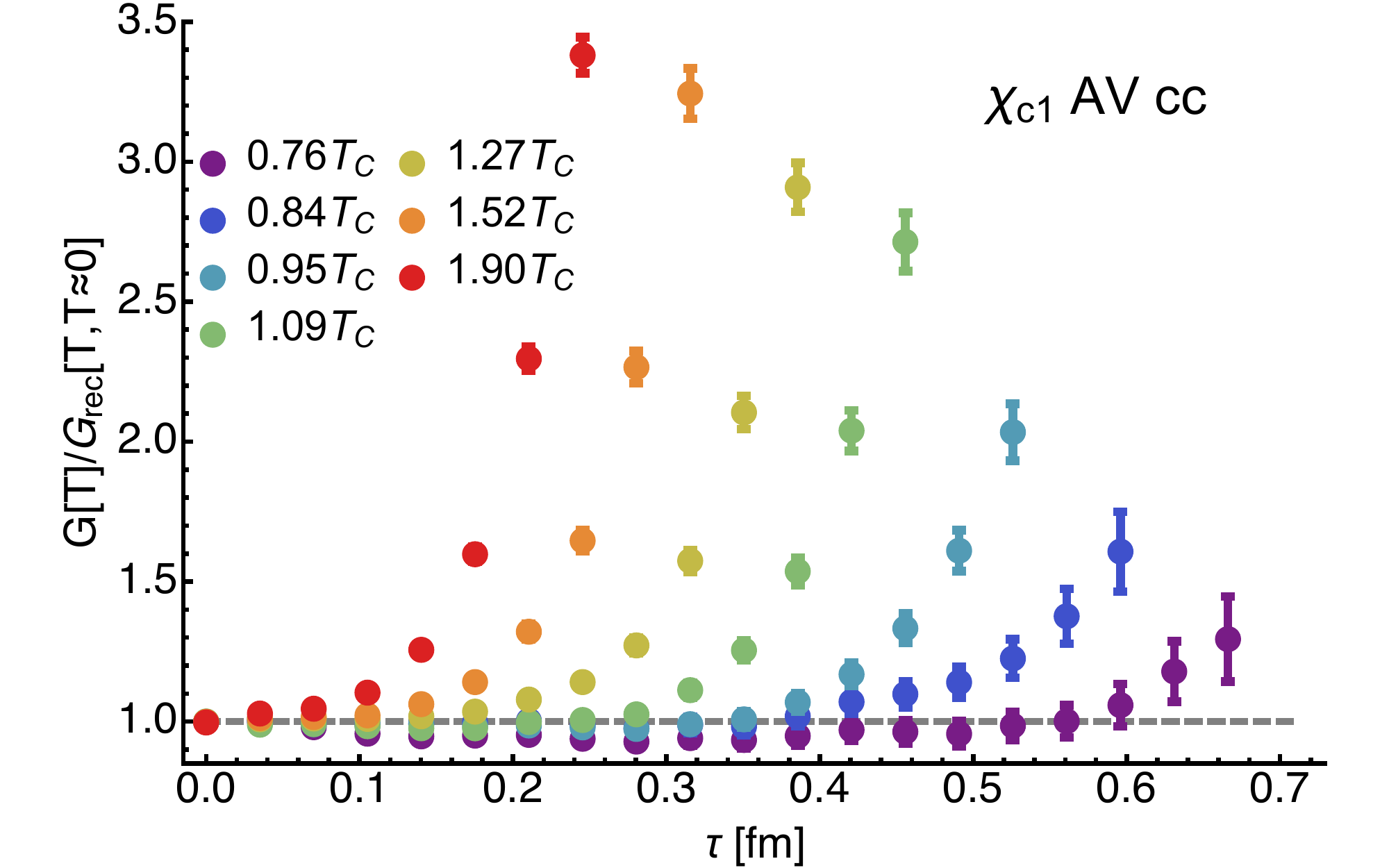}
\caption{Ratio of the in-medium charmonium correlator to the
  corresponding reconstructed correlator for the pseudoscalar (top),
  vector (second to top), scalar (second to bottom) and axial vector (bottom) channels.  Note the difference
  in vertical scale between the top two and the bottom two graphs.}\label{Fig:cc_Grec}
\end{figure}

We first take a look at what information we can glean from the thermal lattice correlators
themselves.  Figure~\ref{Fig:cc_Grec} shows the ratio of the thermal
correlator $G(\tau;T)$ to the reconstructed correlator
$G_r(\tau;T,T_r)$ defined in Eq.~\eqref{eq:Grec-direct}, with
$T_r=0.24T_c$ corresponding to $N_\tau=128$.  In the absence of any
thermal modifications, this ratio equals 1.  We show data for
the pseudoscalar (top), vector (second to top), scalar (second to bottom) and axial vector (bottom) channels.  As is expected, the results for the axial-vector channel are virtually the same as for the scalar channel. Note that in the scalar channel, we have only accessed the correlation function up to $T=1.52$GeV.

In the pseudoscalar channel, we find that the thermal modifications of
the correlator are small, $\lesssim2\%$, at all temperatures. With the current level of
precision no significant in-medium modification can be observed up to $T_c$.
However, for $T>T_c$, we find that the correlator ratio begins to
deviate systematically from unity, and at the highest temperature
($1.9T_c$), the deviation is genuinely significant. Interestingly there is a
qualitative jump between the data at $T=1.52T_c$ and $T=1.90T_c$, the former
following the monotonic upward bending above 1, similar to lower temperatures, while the latter 
lies fully below unity. A rather significant modification of the spectrum must occur in between these
two temperatures.

Comparing this result to the same quantity evaluated in \cite{Ding:2012sp}, we find an interesting
difference. While our ratios show a clear intermediate upward bend between $T_c<T<1.5T_c$, Ref.~\cite{Ding:2012sp}
depicts ratios which decrease monotonically within errors. At around $2T_c$, once we observe the jump in our 
results the Euclidean time dependence again agrees qualitatively. We
note that the reconstructed correlator ratios determined from first
generation FASTSUM ensembles \cite{Skullerud:2014xx} exhibit the same qualitative features as
those of \cite{Ding:2012sp}, suggesting that the difference is largely
due to the limited statistics in the present study.

Let us turn to the vector channel. Similarly to the pseudoscalar case, we again see a clear difference between the data
below and above $T_c$, with $G/G_r$ remaining roughly consistent with
1 below $T_c$ and deviating systematically from 1 above $T_c$.  The
latter deviation manifests itself in a monotonic upward bending, consistent with what has been observed in previous
lattice studies using both relativistic \cite{Ding:2012sp} and non-relativistic formulations \cite{Kim:2015rdi,Kim:2017aio}. The maximum change
in the ratio here lies at $\lesssim5\%$ at $T=1.90T_c$ and contrary to the pseudoscalar channel, no sudden changes in the
correlator occur. The strength of the modification in our case is slightly smaller than what was observed in previous studies.

The P-wave charmonium states behave rather differently compared to the pseudoscalar and vector case, 
A strong modification is present, already
below $T_c$, and may reach up to a factor of $\lesssim3.5$ at $T=1.90T_c$. The change
again takes the form of an upward bending, which increases in strength with temperature. The qualitative
behavior is consistent with findings of previous studies, both based on relativistic and non-relativistic
heavy quarks. The strength of the modification is similar to those observed in the relativistic case but is much more pronounced than in
the non-relativistic study, where it remains below $\lesssim2\%$ for $T\approx 2T_c$. One possible
explanation is related to the appearance of zero modes in the correlator \cite{Umeda:2007hy}, which
are not resolved in the effective theory approach.

The observed upward bending behavior in the correlator ratio has been interpreted in the context of
non-relativistic potential computations to be consistent with the disappearance of excited state peaks
and the steady reduction of the threshold down to lower energies, as temperature rises. Extracting these
intricate changes in the spectrum directly via Bayesian inference is extremely challenging, since the
relevant structures are located relatively close to each other compared to the maximum lattice momentum.
In addition, the small overall change in the ratios of the pseudoscalar and vector channel are only well resolved
for $N_\tau/4<\tau<N_\tau/2$ rendering the number of effective relevant datapoints comparatively small.

\begin{figure}[th!]
\includegraphics[scale=0.35]{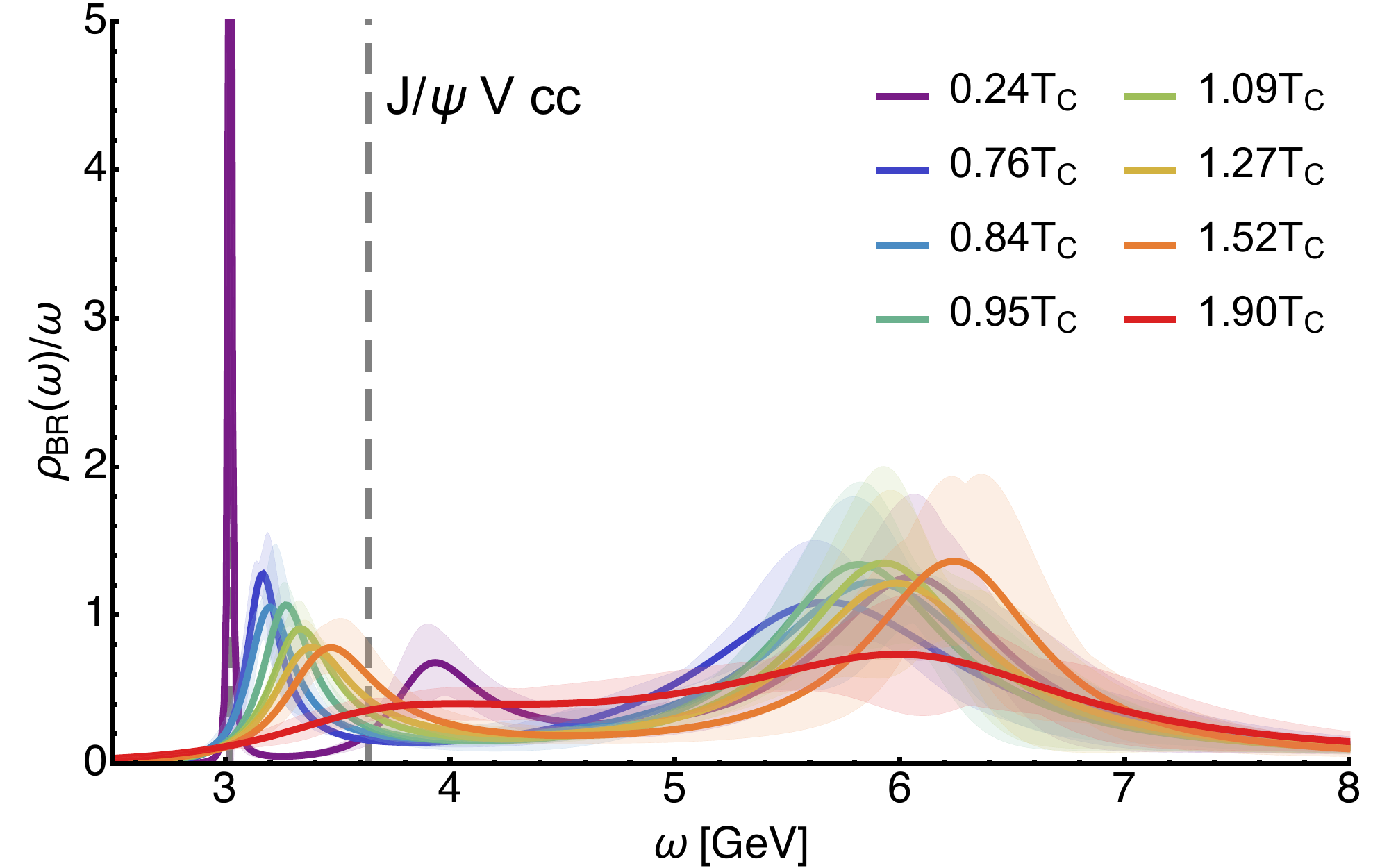}
\includegraphics[scale=0.35]{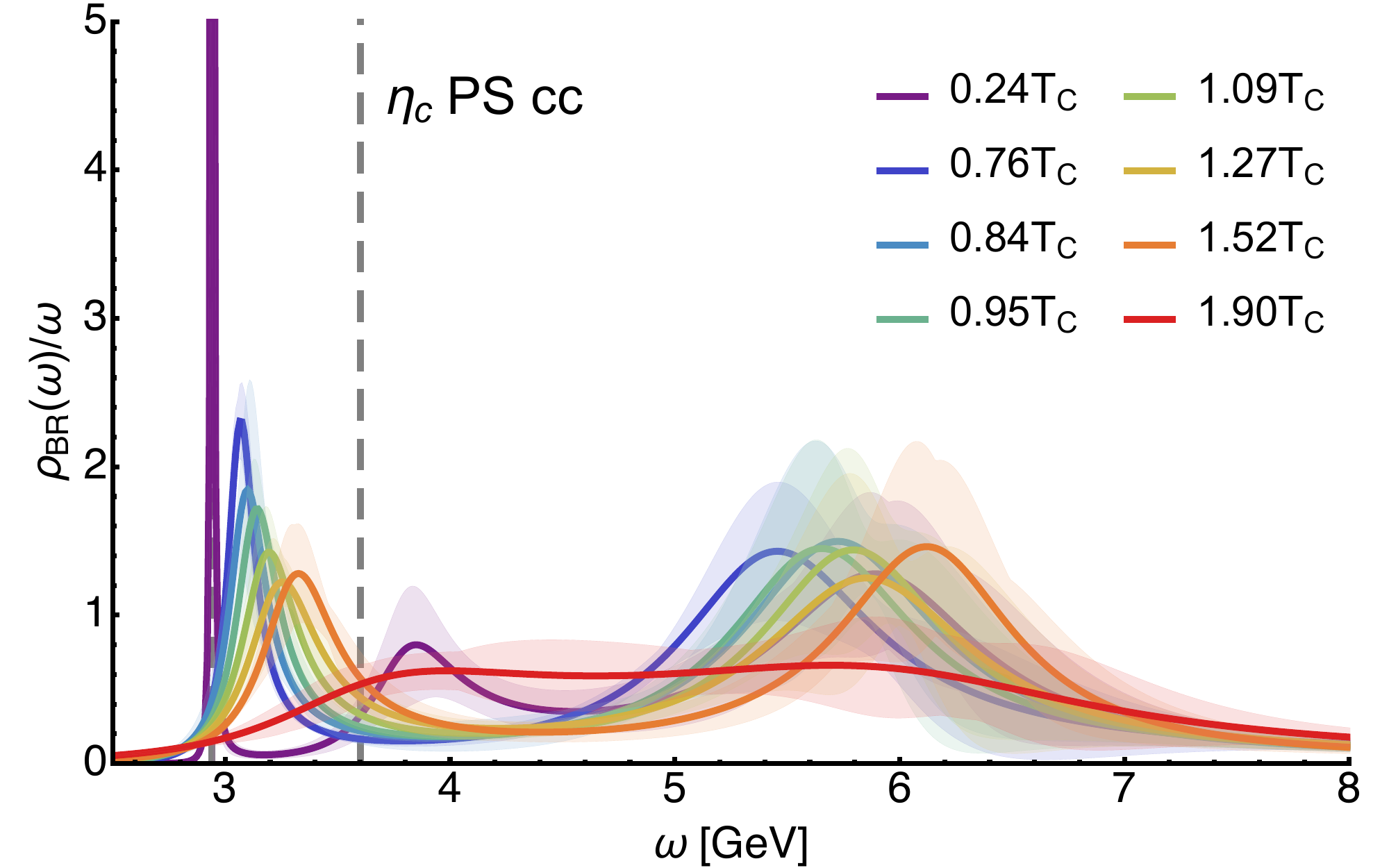}
\caption{Spectral functions for the $c\bar{c}$ vector (top) and
  pseudoscalar (bottom) channels, obtained using the BR method.}\label{Fig:BR_cc_V-PS-FiniteT}
\end{figure}
\begin{figure}[th!]
\includegraphics[scale=0.35]{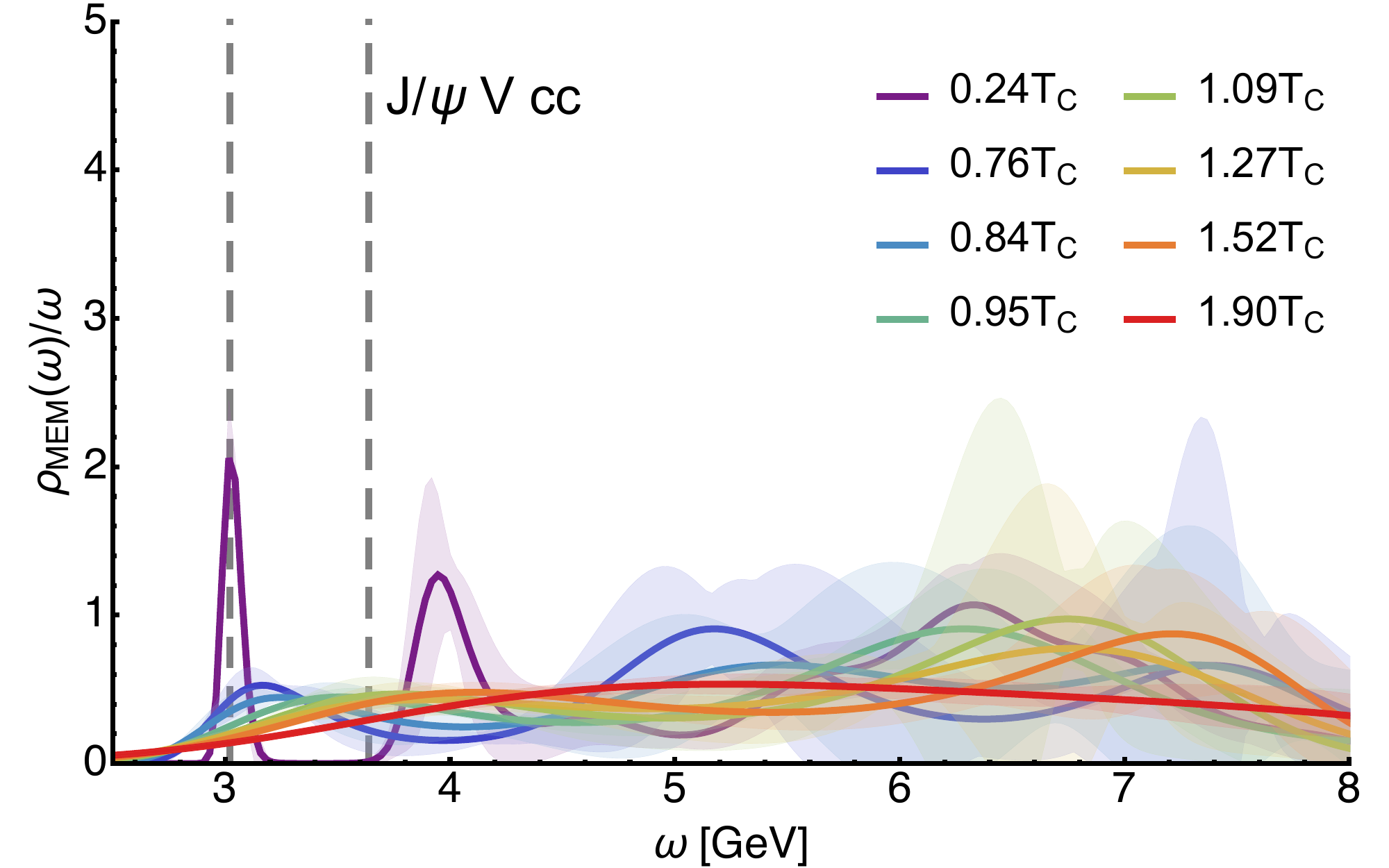}
\includegraphics[scale=0.35]{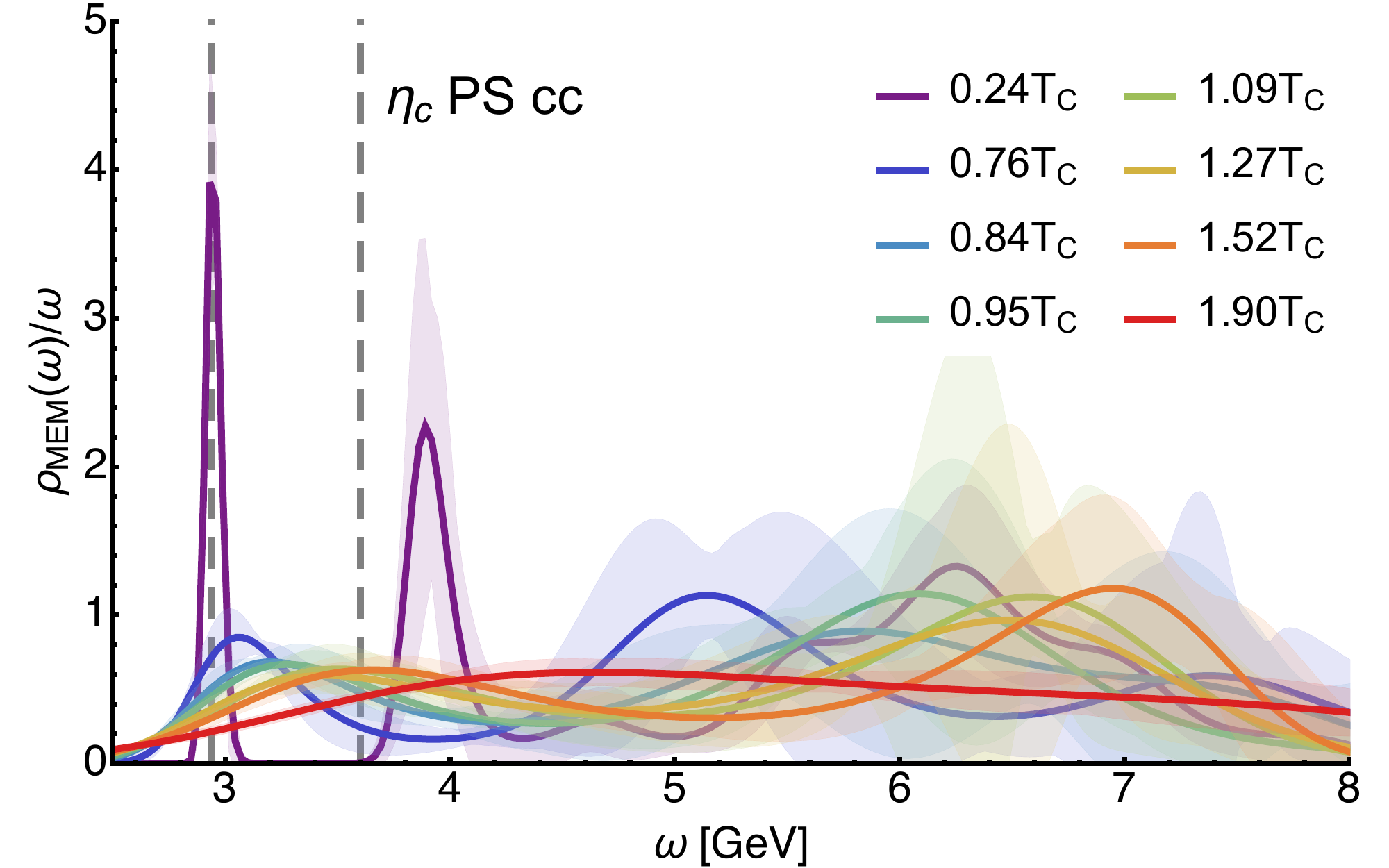}
\caption{Spectral functions for the $c\bar{c}$ vector (top) and
  pseudoscalar (bottom) channels, obtained using the MEM with the Fourier basis.}\label{Fig:BR_cc_V-PS-FiniteTMEM}
\end{figure}

\subsubsection{S-wave ($J/\Psi$ and $\eta_{c}$) spectra}

We begin our investigation of in-medium spectral properties with the
$J/\Psi$ and $\eta_{c}$ particle. In Fig.~\ref{Fig:BR_cc_V-PS-FiniteT}
we show their spectral functions, as obtained by the BR method and in Fig.~\ref{Fig:BR_cc_V-PS-FiniteTMEM} by
MEM with the Fourier basis, each time at eight different temperatures. Errorbands are again taken as the combined variance arising from statistical and default model dependence. With a diminishing number of datapoints at increasing temperatures the default model dependence starts to become the dominating source of error at the $T=1.52T_C$ and $T=1.90T_C$. The gray dashed lines, inserted as reference guide, denote
the positions of the vacuum ground and first excited state peak, as obtained from a variational computation
by the Hadron Spectrum Collaboration \cite{Liu:2012ze}.

\begin{figure}[th!]
\includegraphics[scale=0.35]{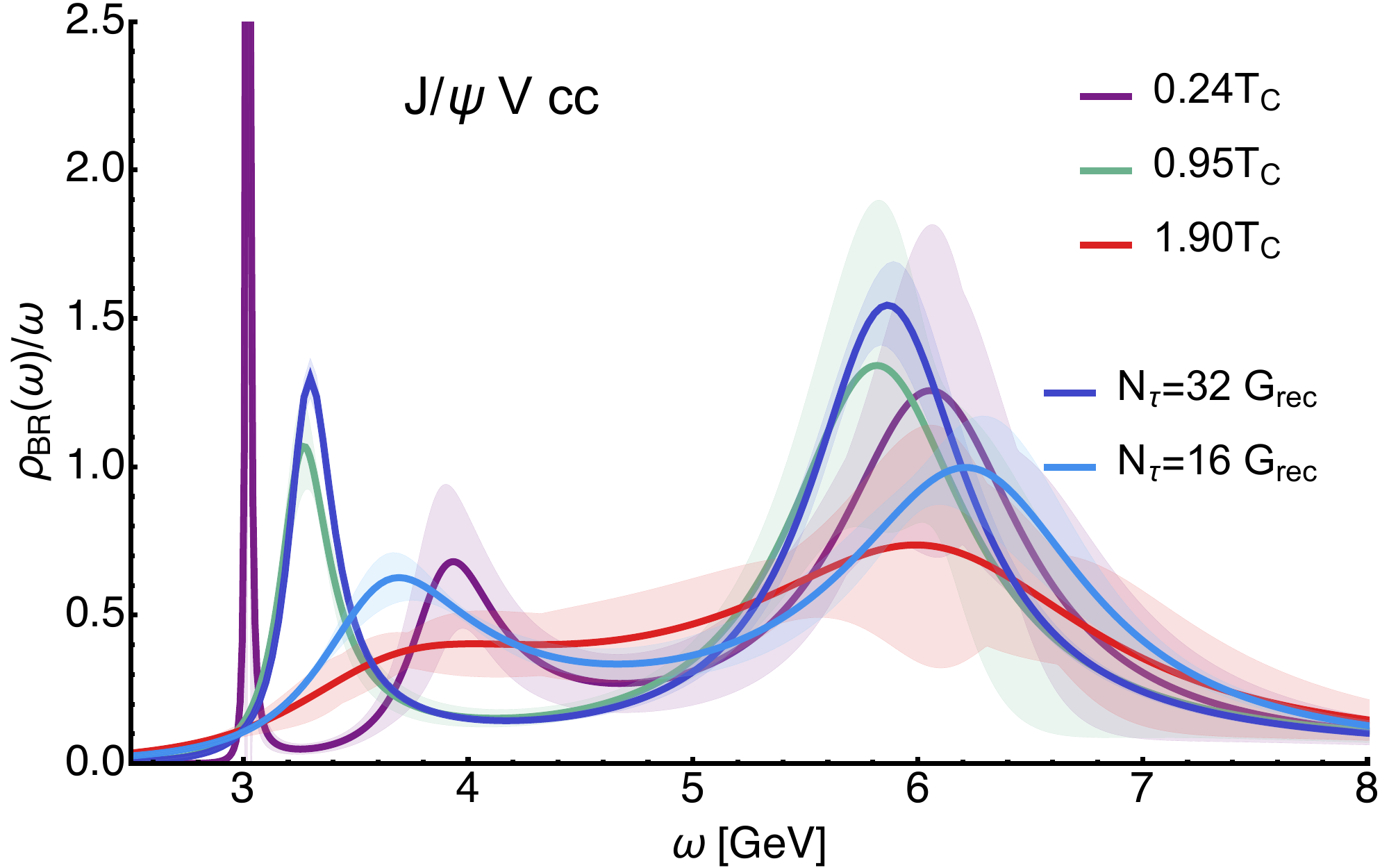}
\includegraphics[scale=0.35]{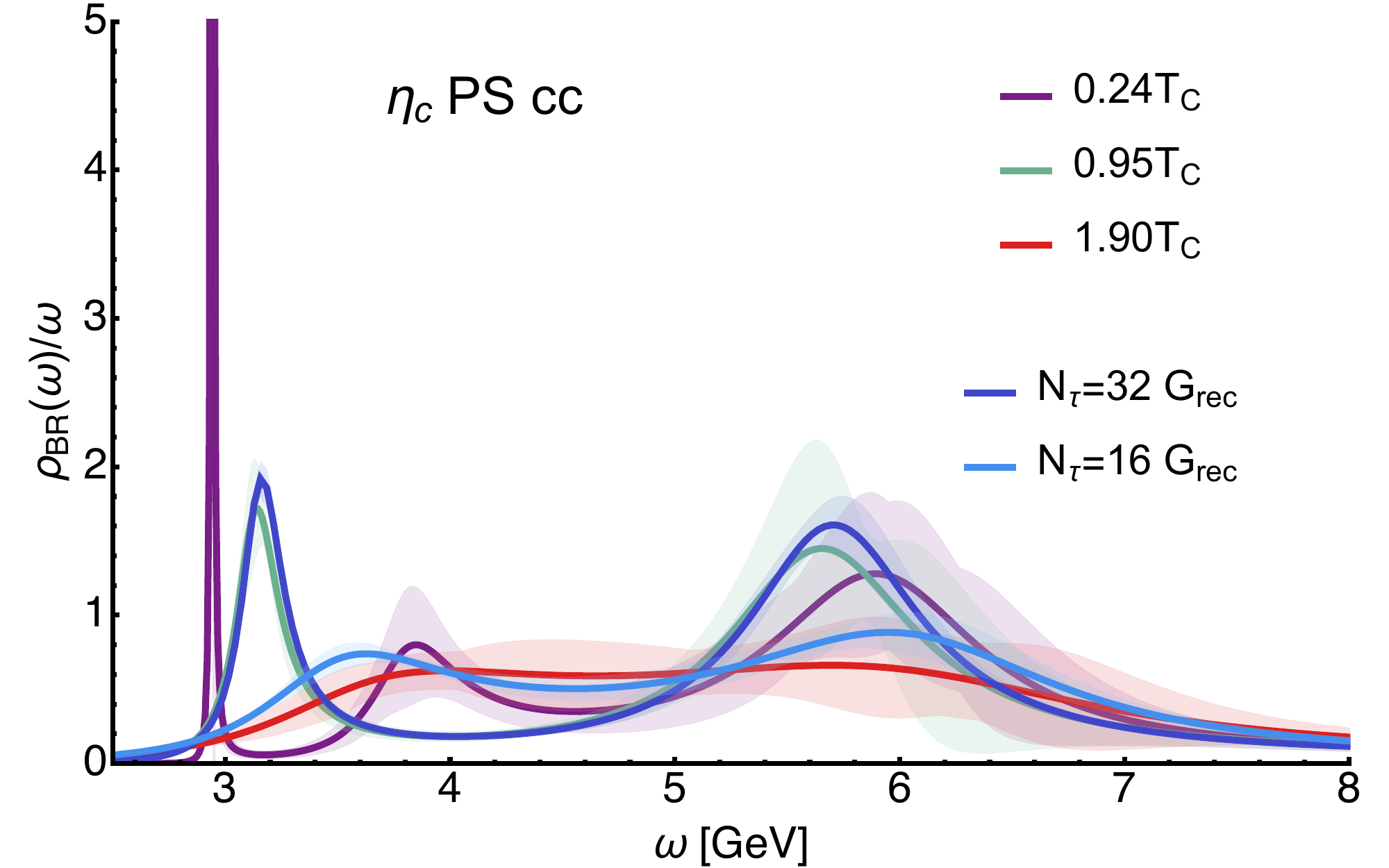}
\caption{Comparison of the BR reconstructed $c\bar{c}$ vector (top) and
  pseudoscalar (bottom) channel $T>0$ spectra with those based on the
  $T\approx0$ reconstructed correlator at the same number of
  datapoints. All reconstructions are obtained using
  $\tau\in[2,N_\tau/2-1]$ and the same statistics. At $0.95T_c$ no significant in-medium
  modification is observed, while at $T=1.9T_c$ there are mild
  indications that the in-medium state has weakened. }\label{Fig:BR_cc_V-PS-Grec}
\end{figure}

\begin{figure}[th!]
\includegraphics[scale=0.35]{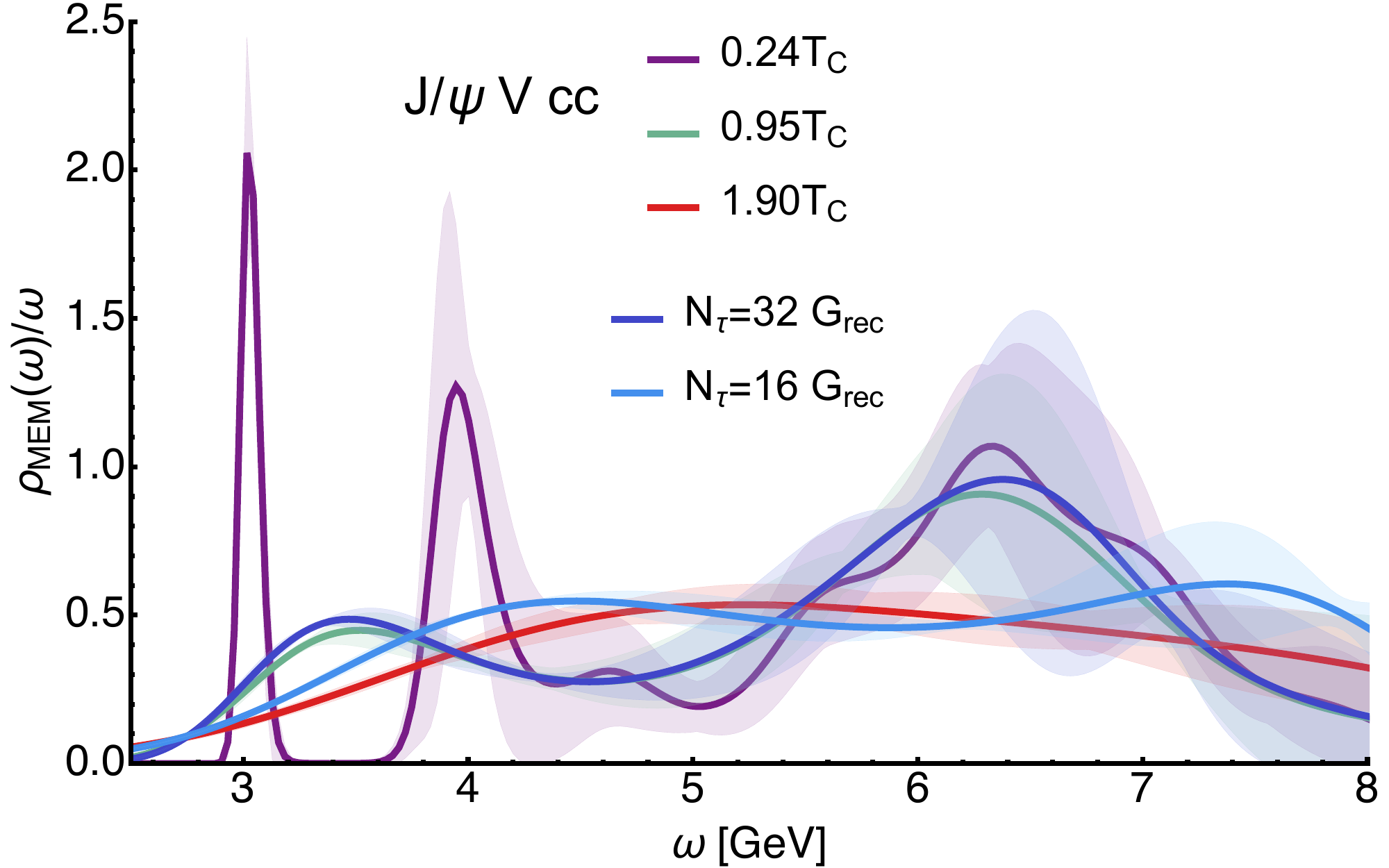}
\includegraphics[scale=0.35]{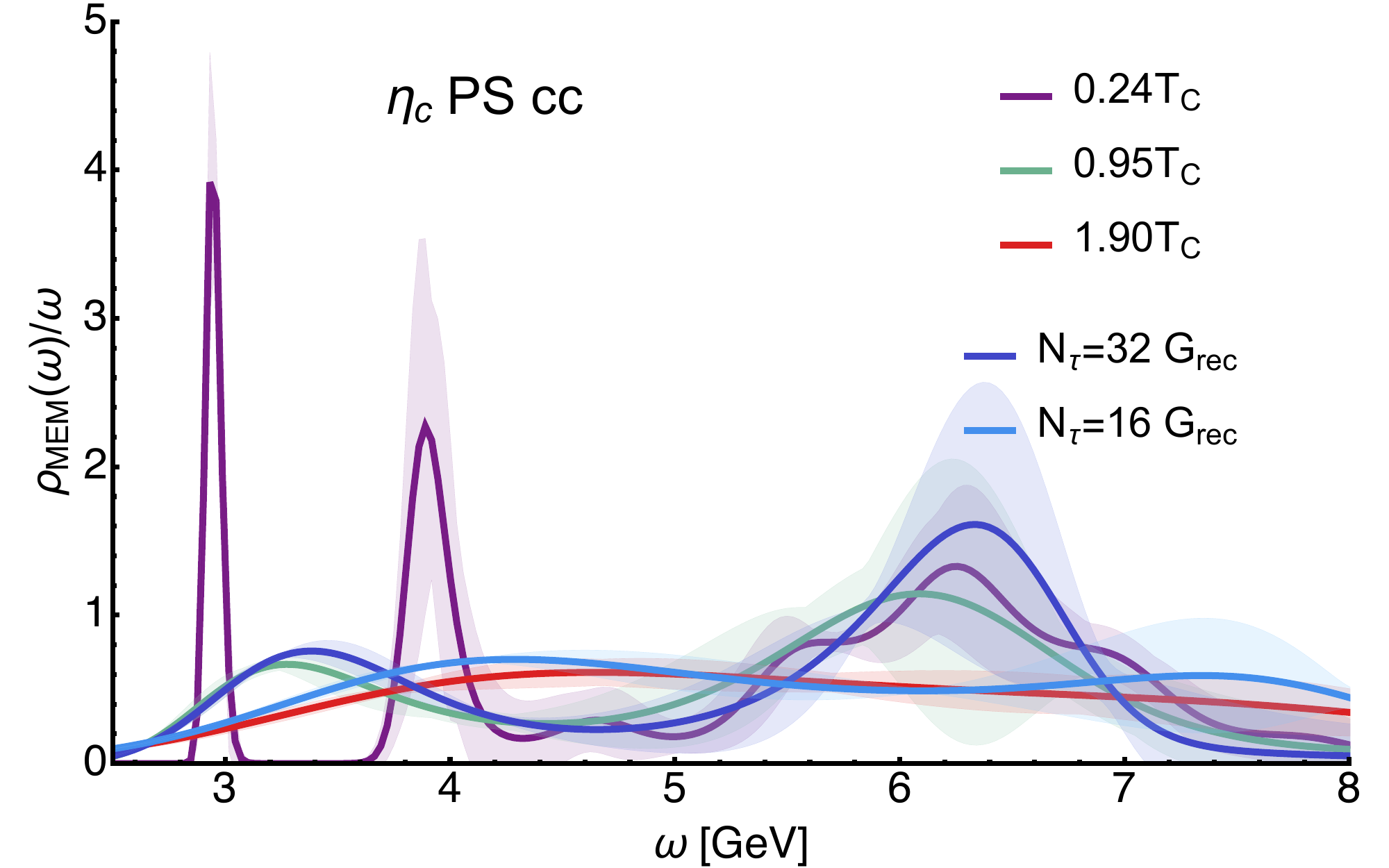}
\caption{As Fig.~\ref{Fig:BR_cc_V-PS-Grec}, using the MEM
  reconstruction.  At $0.95T_c$ no significant in-medium 
  modification is observed, while at $T=1.9T_c$ there are mild
  indications that the in-medium state has weakened.}\label{Fig:BR_cc_V-PS-GrecMEM}
\end{figure}

A first naive inspection of the BR method result by eye tells us that with increasing temperature 
the former ground state structures monotonically move to higher frequencies until above $T\sim1.5T_c$
no discernible structure can be found. The strength of the peak reduces concurrently. The MEM shows qualitatively similar results for
the spectra at $T\lesssim T_c$ but no well pronounced peaks appear above $T_c$. Such differences
between the methods have been observed before in the study of bottomonium in lattice NRQCD \cite{Aarts:2014cda, Kim:2014iga}.
On the one hand the resolution of the MEM reduces with the smaller number of available datapoints
due to a smaller number of admitted basis functions. On the other hand the BR method is known
to be susceptible to ringing artifacts, which may lead to more pronounced peak structures than
actually present in the data.

In light of these different methods artifacts, how can we disentangle them from genuine in-medium effects? It is here
that the reconstructed correlators again play an important role. We already investigated what happens to the reconstruction
of the $T\approx0$ correlator if it is consistently restricted to a smaller number of Euclidean time steps and
found that shifts to higher frequencies and broadening ensued. Thus in Figs.~\ref{Fig:BR_cc_V-PS-Grec} and \ref{Fig:BR_cc_V-PS-GrecMEM}
we compare the reconstructed in-medium spectra at three temperatures corresponding to $N_\tau=128$, $N_\tau=32$ and $N_\tau=16$ with
the reconstruction obtained from the corresponding reconstructed correlator. Fig.~\ref{Fig:BR_cc_V-PS-Grec} depicts the 
outcome based on the BR method, Fig.~\ref{Fig:BR_cc_V-PS-GrecMEM} based on the MEM.

The comparison is illuminating, as it reveals that the shift and broadening observed at $T=0.95T_c$,
contrary to what a simple inspection by eye might suggest, is fully consistent with the vacuum 
spectrum reconstructed from the same number of $N_\tau=32$ datapoints. Only at $T=1.9T_c$ or
equivalently $N_\tau=16$, does the in-medium spectrum show a more strongly washed out behavior than 
the reconstruction from $G_{\rm rec}$. In the vector channel these differences are already significant,
going beyond the combined statistical and systematic errorbars. In the pseudovector channel the difference
is hinted at but is not yet significant.

\subsubsection{P-wave ($\chi_{c0}$ and $\chi_{c1}$) spectra}

\begin{figure}
\includegraphics[scale=0.35]{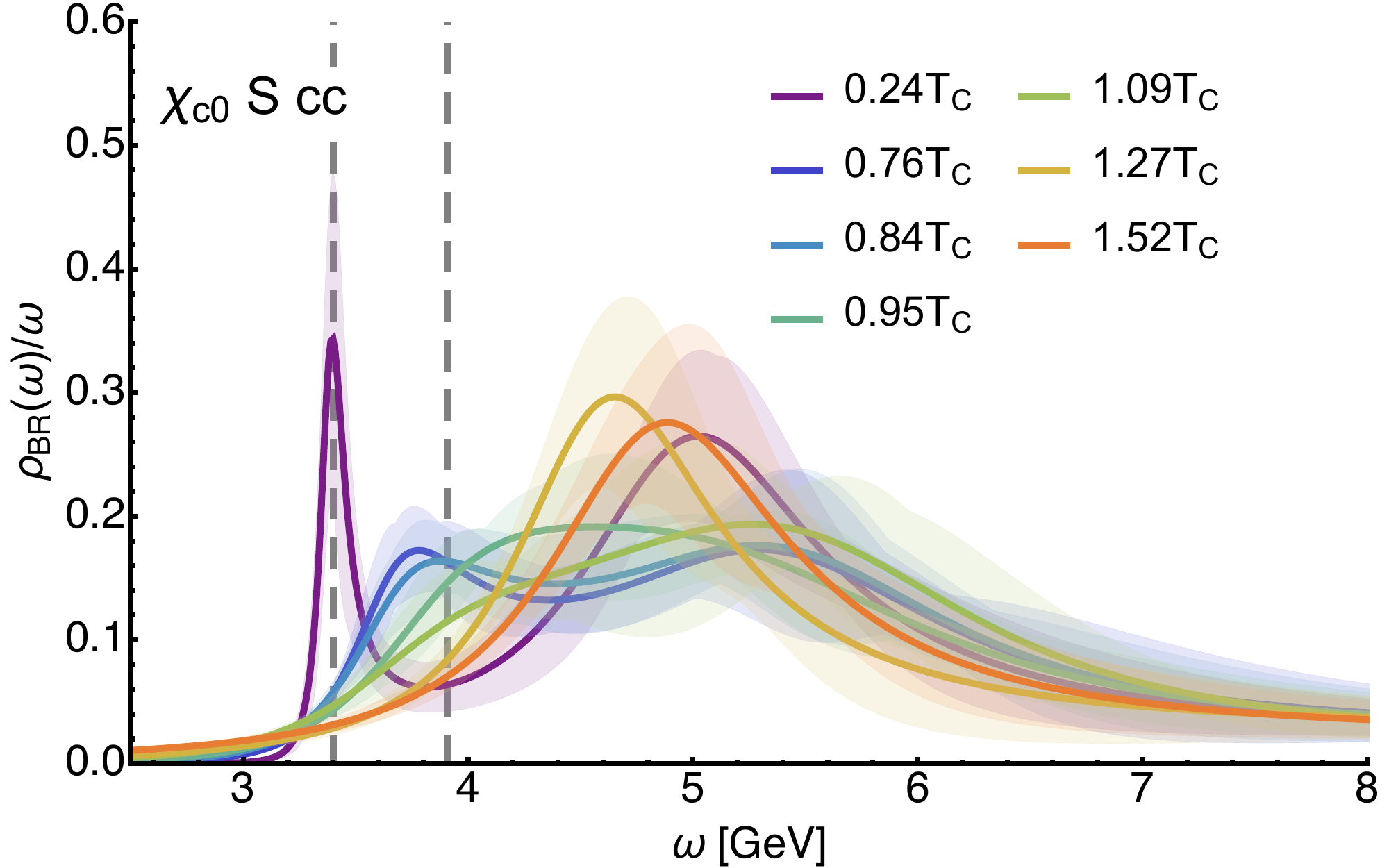}
\includegraphics[scale=0.35]{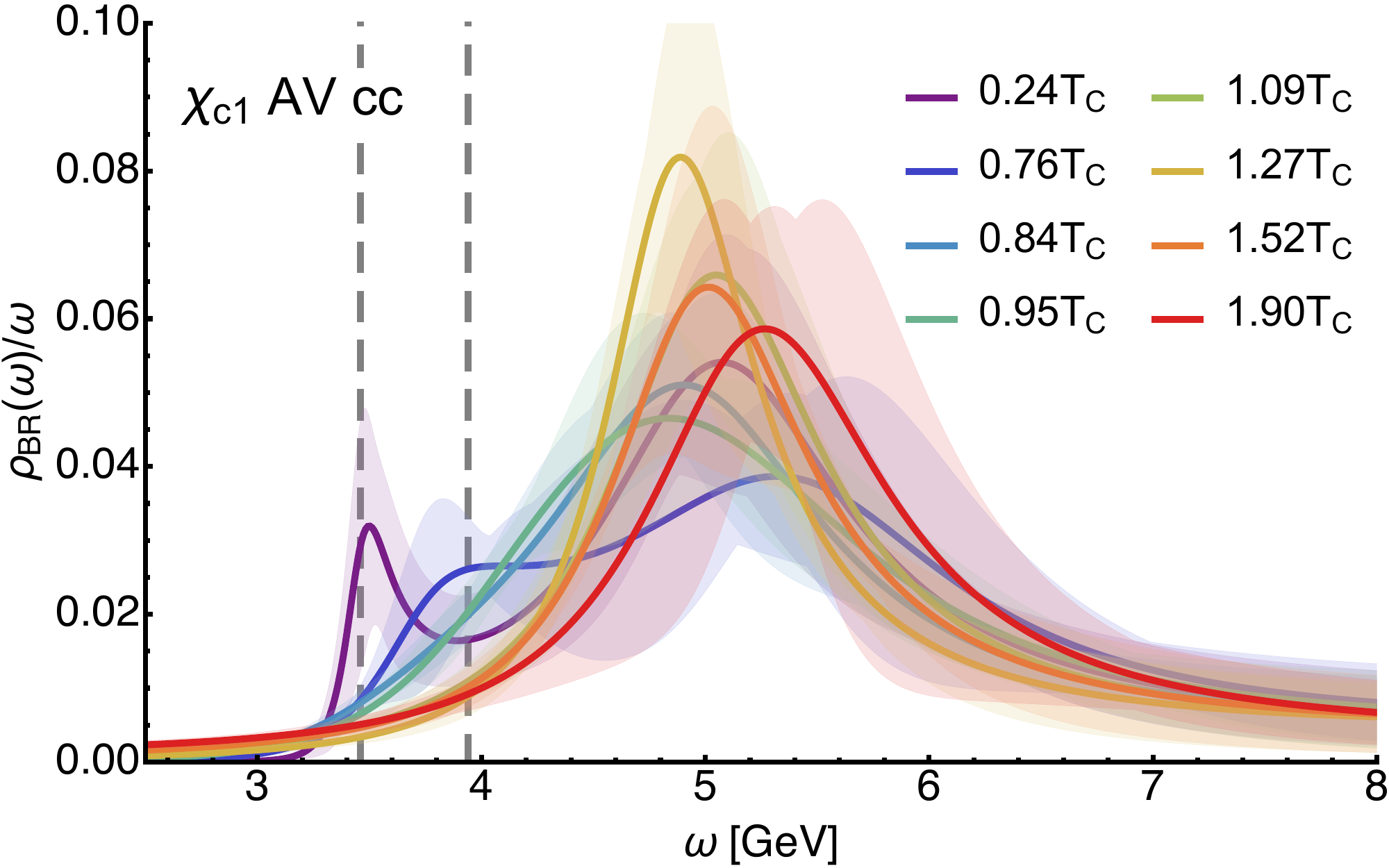}
\caption{Spectral functions for the $c\bar{c}$ scalar (top) and
  axial-vector (bottom) channels, obtained using the BR method.}\label{Fig:BR_cc_S-AV-FiniteT}
  \end{figure}

\begin{figure}
\includegraphics[scale=0.35]{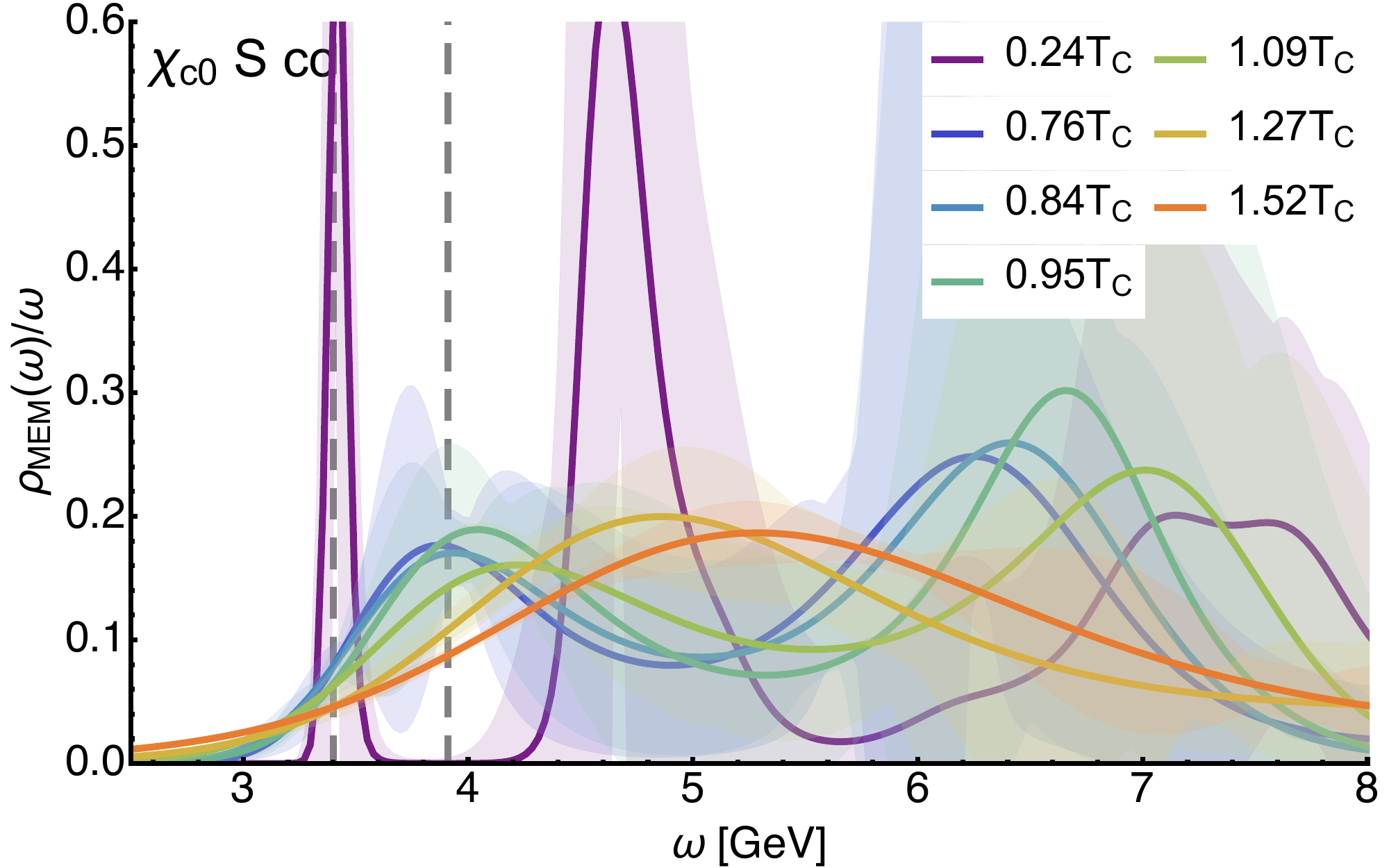}
\includegraphics[scale=0.35]{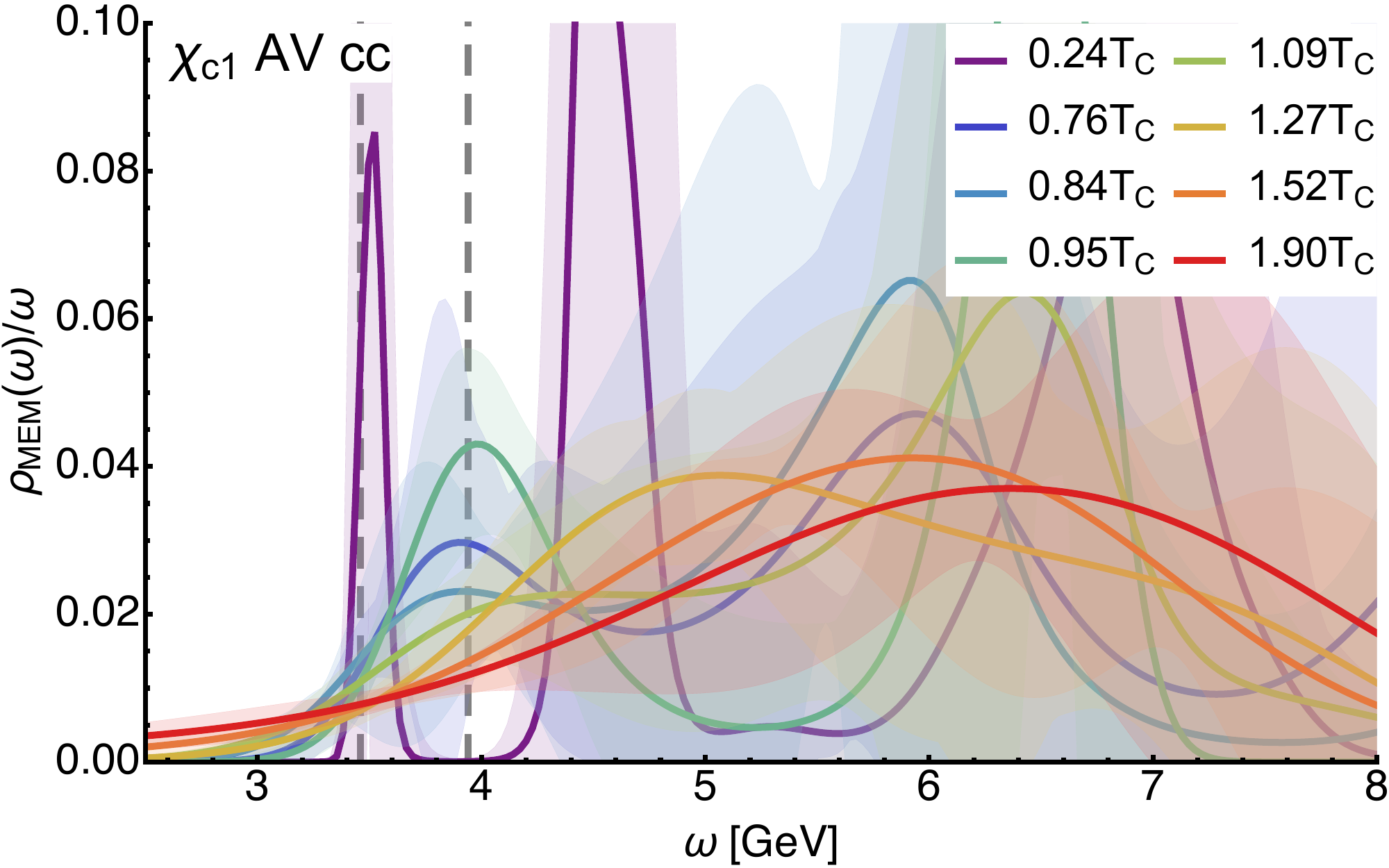}
\caption{Spectral functions for the $c\bar{c}$ scalar (top) and
  axial-vector (bottom) channels, obtained using the MEM with Fourier basis.}\label{Fig:BR_cc_S-AV-FiniteTMEM}
  \end{figure}

We now turn to the P-wave charmonium spectra.
Figure~\ref{Fig:BR_cc_S-AV-FiniteT} and Figure~\ref{Fig:BR_cc_S-AV-FiniteTMEM} show the scalar ($\chi_{c0}$) and
axial-vector ($\chi_{c1}$) spectral functions obtained by the BR
method and by the Fourier basis MEM respectively.

A first inspection by eye suggests that for both channels already at $T=0.76T_c$
there occurs a substantial shift and weakening of the ground state structure.
From $T\approx T_c$ on, no sign of the ground state peak remains. Interestingly
in the $\chi_{c1}$ channel, it is the MEM result that shows slightly more peaked
ground state features below $T_c$ but also with larger uncertainty.


\begin{figure}
\includegraphics[scale=0.35]{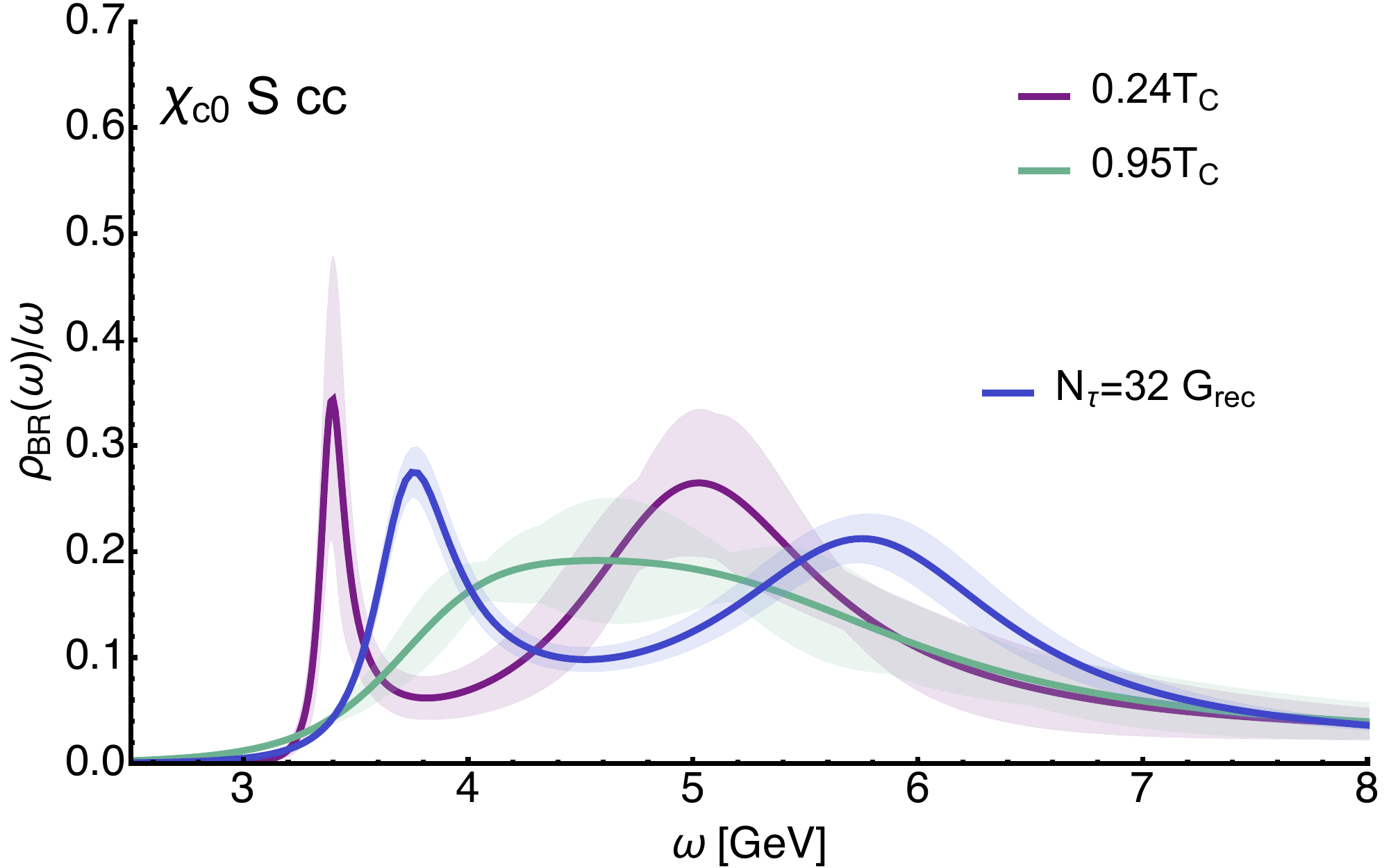}
\includegraphics[scale=0.35]{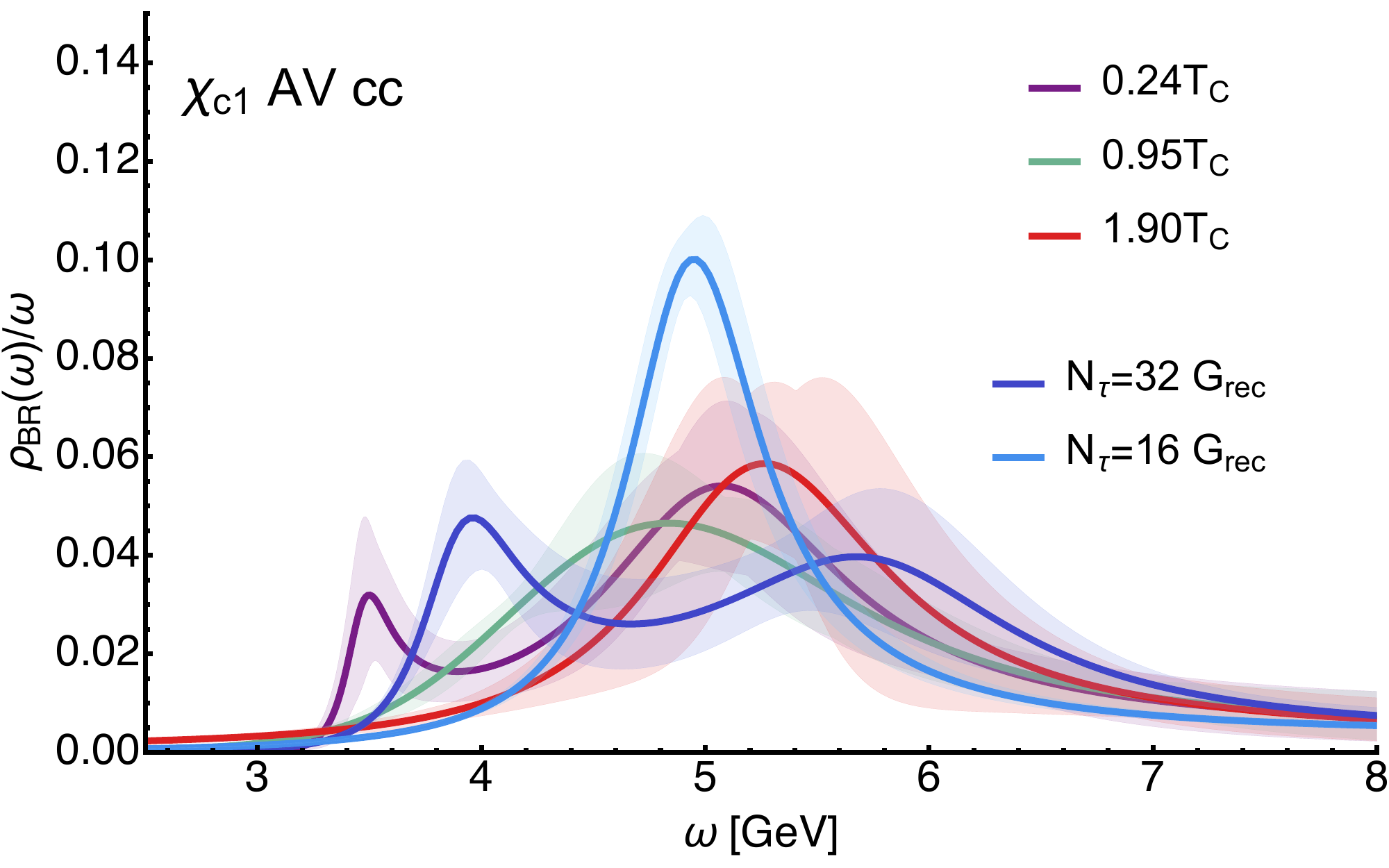}
\caption{Comparison of the BR reconstructed $c\bar{c}$ scalar (top) and
  axial-vector (bottom) channel $T>0$ spectra with those based on the
  $T\approx0$ reconstructed correlator at the same number of
  datapoints. All reconstructions are obtained using
  $\tau\in[2,N_\tau/2-1]$ and the same statistics. Already at
  $0.95T_c$ a significant in-medium
  modification is observed, and no discernible peak structures are found.}\label{Fig:BR_cc_S-AV-Grec}
\end{figure}

\begin{figure}[th!]
\includegraphics[scale=0.35]{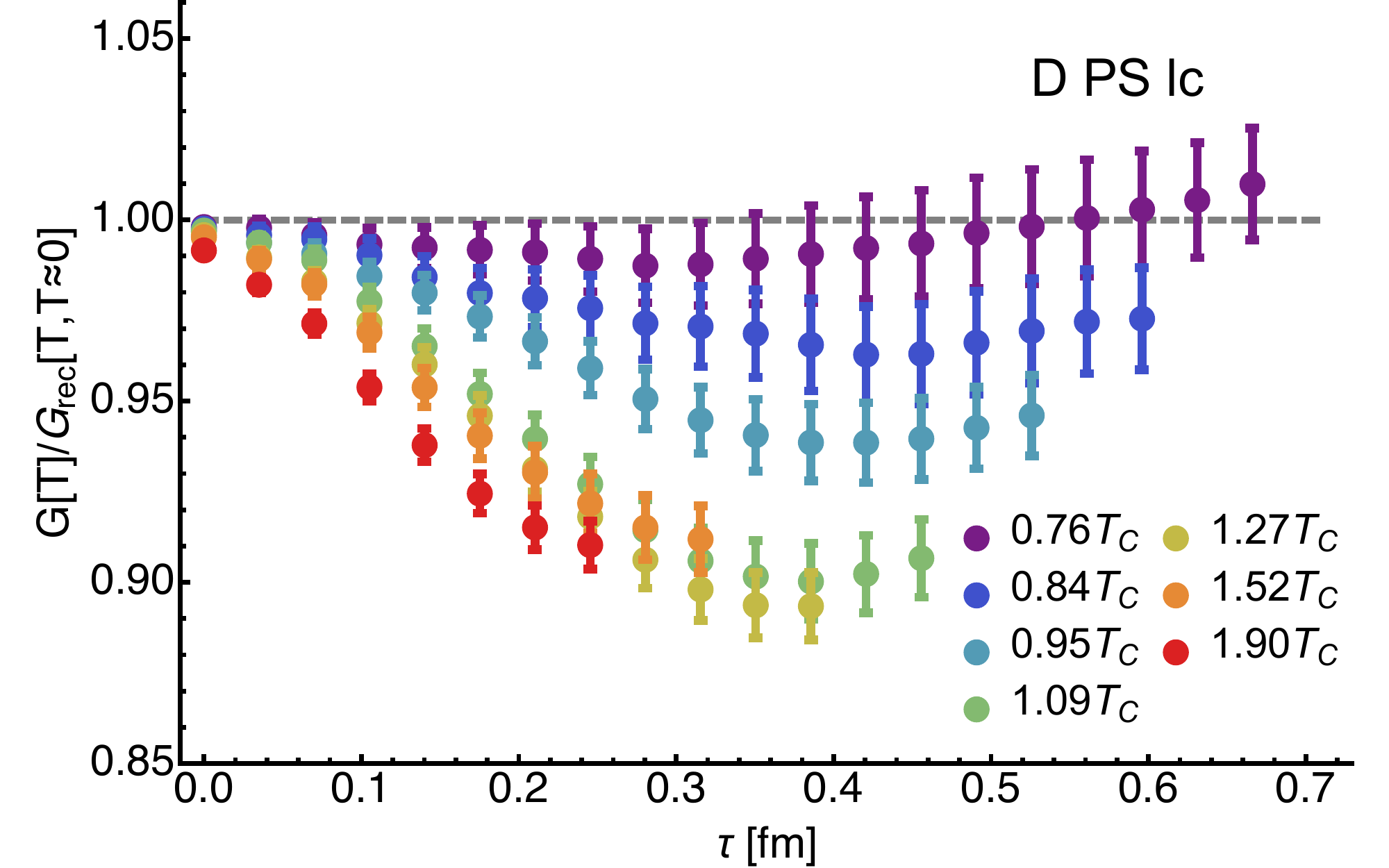}
\includegraphics[scale=0.35]{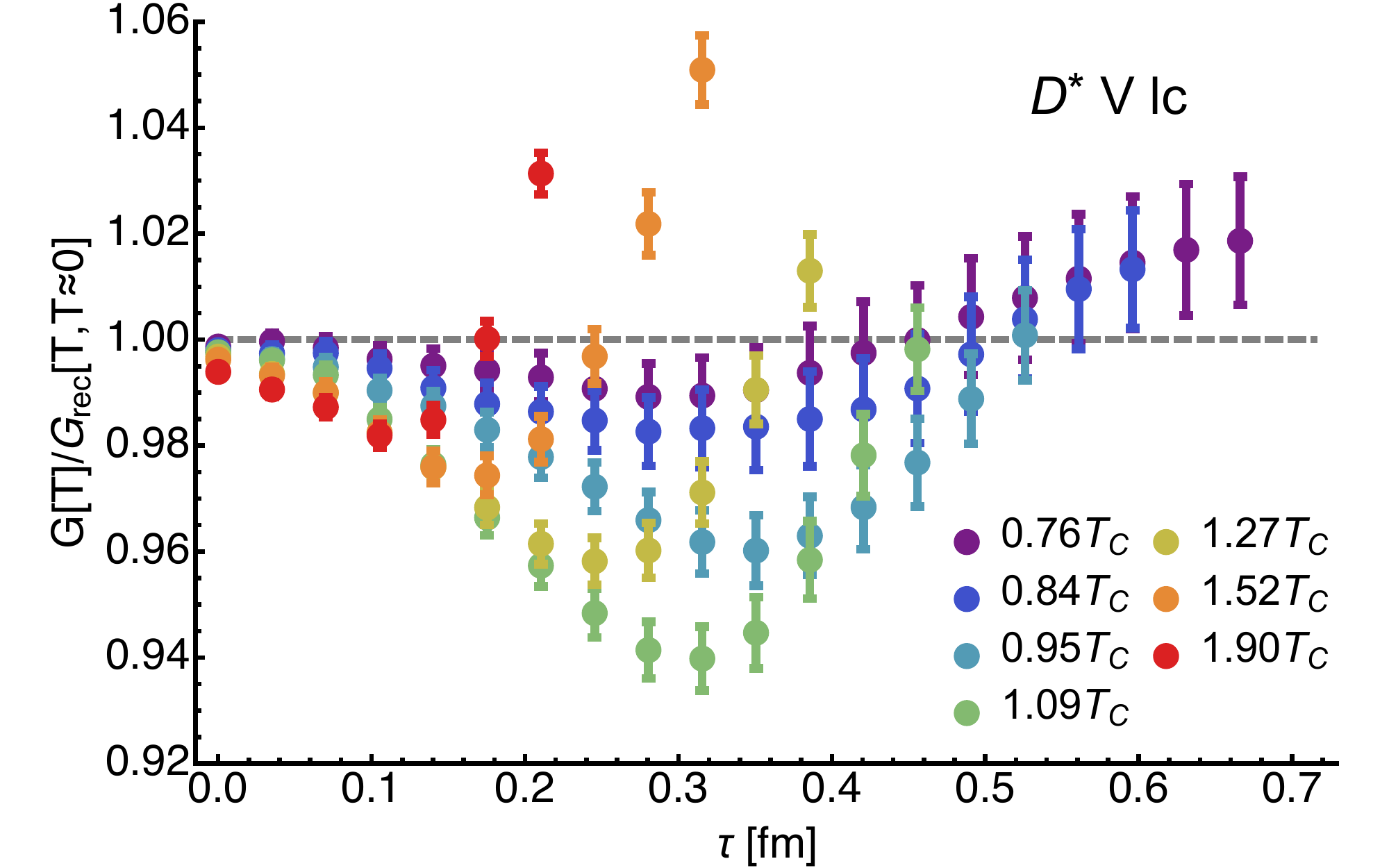}
\includegraphics[scale=0.35]{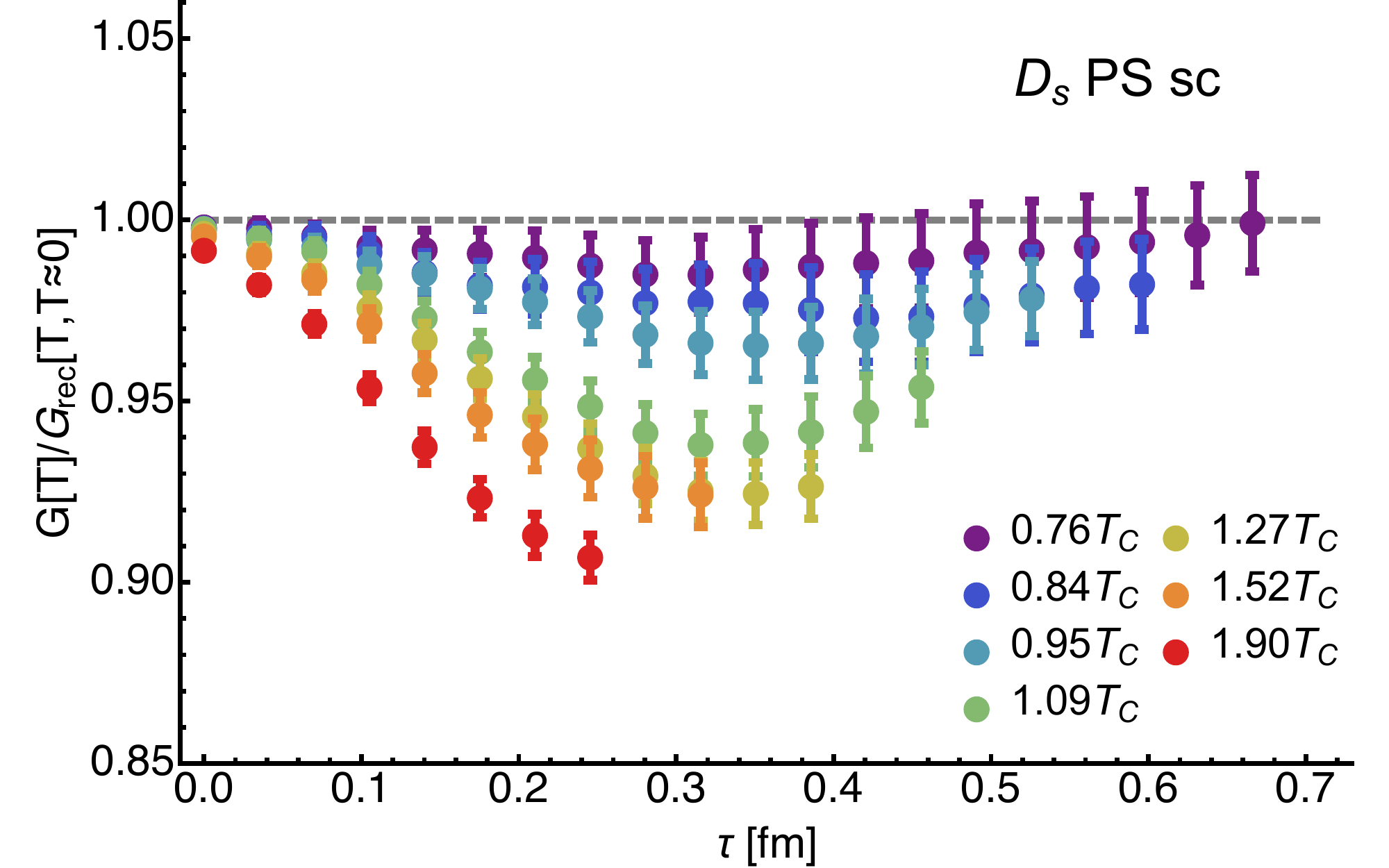}
\includegraphics[scale=0.35]{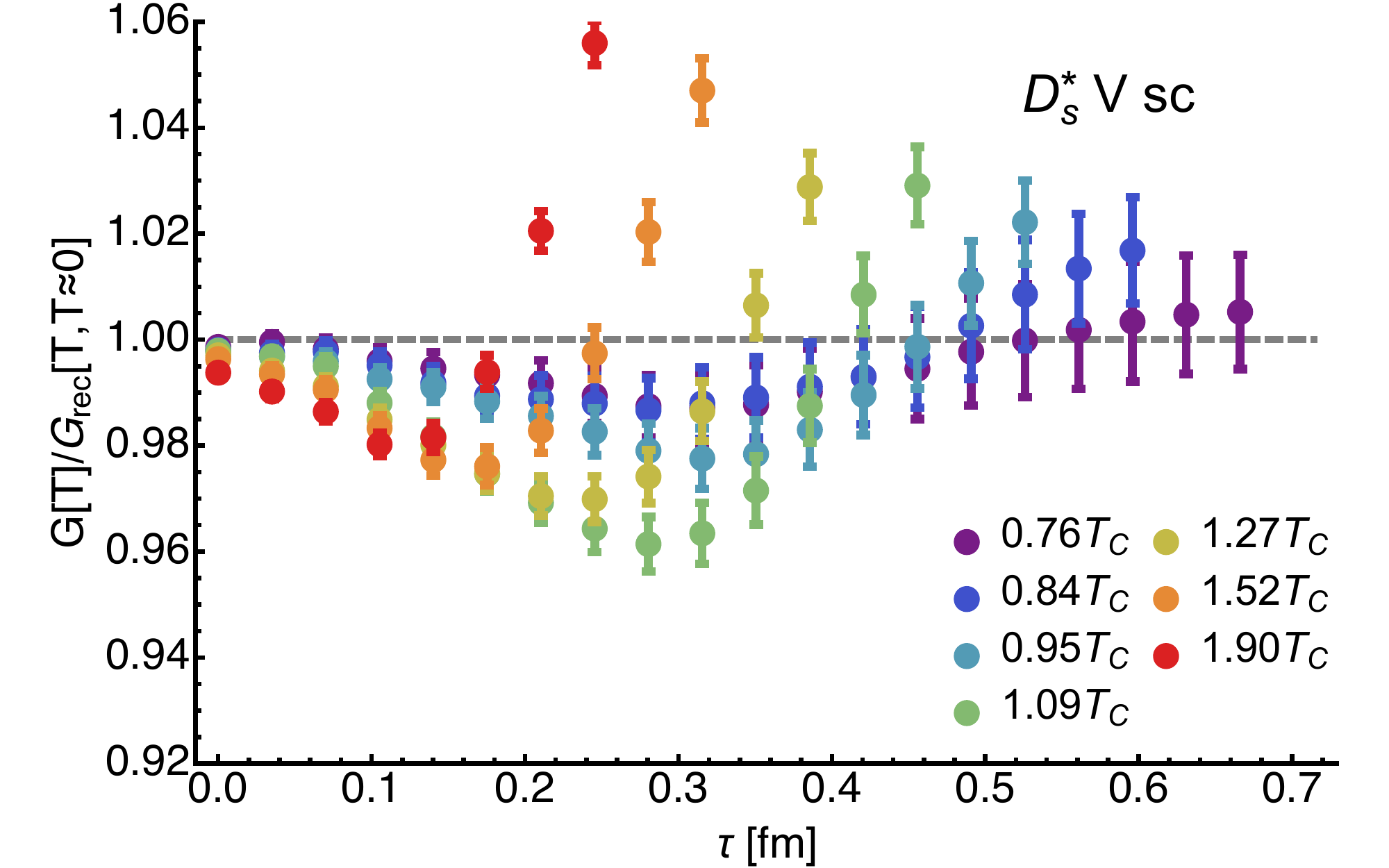}
\caption{Ratio of the in-medium open-charm correlator to the corresponding reconstructed correlator for the $(u,d)+c$ (top two panels) and for $s+c$ (lower two panels). For each combination of quarks both the pseudo-scalar (upper) and the vector (lower) channel are shown.}\label{Fig:CorrRatio_lcsc}
\end{figure}

To ascertain whether this first impression holds true, let us carry out the comparison to the
spectra from the reconstructed correlator again, shown in Fig.~\ref{Fig:BR_cc_S-AV-Grec}.
As was the case with the S-waves, the restricted temporal range leads to an upward shift and
gradual weakening of the ground state peak.  However, at $T=0.95T_c
(N_\tau=32)$, the ground state peak is clearly present in the
reconstructed correlator spectral function, while the in-medium spectral
function obtained from the thermal correlators shows no sign of such a
peak.  Within the current data quality we conclude that the 1P state disappears
from the spectrum below or near $T_c$. Since the MEM comparison for the P-wave
carries very large errorbands, we do not gain any further insight from it and omit it 
from our discussion here.

On the one hand the P-wave channels are those with the smallest signal to noise
ratio, leading to sizable errorbands. On the other hand the magnitude of the
in-medium modification already observed in the correlator ratio is highly significant 
and can be therefore be picked up by the Bayesian reconstruction. We are currently 
increasing the statistics on the P-wave correlators to bring them to a similar level
of relative precision to those of the S-wave to make sure that the relatively early
disappearance of the ground state peak is not simply due to a lower precision.

\subsection{Open Charm at finite temperature}
\label{OpenCharmFiniteT}

After considering the charmonium states, we now turn to the study of 
the open charm mesons. Here we consider the pseudoscalar ($D, D_s$)
and vector ($D^*, D_s^*$) channels, as these are the most relevant for
phenomenology.

\subsubsection{Correlatior ratios}

Up to this point open heavy flavor mesons at finite temperature were solely investigated
by use of spatial correlation functions or bulk observables. Here we present for the first time the actual Euclidean
time correlator ratios, which have a direct connection with the in-medium spectral function.

In figure~\ref{Fig:CorrRatio_lcsc} we show the correlators $G(\tau)$ divided by
the reconstructed correlators $G_r(\tau)$ at the same temperature, for
all four channels.  We observe that the correlator at $T=0.76T_c$
is still consistent with no modifications in all channels, while some nontrivial
trough structure is already hinted at at intermediate $\tau$.

Starting from $T=0.84T_c$ genuine departures from unity emerge. The maximum changes we 
find are around $\lesssim10\%$, which is stronger than what is observed in the corresponding 
charmonium channels, consistent with a lower binding energy. 

The shape of the changes is also distinct from that observed in the charmonium sector. Instead of 
an upward bending that increases with temperature, a deeper and deeper trough emerges
with a minimum at around $\tau\approx0.35$fm. One possible origin of this difference
may lie in the absence of a densely populated regime of excited states, which cannot melt as the temperature increases. Since the potential picture is
not readily applicable for open-heavy flavor, a similarly intuitive explanation as for the modification of charmonium 
is not at hand. We note that the pseudoscalar $D$ meson ratio behavior differs 
clearly from that of pseudoscalar charmonium discussed in the previous section but is
reminiscent of what has been observed for the $\eta_c$ in
\cite{Ding:2012sp} and in the first generation FASTSUM ensembles.

\begin{figure*}[th!]
\includegraphics[scale=0.35]{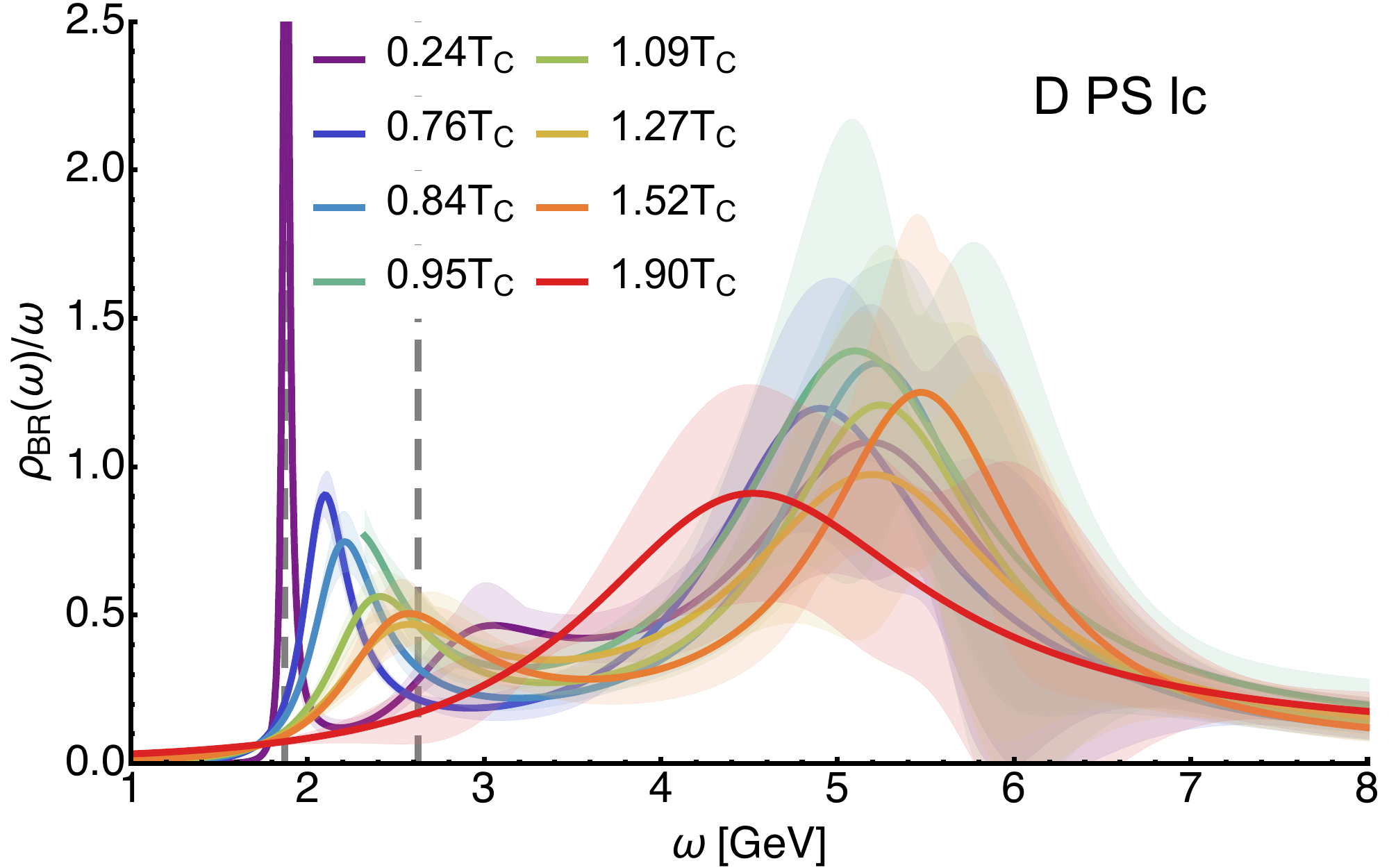}
\includegraphics[scale=0.35]{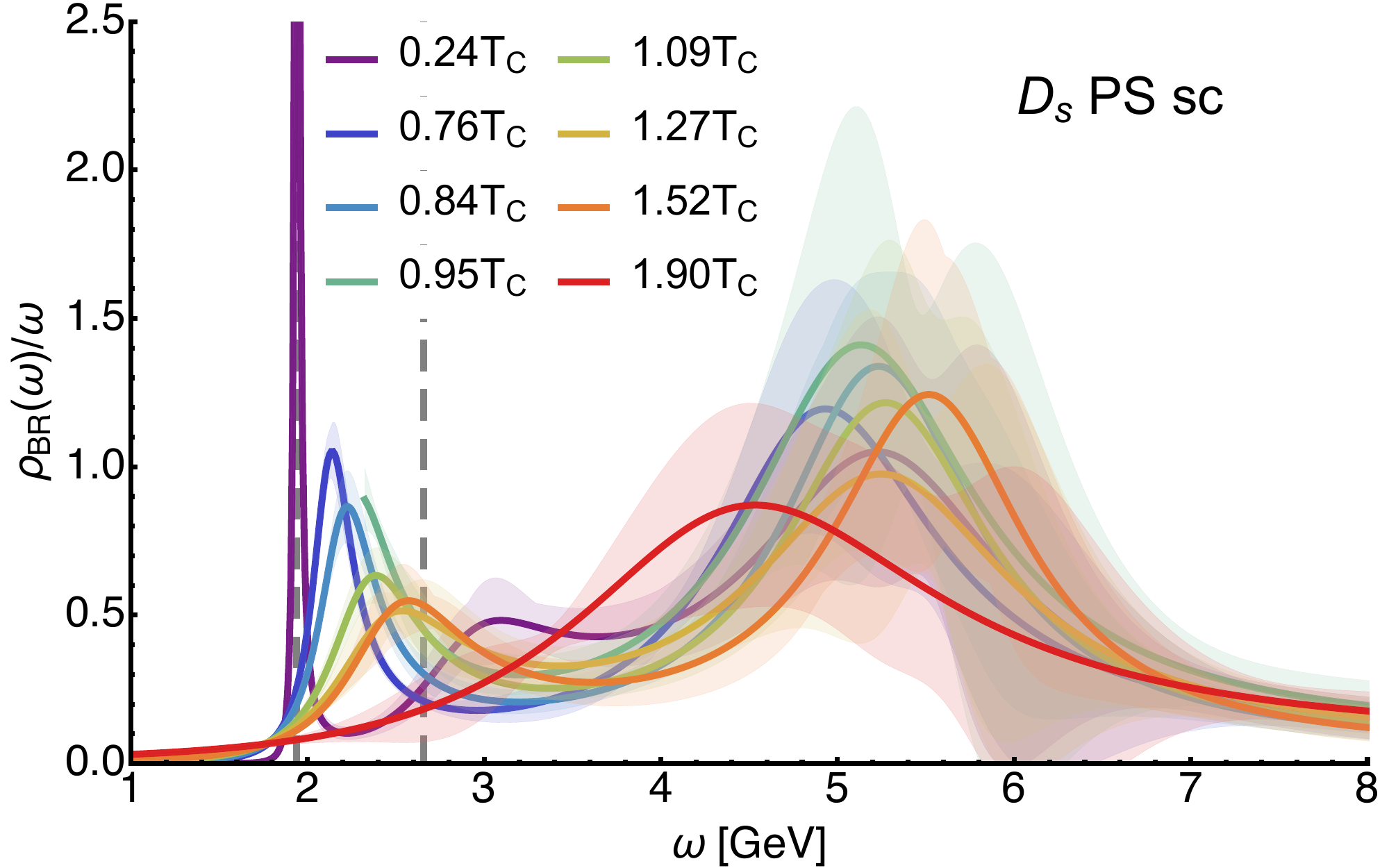}
\includegraphics[scale=0.35]{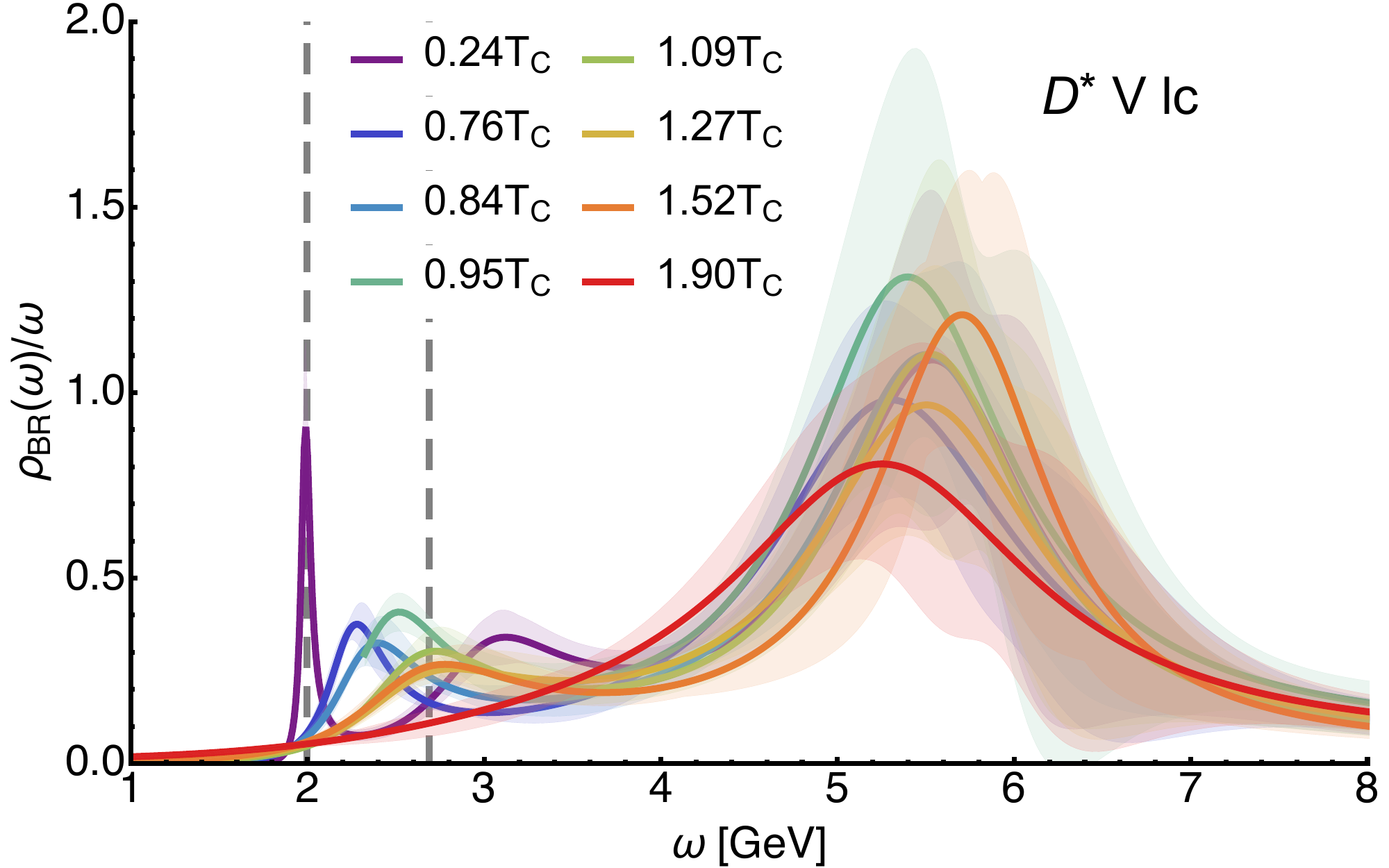}
\includegraphics[scale=0.35]{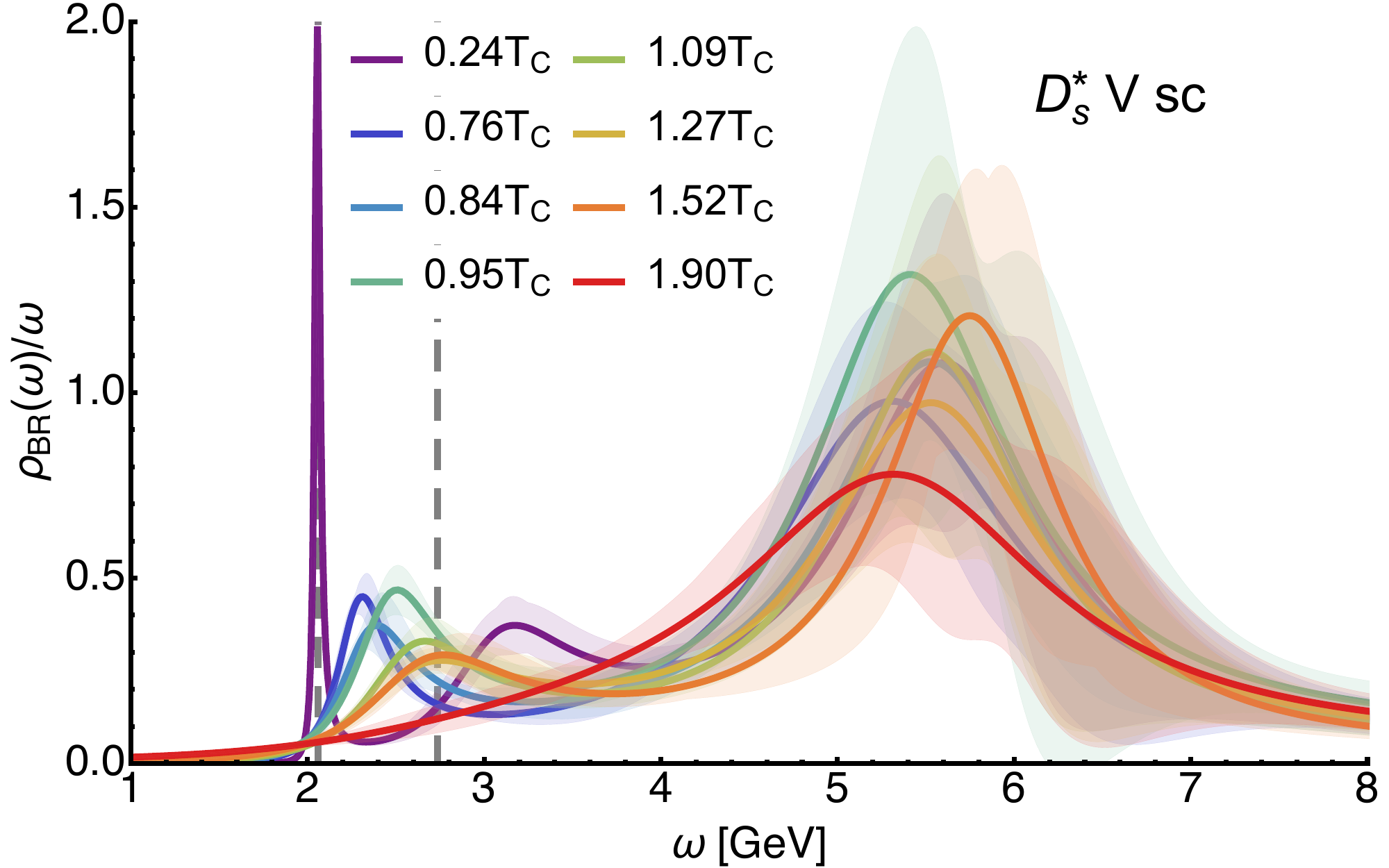}
\caption{Spectral functions for the $lc$ (left) and $sc$ (right)
pseudoscalar channels, obtained using the BR method.  Also shown are the ground
  and first excited state energies determined from a variational
  analysis \cite{Moir:2013ub}.}\label{Fig:BR_lcsc_PS-FiniteT}
\end{figure*}

Up to  around $T_c$ the trough minimum seems to remain almost at the same position 
in all channels. In the $D$ meson case, the qualitative behavior
remains the same also above $T_c$ and the trough deepens. For the
$D^*$ mesons however, a sharp 
upward rise sets in at $\tau/a >N_\tau/4$. 

Let us take over some intuition from the potential computations of quarkonium, where the upward bending arises
once the lower lying peak structures in the spectrum start to be affected by in-medium effects.
While the ground state masses of $D$ and $D^*$ mesons differ by less than 100MeV,
(the latter being heavier than the former) the distinct patterns in the ratio suggest that 
their stability against in-medium effects differs significantly. $D^*$ mesons appear much more
sensitive to thermal fluctuations than $D$ mesons, which {\it in vivo} should lead to 
to a stronger suppression. 

Several heavy-ion experiments, i.e. STAR \cite{Adamczyk:2014uip}, CMS \cite{Sirunyan:2017xss} 
and ALICE \cite{Adam:2015nna,Adam:2015sza,AlicePub1703} have 
conducted measurements of the nuclear suppression factor of open charm mesons. While the former two
collaborations have presented mainly $D$ results, ALICE has determined both $D$ and $D^*$
production yields separately. The currently available data quality
from Run 1 and Run 2 at the LHC
has allowed them to determine $R_{AA}$ at intermediate $p_T$ and a large
number of participants corresponding to the $10-20\%$ centrality class. It is in this regime where the charm quarks have the highest probability 
to become equilibrated with the medium and thus we may attempt to connect to our fully thermal results.

In Run 1 no indications have been found that the $D$ or $D^*$ species show different 
suppression patterns. There is a slight tendency of $D^0$ to show a bit stronger
suppression than the $D^*$ even though the difference is not
significant. The preliminary Run 2
data on the other hand show a tendency for the $D^*$ meson to be more strongly suppressed at 
low $p_T$ compared to the $D$'s, but again due to the relatively large errorbars the effect is
not significant yet. A possible discrepancy between 
our observation of a significantly different behavior of the $D$ and $D^*$ correlator ratios
and the absence of such differences in the observed suppression needs to be further understood.

On the theory side our interpretation of the upward bending of the ratios is based on a naive
intuition borrowed from potential model computations of charmonium and surelys needs refinement. 
A direct determination of the spectral properties of course would be most illuminating, which is
what we will attempt in the following section. 

\begin{figure*}[th!]
\includegraphics[scale=0.35]{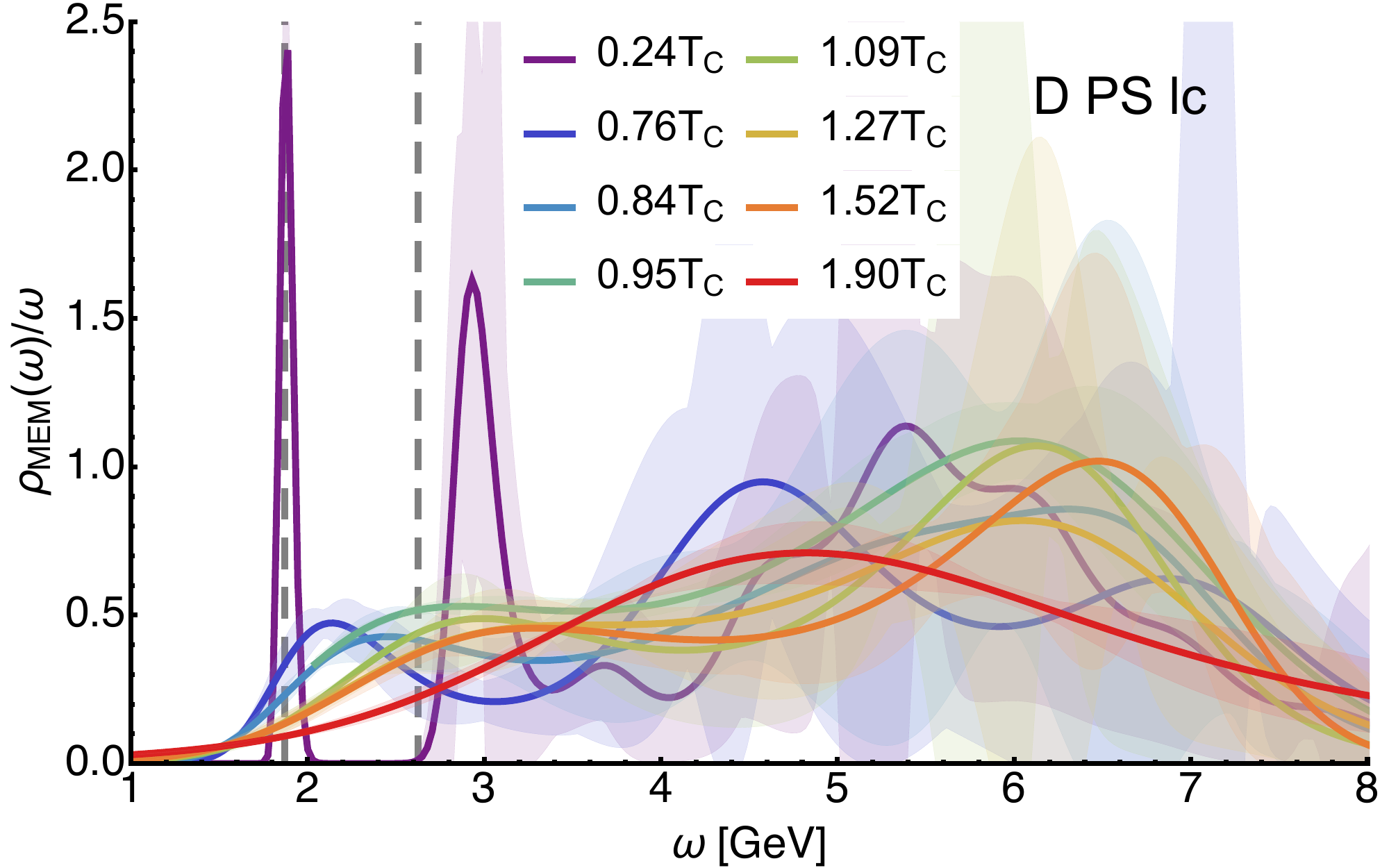}
\includegraphics[scale=0.35]{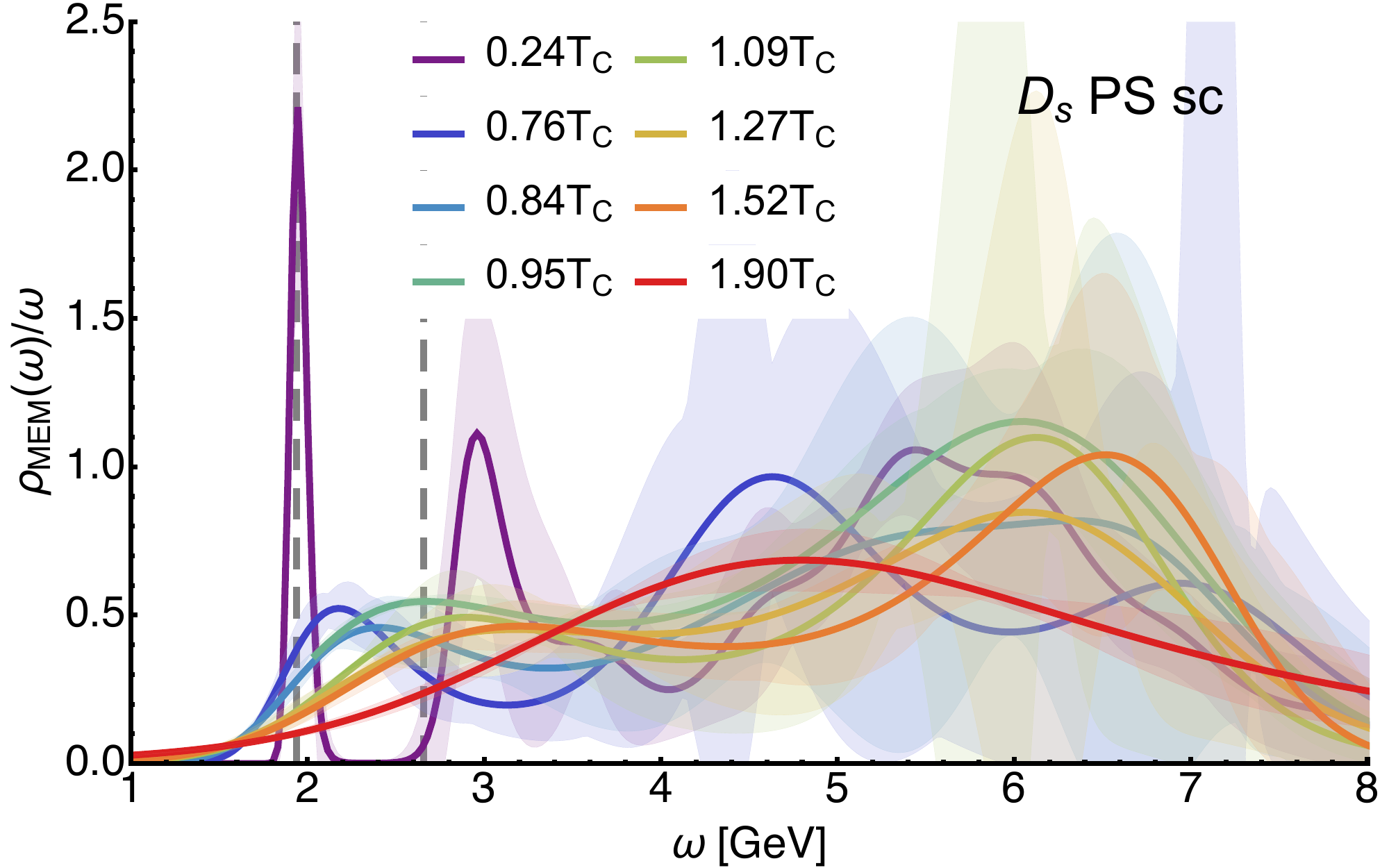}
\includegraphics[scale=0.35]{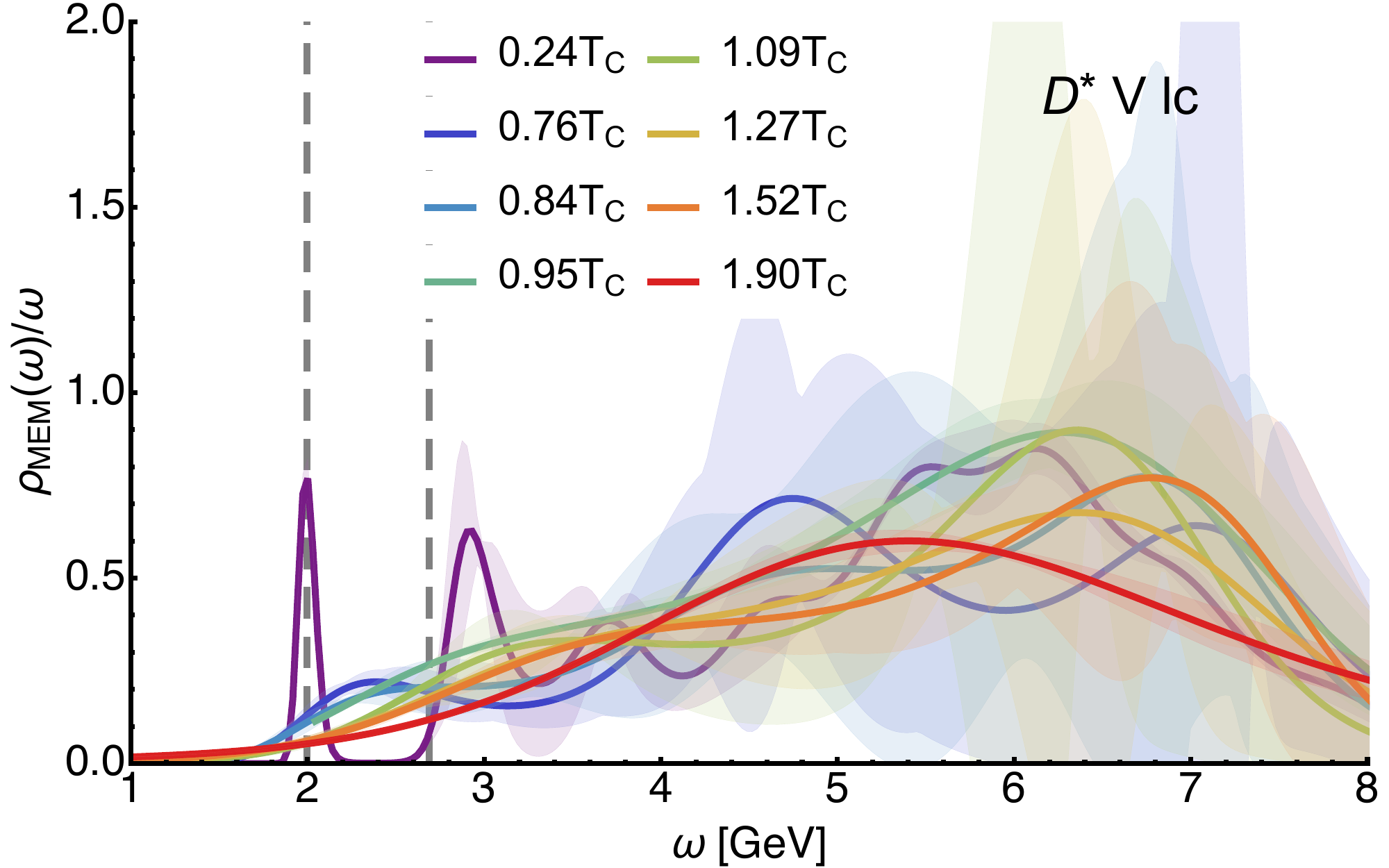}
\includegraphics[scale=0.35]{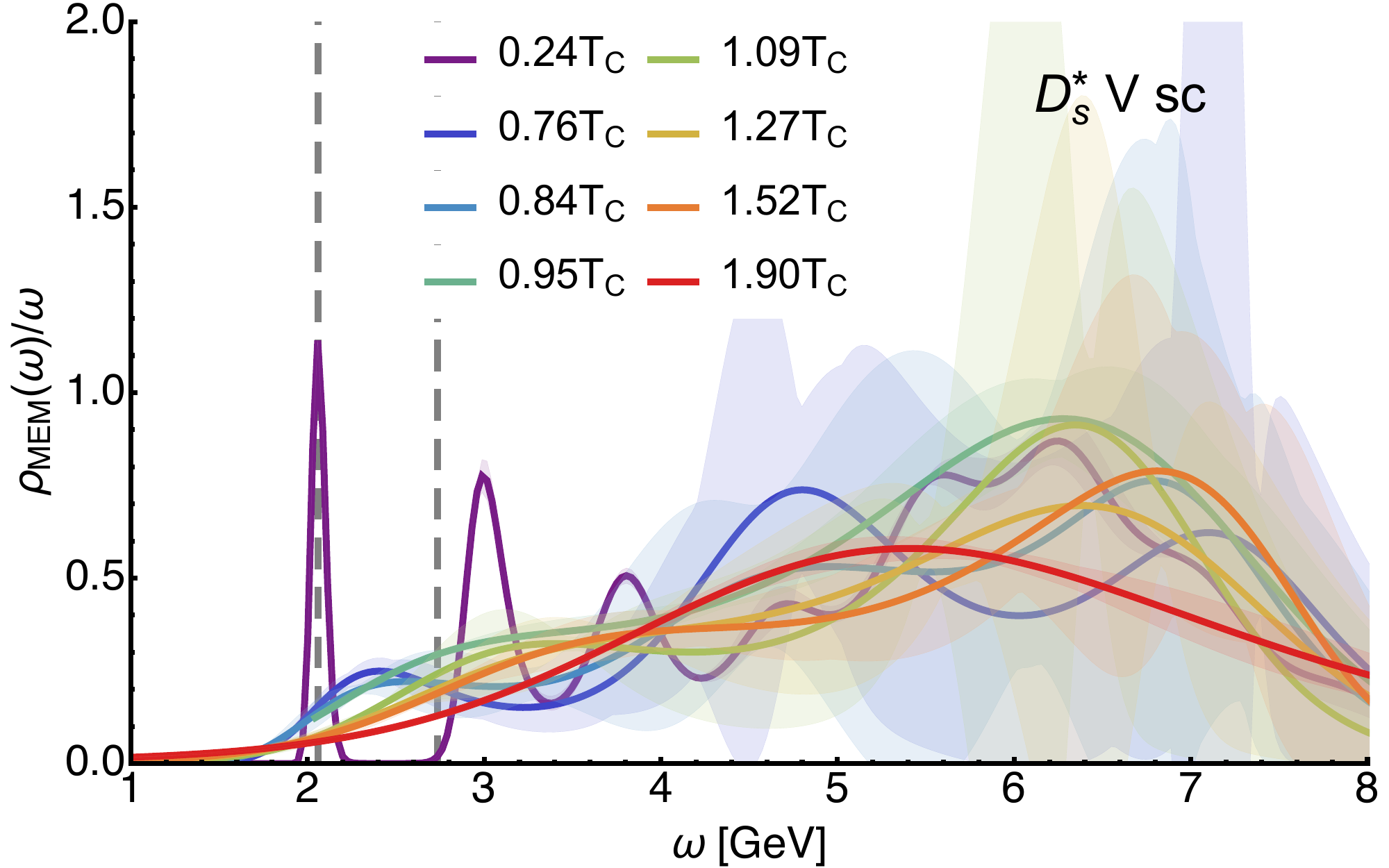}
\caption{Spectral functions for the $lc$ (left) and $sc$ (right)
pseudoscalar channels, obtained  using the MEM with the Fourier basis.  Also shown are the ground
  and first excited state energies determined from a variational
  analysis \cite{Moir:2013ub}.}\label{Fig:BR_lcsc_PS-FiniteTMEM}
\end{figure*}

On the experimental side, the Run 2 data are still being investigated and it will be very interesting to see
whether the tendency of the low $p_T$  suppression of $D^*$ will eventually prevail in the final analysis.

Note also that except for the highest temperature ($T=1.9T_c$), the modifications in the strange--charm 
sector are smaller than those in the light--charm sector, which is consistent with the
the hypothesis that $D_s$ yields may be increased relative to $D$ yields. Experimentally this 
hypothesis is supported by the preliminary ALICE Run 2 analysis, which clearly showed that the $R_{AA}$ for
$D_s$ is larger than the average $R_{AA}$ for $D$ and $D^*$.

\subsubsection{Spectral functions}

We saw that the open-charm correlator ratios behave very different to their charmonium counterparts.
In order to give this theory observation a physical interpretation we need to translate it into a statement about the
spectral properties of open charm mesons. As a first step we may ask whether the in-medium changes of the 
correlator are an indication of an equally different behavior of the ground state spectral structure? To this end 
 we carry out the corresponding Bayesian spectral reconstructions.

In Fig.~\ref{Fig:BR_lcsc_PS-FiniteT} and Fig.~\ref{Fig:BR_lcsc_PS-FiniteTMEM} the results based on the BR method
and the MEM are plotted respectively. The gray dashed lines are the vacuum ground and first excited state masses
obtained from a variational approach in Ref.~\cite{Moir:2013ub}.

Both methods show that below $T_c$, the $D$ mesons exhibit consistently more pronounced structures, compared to their
$D^*$ cousins. The naive inspection by eye again finds a broadening and shifting of peaks with temperature.
The BR method exhibits remnant peak structures up to $T\approx
1.5T_c$, while no sign of 
any remnant structure appears at $T=1.90T_c$. The MEM on the other hand 
shows overall more washed out structures, so that at $T>T_c$ one is hard pressed to identify
a genuine peak. For both methods we also see that above $T>T_c$ the strength of the  $D$ meson structures is slightly stronger
than those of $D^*$, which is consistent with the observation of stronger deviations from unity in the $D^*$ correlator ratio.

\begin{figure*}
\includegraphics[scale=0.35]{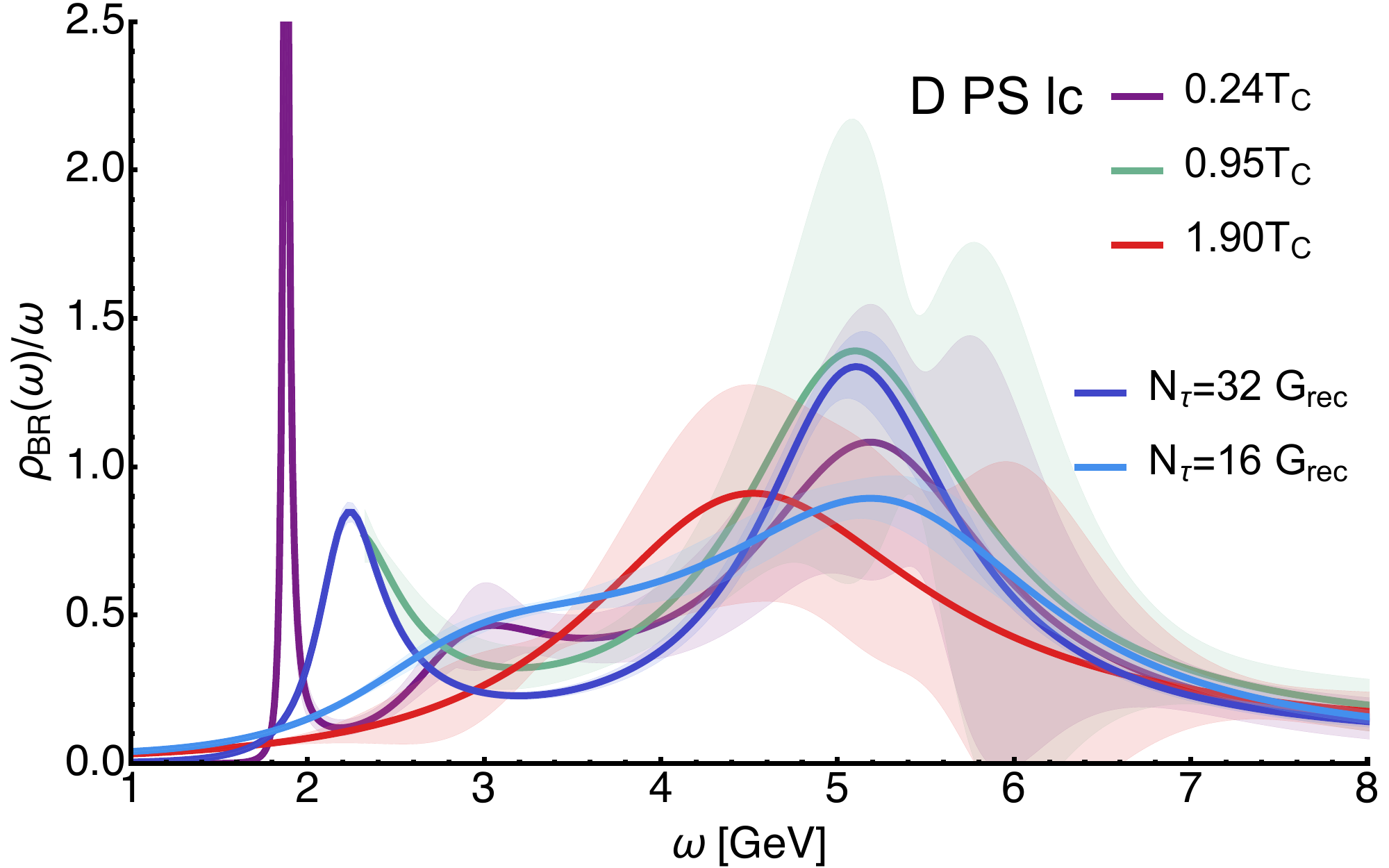}
\includegraphics[scale=0.35]{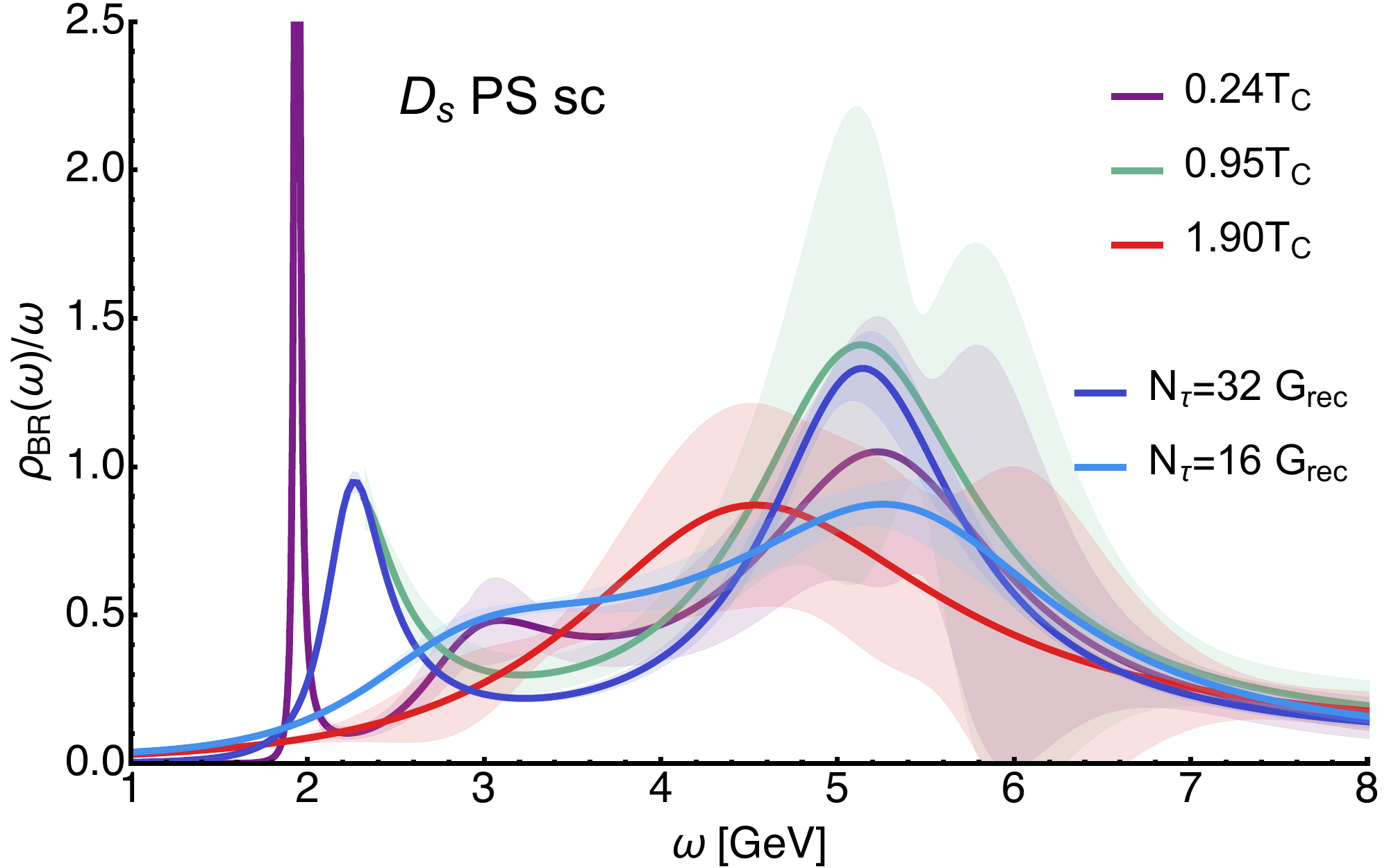}
\includegraphics[scale=0.35]{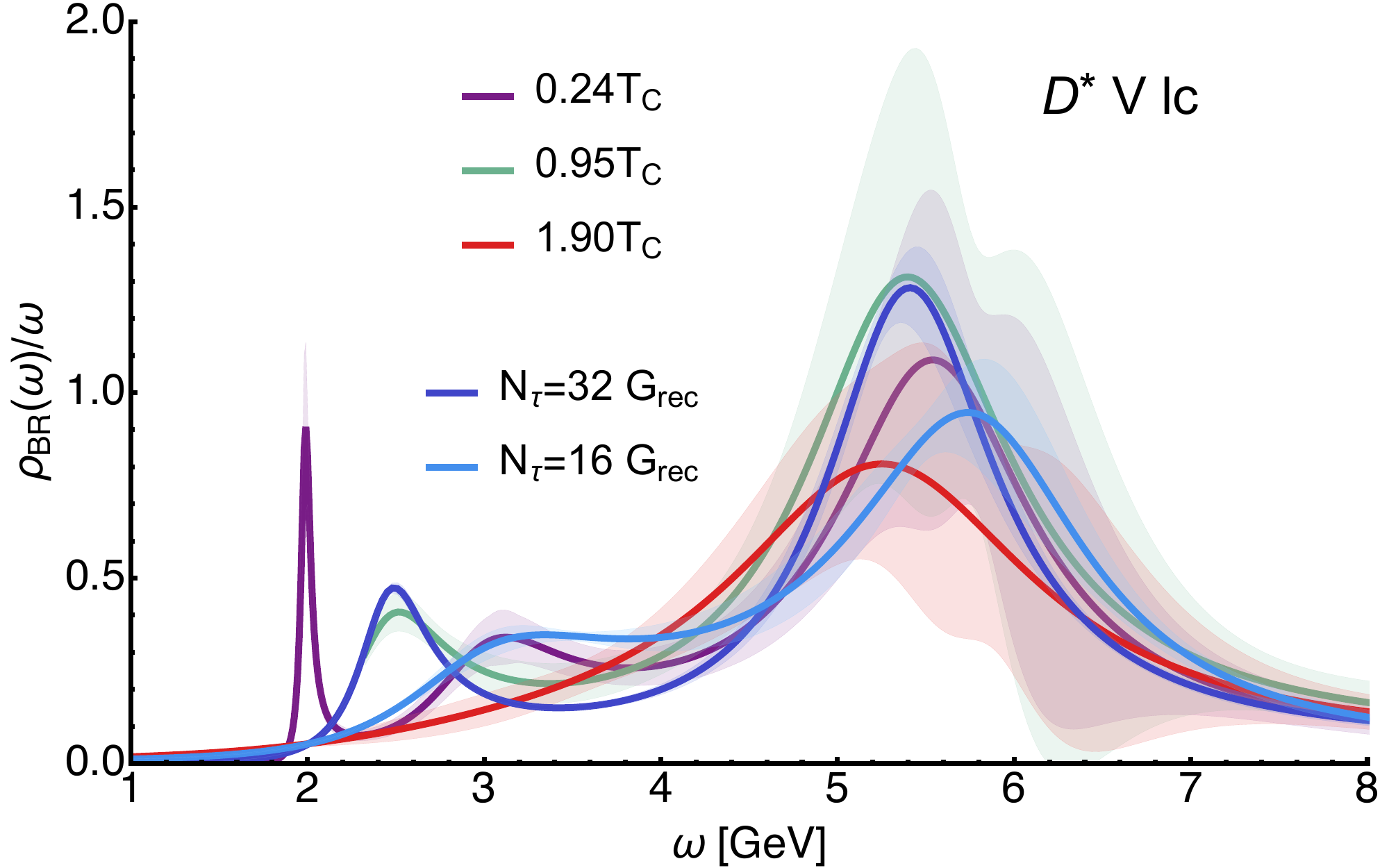}
\includegraphics[scale=0.35]{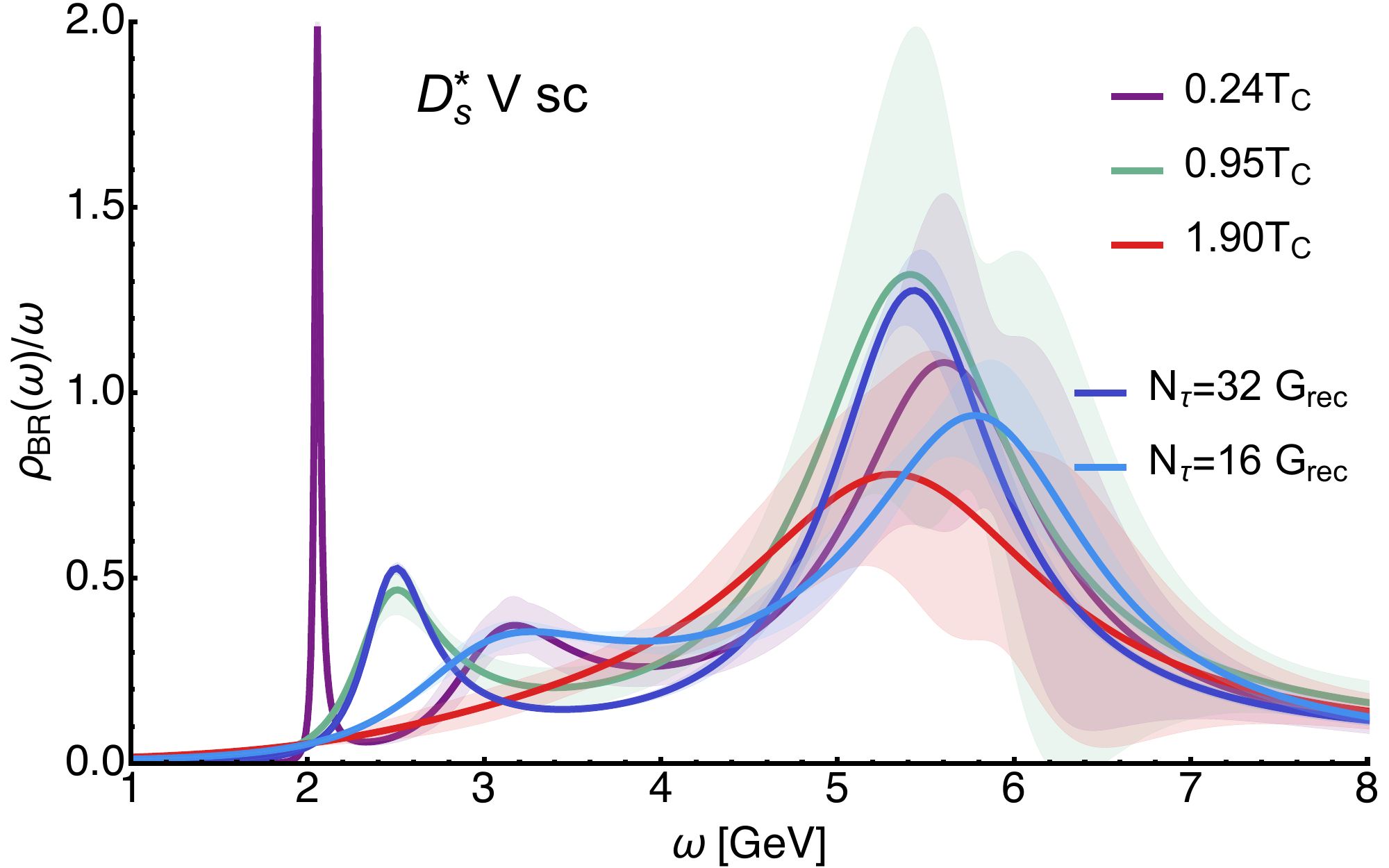}
\caption{Comparison of the BR reconstructed open-charm $T>0$ spectra with those based on the
  $T\approx0$ reconstructed correlator at the same number of
  datapoints.  The upper panels show the pseudoscalar channel; the
  lower panels are the vector channel. The $lc$ channels are on the
  left, and the $sc$ channels on the right.  All reconstructions are obtained using
  $\tau\in[2,N_\tau/2-1]$. At $0.95T_c$ no significant in-medium
  modification is observed, while at $T=1.9T_c$ there is no evidence
  of any surviving bound state. }\label{Fig:BR_lcsc_V-PS-Grec}
\end{figure*}

\begin{figure*}
\includegraphics[scale=0.35]{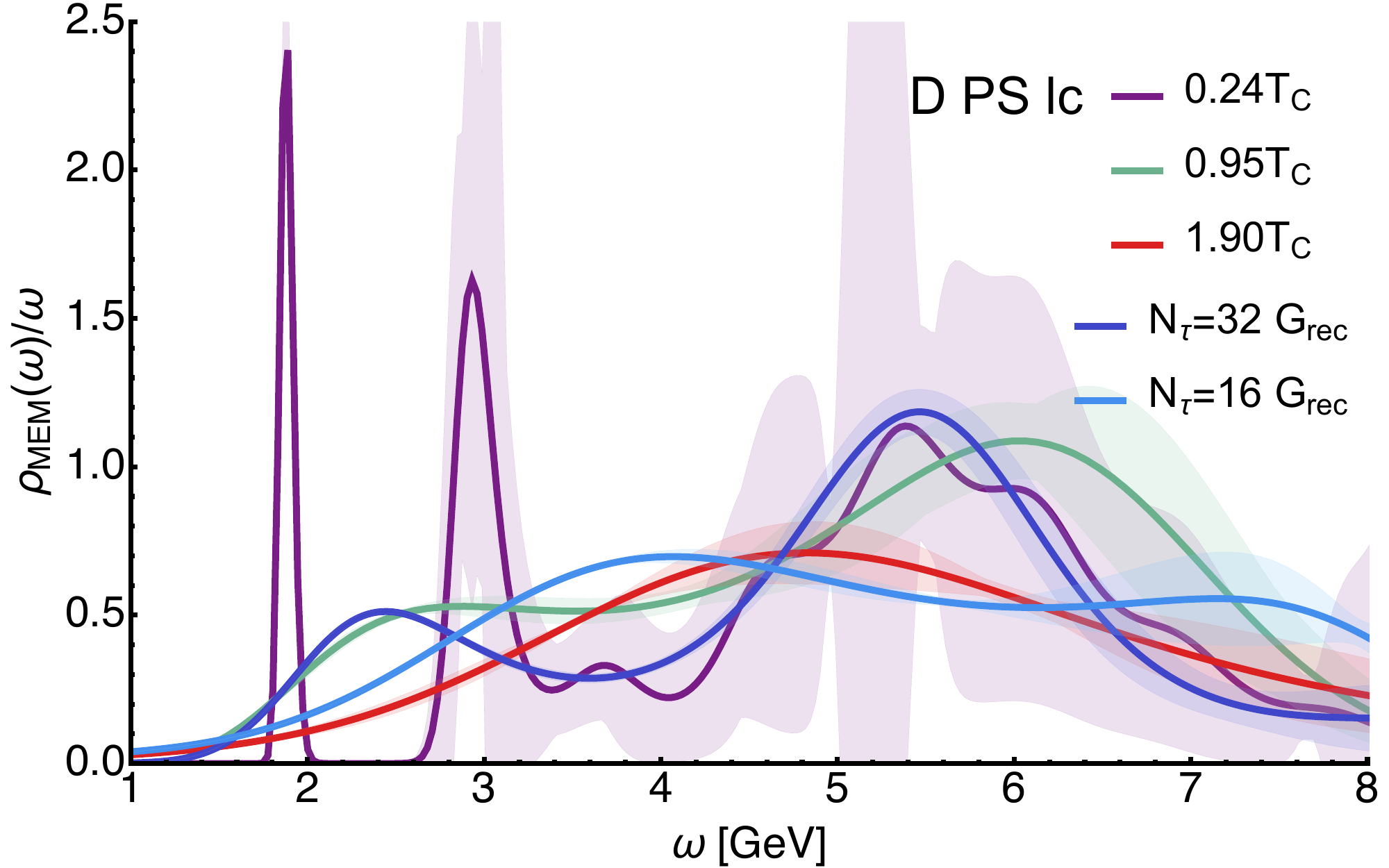}
\includegraphics[scale=0.35]{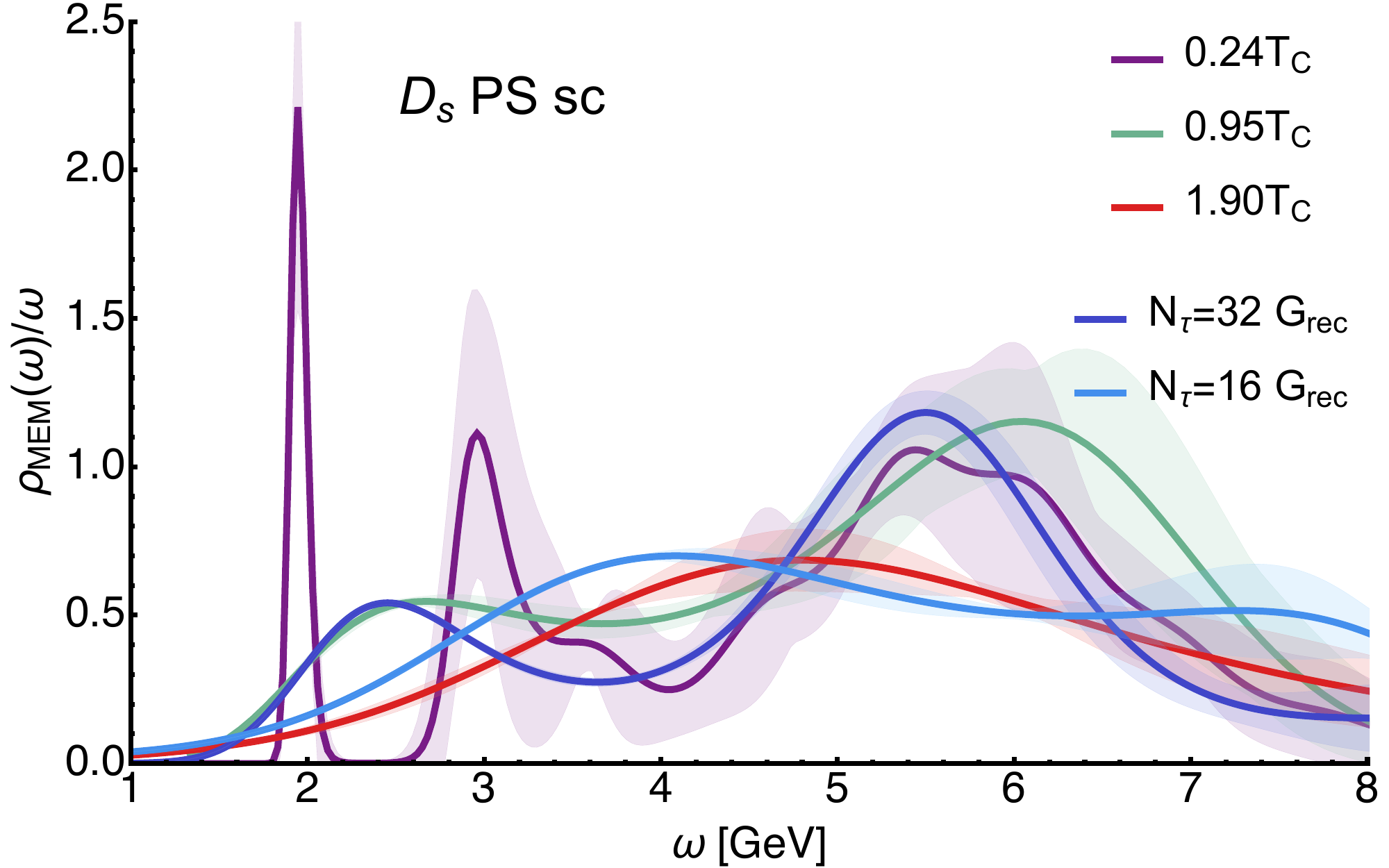}
\includegraphics[scale=0.35]{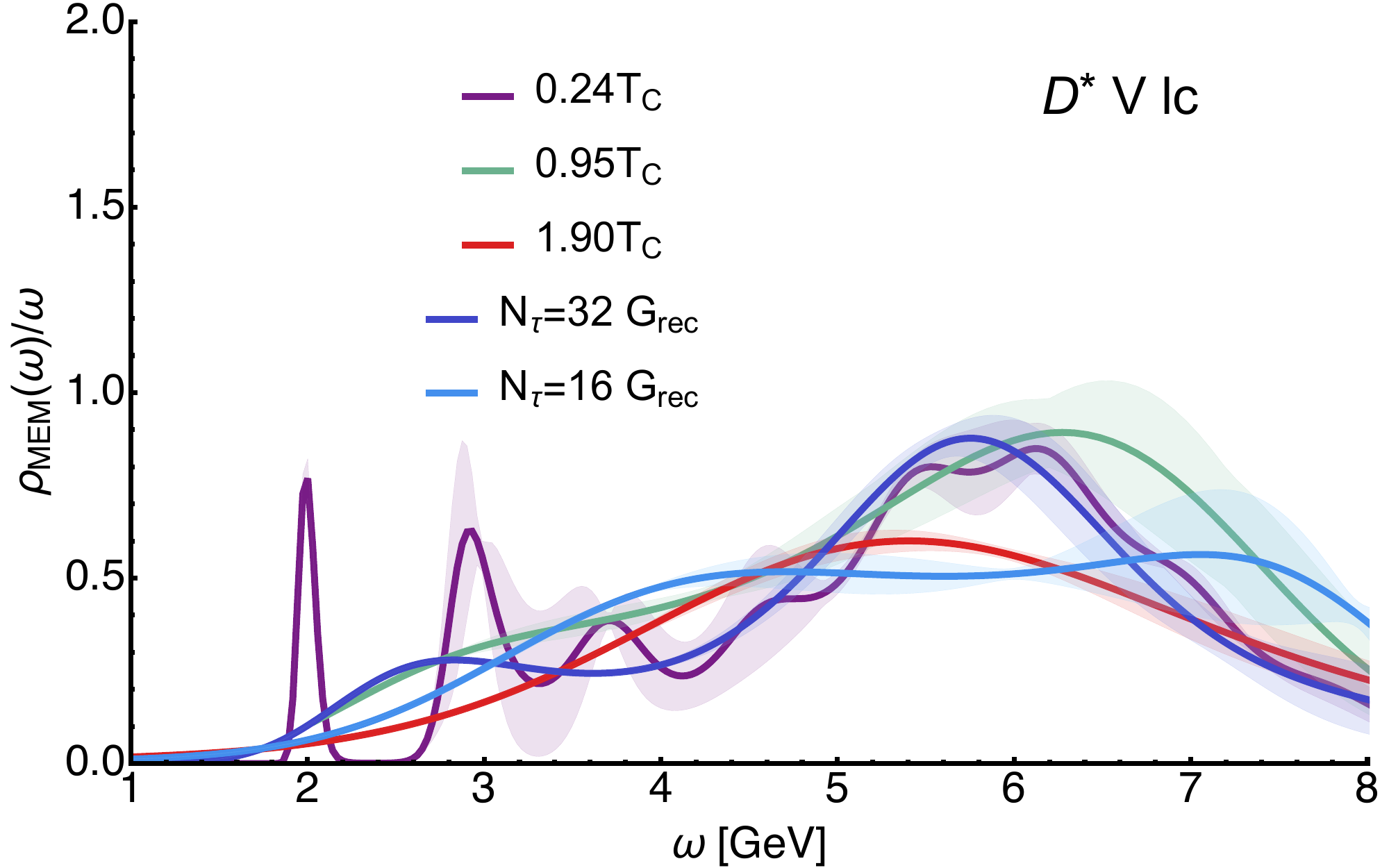}
\includegraphics[scale=0.35]{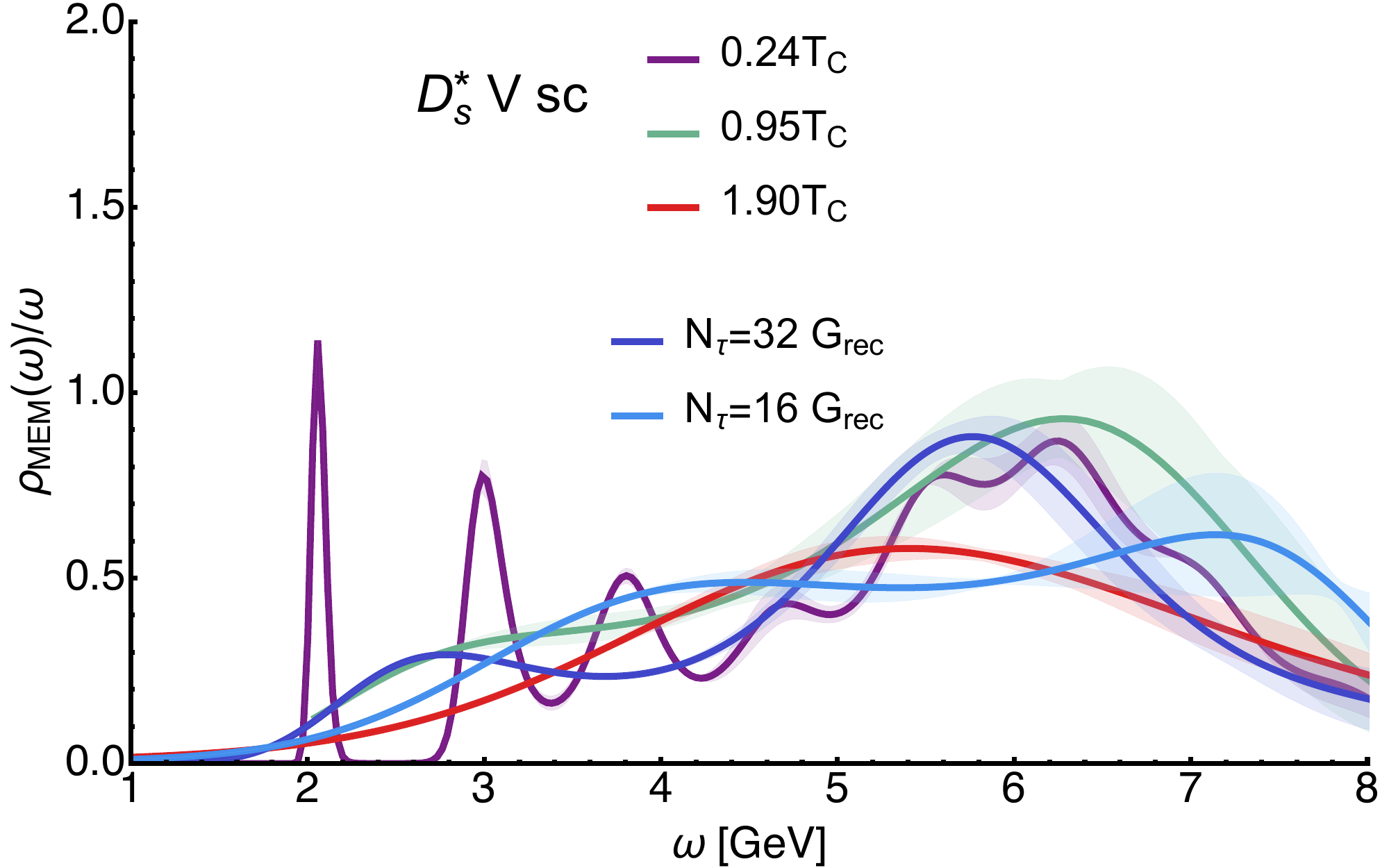}
\caption{As Fig.~\ref{Fig:BR_lcsc_V-PS-Grec}, using the MEM reconstruction. At $0.95T_c$ no significant in-medium
  modification is observed, while at $T=1.9T_c$ there is no evidence
  of any surviving bound state. }\label{Fig:BR_lcsc_V-PS-GrecMEM}
\end{figure*}

The mandatory comparison with the spectra coming from the
corresponding reconstructed correlators, shown in
Figs.~\ref{Fig:BR_lcsc_V-PS-Grec}, \ref{Fig:BR_lcsc_V-PS-GrecMEM},
tells us that in agreement with the correlator ratios, up to $T_c$ no significant modification of the ground state occurs,
while at $1.9T_c$ neither $(u,d)+c$ nor $s+c$ channels show any sign of a ground state remnant.
The currently available data quality however does not yet allow us to distinguish whether 
also in terms of the ground state spectral peak $D^*$ mesons are more susceptible to in-medium
effects than the $D$'s.  Within our current accuracy, no significant
difference is found between the spectral functions of $D$ and $D_s$ mesons.

\section{Summary and Conclusion}
\label{summary}

We have presented a combined analysis of charmonium and open-charm meson properties in 
a thermal medium based on fully relativistic lattice discretized heavy quarks. On the second generation 
FASTSUM ensembles we scrutinized the
ratios of the in-medium correlators to the so called reconstructed correlators, as well as the corresponding
in-medium spectral functions obtained from both the BR and the Maximum Entropy methods. Method
artifacts related to the diminishing of the available number of correlator points in Euclidean time at higher
temperatures were crosschecked by reconstructing at the same time the low temperature spectrum
from the reconstructed correlator itself.

For the charmonium S-wave particles $\eta_c$ and $J/\Psi$ we found that the correlator ratios show
a qualitatively and quantitatively very similar behavior to that observed in studies deploying non-relativistic 
QCD. At temperatures above $T_c$, where the in-medium modification is statistically 
significant, an upward bending occurs with strength increasing with temperature. One exception is the
$\eta_c$ channel at $T=1.9T_c$, where compared to $T=1.5T_c$ a significant change appears, the ratio
moves from bending above unity to below unity. Interestingly, only at
this high temperature is the ratio consistent
with what has been observed in previous studies of relativistic charmonium, where the upward bending 
in the $\eta_c$ channel was absent. The other difference to previous relativistic charmonium studies lies in the
strength of the vector channel deviations from unity, which in our case are much milder than those observed previously.

The $\chi_{c0}$ and $\chi_{c1}$ channel ratios show a much stronger in-medium modification. Even though
the absolute errors in these channels are much larger than for the S-waves, the deviation from unity is
easily discernible. The strong upward bend is consistent with previous findings in relativistic studies
but much larger than the corresponding one observed in non-relativistic QCD.

The Bayesian spectral reconstructions show qualitatively similar results: peaks seem to shift in frequency
and broaden with increasing temperature. In general, the BR method shows stronger peaked features 
than the MEM, which is understood to arise from the restricted search space of the latter and the susceptibility of the
former to numerical ringing. To distill genuine in-medium effects, we compare the in-medium spectra to the
ones obtained from the reconstructed correlator. Here also the BR method and MEM give consistent results,
showing that around $T=T_c$ no in-medium modification is visible in the S-wave, while significant weakening 
of the peak structure occurs for the P-waves. At $T=1.9T_c$ also in the S-wave channel a weakening of the
peak structure can be observed.

For the $D$ and $D^*$ mesons we carry out the same analysis starting with the correlators. This is the first time that
instead of spatial correlation functions, the actual Euclidean time correlators are considered, which are related to 
the in-medium spectral function via a single integral. Already below $T_c$ significant in-medium modifications are observed, which instead
of manifesting themselves in an upward bend, take the form of a deeper and deeper trough. The absence of a densely populated
excited states regime is proposed as reason for the difference to the charmonium ratios. Above $T_c$ the $D$ and $D^*$
mesons start  to show strong differences. The former continue monotonically to deepen the trough, while the latter 
show a sudden rise of the ratios, which hints at the $D^*$ ground state peak starting to be affected earlier than for the $D$.

The spectral reconstructions, while being consistent with the behavior observed in the correlator ratios below $T_c$, are not
yet precise enough to distinguish any differences between the modification of the $D$ and $D^*$ ground state. In all considered channels
there is no significant ground state modification visible at $T_c$, while at $1.9T_c$ no more structure appears to remain.

In terms of phenomenological relevance the open charm results provide two main insights: We find that even though the 
$D$ and $D^*$ mesons only differ by around 100MeV in vacuum binding energy, the latter show a much stronger in-medium
modification above $T_c$ in the correlator ratio than the former. Taking intuition from potential model computations
of charmonium, such behavior would translate into stronger suppression for the $D^*$. On the other hand none of the
currently available open charm data are able to resolve any differences between the different  $D$ and $D^*$
nuclear modification factors. The newest preliminary results from Run 2 by ALICE may contain a (not yet significant) hint 
towards stronger $D^*$ suppression at very low $p_T$ though. Further experimental efforts into reducing the uncertainty of the
measured $R_{AA}$ in $Pb+Pb$ collisions are thus very welcome.

The second insight is related to the systematically smaller medium
modifications observed in the $D_s$ compared to the $D$
correlators. Consistent with intuition, the system with the heavier of the
light quarks appears more strongly bound and thus more stable.
In turn the suppression of the $D_s$ in a purely thermal setting is expected to be weaker than that for the $D$. In the new Run 2 
results from ALICE this ordering of suppression among different heavy-flavor mesons with and without s-quarks is 
clearly observed in the corresponding $R_{AA}$.

In order to investigate in more detail the different behavior of individual in-medium $D$ and $D^*$ meson states above $T_c$, 
hinted at by their correlator ratios, we will need to significantly improve the robustness of the spectral reconstruction. 
Before the arrival of the third generation of FASTSUM ensembles, which will feature twice the number of Euclidean datapoints, 
we are currently working on increasing the statistics of the meson correlator measurements on the second generation ensembles, 
which at the same time will help to further constrain validity of a disappearance of the
P-wave charmonium ground state peak at around $T_c$.

This study has been carried out at a single lattice spacing and with
relatively heavy up and down quarks.  The quark mass effects will be
investigated in the near future when a new ensemble with
$m_\pi\approx230$~MeV becomes available.  This is not expected to have
a substantial impact on the charmonium sector, but may allow us to
determine with more certainty whether there is a difference in the
behaviour of $D$ and $D_s$ mesons.  The relatively large lattice
spatial lattice spacing together with the use of a tree-level spatial
clover coefficient was found in \cite{Liu:2012ze,Moir:2013ub} to result in an
underestimate of the S-wave hyperfine splitting in both charmonium and
open-charm systems of 20-40~MeV, which is beyond the precision of the
current study.  No significant volume dependence was found at zero
temperature in these studies, but finite volume effects will in the
future be studied directly by including results also for a larger
volume of $\sim(4\,\mathrm{fm})^3$.  In addition to this, it is
planned to improve on the current study by halving the temporal
lattice spacing, hence reducing lattice artefacts while doubling the
number of available data points.

\acknowledgments
A.R. thanks S. Masciocchi for advice on the experimental literature on
D-mesons and acknowledges support by the DFG Collaborative Research
Centre ``SFB 1225 (ISOQUANT)''.  A.K. was supported by the Irish
Research Council.  J.I.S. acknowledges support from SFI grant
08-RFP-PHY1462 during the early stages of this research.  This work
has been carried out using computational resources provided by the
Irish Centre for High End Computing and the STFC funded DiRAC
facility.


\begin{thebibliography}{0}

\bibitem{Andronic:2015wma} 
  A.~Andronic {\it et al.},
  Eur.\ Phys.\ J.\ C {\bf 76}, no. 3, 107 (2016)
  doi:10.1140/epjc/s10052-015-3819-5
  [arXiv:1506.03981 [nucl-ex]].
  
\bibitem{Aarts:2016hap} 
  G.~Aarts {\it et al.},
  Eur.\ Phys.\ J.\ A {\bf 53}, no. 5, 93 (2017)
  doi:10.1140/epja/i2017-12282-9
  [arXiv:1612.08032 [nucl-th]].

\bibitem{Prino:2016cni} 
  F.~Prino and R.~Rapp,
  J.\ Phys.\ G {\bf 43}, no. 9, 093002 (2016)
  doi:10.1088/0954-3899/43/9/093002
  [arXiv:1603.00529 [nucl-ex]].

\bibitem{Uphoff:2012gb} 
  J.~Uphoff, O.~Fochler, Z.~Xu and C.~Greiner,
  Phys.\ Lett.\ B {\bf 717}, 430 (2012)
  doi:10.1016/j.physletb.2012.09.069
  [arXiv:1205.4945 [hep-ph]].

\bibitem{Xu:2017obm} 
  Y.~Xu, S.~Cao, M.~Nahrgang, J.~E.~Bernhard and S.~A.~Bass,
  arXiv:1710.00807 [nucl-th].

\bibitem{Ozvenchuk:2014rpa} 
  V.~Ozvenchuk, J.~M.~Torres-Rincon, P.~B.~Gossiaux, L.~Tolos and J.~Aichelin,
  Phys.\ Rev.\ C {\bf 90}, 054909 (2014)
  doi:10.1103/PhysRevC.90.054909
  [arXiv:1408.4938 [hep-ph]].

\bibitem{CaronHuot:2007gq} 
  S.~Caron-Huot and G.~D.~Moore,
  Phys.\ Rev.\ Lett.\  {\bf 100}, 052301 (2008)
  doi:10.1103/PhysRevLett.100.052301
  [arXiv:0708.4232 [hep-ph]].

\bibitem{Alberico:2013bza} 
  W.~M.~Alberico, A.~Beraudo, A.~De Pace, A.~Molinari, M.~Monteno, M.~Nardi, F.~Prino and M.~Sitta,
  Eur.\ Phys.\ J.\ C {\bf 73}, 2481 (2013)
  doi:10.1140/epjc/s10052-013-2481-z
  [arXiv:1305.7421 [hep-ph]].

\bibitem{Kovtun:2003wp} 
  P.~Kovtun, D.~T.~Son and A.~O.~Starinets,
  JHEP {\bf 0310}, 064 (2003)
  doi:10.1088/1126-6708/2003/10/064
  [hep-th/0309213].

\bibitem{Horowitz:2012cf} 
  W.~A.~Horowitz,
  Nucl.\ Phys.\ A {\bf 904-905}, 186c (2013)
  doi:10.1016/j.nuclphysa.2013.01.061
  [arXiv:1210.8330 [nucl-th]].

\bibitem{Francis:2015daa} 
  A.~Francis, O.~Kaczmarek, M.~Laine, T.~Neuhaus and H.~Ohno,
  Phys.\ Rev.\ D {\bf 92}, no. 11, 116003 (2015)
  doi:10.1103/PhysRevD.92.116003
  [arXiv:1508.04543 [hep-lat]].
  
\bibitem{Djordjevic:2013xoa} 
  M.~Djordjevic and M.~Djordjevic,
  Phys.\ Lett.\ B {\bf 734}, 286 (2014)
  doi:10.1016/j.physletb.2014.05.053
  [arXiv:1307.4098 [hep-ph]].
  
\bibitem{Riek:2010py} 
  F.~Riek and R.~Rapp,
  New J.\ Phys.\  {\bf 13}, 045007 (2011)
  doi:10.1088/1367-2630/13/4/045007
  [arXiv:1012.0019 [nucl-th]].

\bibitem{Song:2015sfa} 
  T.~Song, H.~Berrehrah, D.~Cabrera, J.~M.~Torres-Rincon, L.~Tolos, W.~Cassing and E.~Bratkovskaya,
  Phys.\ Rev.\ C {\bf 92}, no. 1, 014910 (2015)
  doi:10.1103/PhysRevC.92.014910
  [arXiv:1503.03039 [nucl-th]].

\bibitem{Brambilla:2004jw} 
  N.~Brambilla, A.~Pineda, J.~Soto and A.~Vairo,
  Rev.\ Mod.\ Phys.\  {\bf 77}, 1423 (2005) [hep-ph/0410047].

\bibitem{Laine:2006ns}
  M.~Laine, O.~Philipsen, P.~Romatschke and M.~Tassler,
  JHEP {\bf 0703}, 054 (2007)

\bibitem{Brambilla:2008cx} 
  N.~Brambilla, J.~Ghiglieri, A.~Vairo, P.~Petreczky,
  Phys.\ Rev.\ D {\bf 78}, 014017 (2008) [arXiv:0804.0993 [hep-ph]].


\bibitem{Andronic:2008gm} 
  A.~Andronic, P.~Braun-Munzinger, K.~Redlich and J.~Stachel,
  J.\ Phys.\ G {\bf 35}, 104155 (2008)
  doi:10.1088/0954-3899/35/10/104155
  [arXiv:0805.4781 [nucl-th]].

\bibitem{Fuchs:2004fh} 
  C.~Fuchs, B.~V.~Martemyanov, A.~Faessler and M.~I.~Krivoruchenko,
  Phys.\ Rev.\ C {\bf 73}, 035204 (2006)
  doi:10.1103/PhysRevC.73.035204
  [nucl-th/0410065].

\bibitem{Abelev:2013lca} 
  B.~Abelev {\it et al.} [ALICE Collaboration],
  Phys.\ Rev.\ Lett.\  {\bf 111}, 102301 (2013)
  doi:10.1103/PhysRevLett.111.102301
  [arXiv:1305.2707 [nucl-ex]].

\bibitem{ALICE:2013xna} 
  E.~Abbas {\it et al.} [ALICE Collaboration],
  Phys.\ Rev.\ Lett.\  {\bf 111}, 162301 (2013)
  doi:10.1103/PhysRevLett.111.162301
  [arXiv:1303.5880 [nucl-ex]].
  
\bibitem{Acharya:2017tgv} 
  S.~Acharya {\it et al.} [ALICE Collaboration],
  arXiv:1709.05260 [nucl-ex].

\bibitem{Abelev:2014hha} 
  B.~B.~Abelev {\it et al.} [ALICE Collaboration],
  Phys.\ Rev.\ Lett.\  {\bf 113}, no. 23, 232301 (2014)
  doi:10.1103/PhysRevLett.113.232301
  [arXiv:1405.3452 [nucl-ex]].

\bibitem{De:2016wmf} 
  S.~De [ALICE Collaboration],
  J.\ Phys.\ Conf.\ Ser.\  {\bf 770}, no. 1, 012006 (2016)
  doi:10.1088/1742-6596/770/1/012006
  [arXiv:1609.02862 [nucl-ex]].
  
 
\bibitem{Burnier:2015tda} 
  Y.~Burnier, O.~Kaczmarek and A.~Rothkopf,
  JHEP {\bf 1512}, 101 (2015)
  doi:10.1007/JHEP12(2015)101
  [arXiv:1509.07366 [hep-ph]].

\bibitem{Burnier:2016kqm} 
  Y.~Burnier, O.~Kaczmarek and A.~Rothkopf,
  JHEP {\bf 1610}, 032 (2016)
  doi:10.1007/JHEP10(2016)032
  [arXiv:1606.06211 [hep-ph]].

 
\bibitem{Aarts:2010ek} 
  G.~Aarts, S.~Kim, M.~P.~Lombardo, M.~B.~Oktay, S.~M.~Ryan, D.~K.~Sinclair and J.-I.~Skullerud,
  Phys.\ Rev.\ Lett.\  {\bf 106}, 061602 (2011)
  doi:10.1103/PhysRevLett.106.061602
  [arXiv:1010.3725 [hep-lat]].

\bibitem{Aarts:2011sm} 
  G.~Aarts, C.~Allton, S.~Kim, M.~P.~Lombardo, M.~B.~Oktay, S.~M.~Ryan, D.~K.~Sinclair and J.~I.~Skullerud,
  JHEP {\bf 1111}, 103 (2011)
  doi:10.1007/JHEP11(2011)103
  [arXiv:1109.4496 [hep-lat]].

\bibitem{Aarts:2012ka} 
  G.~Aarts, C.~Allton, S.~Kim, M.~P.~Lombardo, M.~B.~Oktay, S.~M.~Ryan, D.~K.~Sinclair and J.~I.~Skullerud,
  JHEP {\bf 1303}, 084 (2013)
  doi:10.1007/JHEP03(2013)084
  [arXiv:1210.2903 [hep-lat]].

\bibitem{Aarts:2013kaa} 
  G.~Aarts, C.~Allton, S.~Kim, M.~P.~Lombardo, S.~M.~Ryan and J.-I.~Skullerud,
  JHEP {\bf 1312}, 064 (2013)
  doi:10.1007/JHEP12(2013)064
  [arXiv:1310.5467 [hep-lat]].

\bibitem{Aarts:2014cda} 
  G.~Aarts, C.~Allton, T.~Harris, S.~Kim, M.~P.~Lombardo, S.~M.~Ryan and J.~I.~Skullerud,
  JHEP {\bf 1407}, 097 (2014)
  doi:10.1007/JHEP07(2014)097
  [arXiv:1402.6210 [hep-lat]].

\bibitem{Kim:2014iga} 
  S.~Kim, P.~Petreczky and A.~Rothkopf,
  Phys.\ Rev.\ D {\bf 91}, 054511 (2015)
  doi:10.1103/PhysRevD.91.054511
  [arXiv:1409.3630 [hep-lat]].

\bibitem{Kim:2015rdi} 
  S.~Kim, P.~Petreczky and A.~Rothkopf,
  Nucl.\ Phys.\ A {\bf 956}, 713 (2016)
  doi:10.1016/j.nuclphysa.2015.12.011
  [arXiv:1512.05289 [hep-lat]].

\bibitem{Kim:2017aio} 
  S.~Kim, P.~Petreczky and A.~Rothkopf,
  Nucl.\ Phys.\ A {\bf 967}, 724 (2017)
  doi:10.1016/j.nuclphysa.2017.04.010
  [arXiv:1704.05221 [hep-lat]].

\bibitem{Umeda:2007hy} 
  T.~Umeda,
  Phys.\ Rev.\ D {\bf 75}, 094502 (2007)
  doi:10.1103/PhysRevD.75.094502
  [hep-lat/0701005].

\bibitem{Burnier:2014ssa} 
  Y.~Burnier, O.~Kaczmarek and A.~Rothkopf,
  Phys.\ Rev.\ Lett.\  {\bf 114}, no. 8, 082001 (2015)
  doi:10.1103/PhysRevLett.114.082001
  [arXiv:1410.2546 [hep-lat]].

\bibitem{Ding:2017rty} 
  H.~T.~Ding, O.~Kaczmarek, A.~L.~Kruse, H.~Ohno and H.~Sandmeyer,
  arXiv:1710.08858 [hep-lat].

\bibitem{Shu:2015tva} 
  H.~T.~Shu, H.~T.~Ding, O.~Kaczmarek, S.~Mukherjee and H.~Ohno,
  PoS LATTICE {\bf 2015}, 180 (2016)
  [arXiv:1510.02901 [hep-lat]].

\bibitem{Ikeda:2016czj} 
  A.~Ikeda, M.~Asakawa and M.~Kitazawa,
  Phys.\ Rev.\ D {\bf 95}, no. 1, 014504 (2017)
  doi:10.1103/PhysRevD.95.014504
  [arXiv:1610.07787 [hep-lat]].

\bibitem{Asakawa:2003re} 
  M.~Asakawa and T.~Hatsuda,
  Phys.\ Rev.\ Lett.\  {\bf 92}, 012001 (2004)
  doi:10.1103/PhysRevLett.92.012001
  [hep-lat/0308034].

\bibitem{Ding:2012sp} 
  H.~T.~Ding, A.~Francis, O.~Kaczmarek, F.~Karsch, H.~Satz and W.~Soeldner,
  Phys.\ Rev.\ D {\bf 86}, 014509 (2012)
  doi:10.1103/PhysRevD.86.014509
  [arXiv:1204.4945 [hep-lat]].

\bibitem{Aarts:2007pk} 
  G.~Aarts, C.~Allton, M.~B.~Oktay, M.~Peardon and J.~I.~Skullerud,
  Phys.\ Rev.\ D {\bf 76}, 094513 (2007)
  doi:10.1103/PhysRevD.76.094513
  [arXiv:0705.2198 [hep-lat]].

\bibitem{Borsanyi:2014vka} 
  S.~Borsanyi {\it et al.},
  JHEP {\bf 1404}, 132 (2014)
  doi:10.1007/JHEP04(2014)132
  [arXiv:1401.5940 [hep-lat]].

\bibitem{Bazavov:2014yba} 
  A.~Bazavov {\it et al.},
  Phys.\ Lett.\ B {\bf 737}, 210 (2014)
  doi:10.1016/j.physletb.2014.08.034
  [arXiv:1404.4043 [hep-lat]].

\bibitem{Bazavov:2014cta} 
  A.~Bazavov, F.~Karsch, Y.~Maezawa, S.~Mukherjee and P.~Petreczky,
  Phys.\ Rev.\ D {\bf 91}, no. 5, 054503 (2015)
  doi:10.1103/PhysRevD.91.054503
  [arXiv:1411.3018 [hep-lat]].

\bibitem{Mukherjee:2015mxc} 
  S.~Mukherjee, P.~Petreczky and S.~Sharma,
  Phys.\ Rev.\ D {\bf 93}, no. 1, 014502 (2016)
  doi:10.1103/PhysRevD.93.014502
  [arXiv:1509.08887 [hep-lat]].

\bibitem{Kelly:2016apt} 
  A.~Kelly and J.~I.~Skullerud,
  PoS LATTICE {\bf 2016}, 082 (2016).

\bibitem{Kelly:2017opi} 
  A.~Kelly and J.~I.~Skullerud,
  EPJ Web Conf.\  {\bf 137}, 07025 (2017)
  doi:10.1051/epjconf/201713707025
  [arXiv:1701.09005 [hep-lat]].
  
  
\bibitem{Hilger:2008jg} 
  T.~Hilger, R.~Thomas and B.~Kampfer,
  Phys.\ Rev.\ C {\bf 79}, 025202 (2009)
  doi:10.1103/PhysRevC.79.025202
  [arXiv:0809.4996 [nucl-th]].

\bibitem{Wang:2015uya} 
  Z.~G.~Wang,
  Phys.\ Rev.\ C {\bf 92}, no. 6, 065205 (2015)
  doi:10.1103/PhysRevC.92.065205
  [arXiv:1501.05093 [hep-ph]].

\bibitem{Suzuki:2015est} 
  K.~Suzuki, P.~Gubler and M.~Oka,
  Phys.\ Rev.\ C {\bf 93}, no. 4, 045209 (2016)
  doi:10.1103/PhysRevC.93.045209
  [arXiv:1511.04513 [hep-ph]].
  
 
\bibitem{Aarts:2014nba} 
  G.~Aarts, C.~Allton, A.~Amato, P.~Giudice, S.~Hands and J.~I.~Skullerud,
  JHEP {\bf 1502}, 186 (2015)
  doi:10.1007/JHEP02(2015)186
  [arXiv:1412.6411 [hep-lat]].

\bibitem{Edwards:2008ja} 
  R.~G.~Edwards, B.~Joo and H.~W.~Lin,
  Phys.\ Rev.\ D {\bf 78}, 054501 (2008)
  doi:10.1103/PhysRevD.78.054501
  [arXiv:0803.3960 [hep-lat]].

\bibitem{Liu:2012ze} 
  L.~Liu {\it et al.} [Hadron Spectrum Collaboration],
  JHEP {\bf 1207}, 126 (2012)
  doi:10.1007/JHEP07(2012)126
  [arXiv:1204.5425 [hep-ph]].

\bibitem{Moir:2013ub} 
  G.~Moir, M.~Peardon, S.~M.~Ryan, C.~E.~Thomas and L.~Liu,
  JHEP {\bf 1305}, 021 (2013)
  doi:10.1007/JHEP05(2013)021
  [arXiv:1301.7670 [hep-ph]].

\bibitem{Edwards:2004sx} 
  R.~G.~Edwards {\it et al.} [SciDAC and LHPC and UKQCD Collaborations],
  Nucl.\ Phys.\ Proc.\ Suppl.\  {\bf 140}, 832 (2005)
  doi:10.1016/j.nuclphysbps.2004.11.254
  [hep-lat/0409003].

\bibitem{Jarrell:1996}  
  J.~Skilling, S.F.~Gull, 
  Lecture Notes-Monograph Series 20, 341 (1991);
  M.~Jarrell and J.E.~Gubernatis, 
  Physics Reports, {\bf 269}, 133-195, (1996);

\bibitem{Bishop:2007}
C.M.~Bishop,
Pattern Recognition and Machine Learning,
Springer, 2nd ed. (2007).

\bibitem{Asakawa:2000tr} 
  M.~Asakawa, T.~Hatsuda and Y.~Nakahara,
  Prog.\ Part.\ Nucl.\ Phys.\  {\bf 46}, 459 (2001).
\bibitem{Jakovac:2006sf} 
  A.~Jakovac, P.~Petreczky, K.~Petrov and A.~Velytsky,
  Phys.\ Rev.\ D {\bf 75}, 014506 (2007).
\bibitem{Nickel:2006mm} 
  D.~Nickel,
  Annals Phys.\  {\bf 322}, 1949 (2007).
\bibitem{Rothkopf:2011}  
  A.~Rothkopf,
  J.\ Comput.\ Phys.\  {\bf 238}, 106 (2013).

\bibitem{Burnier:2013nla} 
  Y.~Burnier and A.~Rothkopf,
  Phys.\ Rev.\ Lett.\  {\bf 111}, 182003 (2013).

\cite{Bryan}
\bibitem{Bryan}
R.~K.~.Bryan, Eur.Bioph.J. {\bf 18} 165 (1990)

\bibitem{Rothkopf:2012vv} 
  A.~Rothkopf,
  PoS LATTICE {\bf 2012}, 100 (2012)
  [arXiv:1208.5162 [physics.comp-ph]].

\bibitem{Skullerud:2014xx}
  J.-I.~Skullerud, talk at \emph{XI Quark Confinement and the Hadron
    Spectrum}, St.\ Petersburg, Russia, 8--12 September 2014.

\bibitem{Adamczyk:2014uip} 
  L.~Adamczyk {\it et al.} [STAR Collaboration],
  Phys.\ Rev.\ Lett.\  {\bf 113}, no. 14, 142301 (2014)
  doi:10.1103/PhysRevLett.113.142301
  [arXiv:1404.6185 [nucl-ex]].
  
\bibitem{Sirunyan:2017xss} 
  A.~M.~Sirunyan {\it et al.} [CMS Collaboration],
  arXiv:1708.04962 [nucl-ex].

\bibitem{Adam:2015nna} 
  J.~Adam {\it et al.} [ALICE Collaboration],
  JHEP {\bf 1511}, 205 (2015)
  Addendum: [JHEP {\bf 1706}, 032 (2017)]
  doi:10.1007/JHEP11(2015)205, 10.1007/JHEP06(2017)032
  [arXiv:1506.06604 [nucl-ex]].
  
\bibitem{Adam:2015sza} 
  J.~Adam {\it et al.} [ALICE Collaboration],
  JHEP {\bf 1603}, 081 (2016)
  doi:10.1007/JHEP03(2016)081
  [arXiv:1509.06888 [nucl-ex]].
  
\bibitem{AlicePub1703}
[ALICE Collaboration],
ALICE-PUBLIC-2017-003
\url{https://cds.cern.ch/record/2265109}
%





\end{thebibliography}
\end{document}